%% file: mainfile.tex
\documentclass{article}[11pt]
\usepackage[margin=1in]{geometry}
\usepackage{amssymb,amsmath,amsthm}
\usepackage{amsfonts}
\usepackage{dsfont}
\usepackage{verbatim,hyperref}

\usepackage{algorithm}
\usepackage[noend]{algpseudocode}

\usepackage{color}

\newtheorem{lemma}{Lemma}
\newtheorem{claim}{Claim}
\newtheorem{theorem}{Theorem}
\newtheorem{proposition}{Proposition}

\newtheorem{corollary}{Corollary}
\newtheorem{fact}{Fact}
\newtheorem{remark}{Remark}

{\makeatletter
	\gdef\xxxmark{%
		\expandafter\ifx\csname @mpargs\endcsname\relax 
		\expandafter\ifx\csname @captype\endcsname\relax 
		\marginpar{xxx}
		\else
		xxx 
		\fi
		\else
		xxx 
		\fi}
	\gdef\xxx{\@ifnextchar[\xxx@lab\xxx@nolab}
	\long\gdef\xxx@lab[#1]#2{{\bf [\xxxmark #2 ---{\sc #1}]}}
	\long\gdef\xxx@nolab#1{{\bf [\xxxmark #1]}}
}

\newcommand{\e}{\epsilon}
\newcommand{\poly}{\text{poly}}
\newcommand{\partitiontesting}{{\bf PartitionTesting }}
\newcommand{\clusterability}{{\bf Clusterability}}
\newcommand{\noisyparities}{{\bf NoisyParities }}

\theoremstyle{definition}
\newtheorem{definition}{Definition}

\title{Testing Graph Clusterability: Algorithms and Lower Bounds}
\author{Ashish Chiplunkar\\EPFL \\ashish.chiplunkar@gmail.com \and Michael Kapralov\\EPFL \\michael.kapralov@epfl.ch \and Sanjeev Khanna\\ University of Pennsylvania \\sanjeev@cis.upenn.edu \and Aida Mousavifar\\EPFL \\ aidasadat.mousavifar@epfl.ch \and Yuval Peres \\ Microsoft Research Redmond\\peres@microsoft.com}
\date{}

\begin{document}

\maketitle

\begin{abstract}
We consider the problem of testing graph cluster structure: given access to a graph $G=(V, E)$, can we quickly determine whether the graph can be partitioned into a few clusters with good inner conductance, or is far from any such graph? This is a generalization of the well-studied problem of testing graph expansion, where one wants to distinguish between the graph having good expansion (i.e.\ being a good single cluster) and the graph having a sparse cut (i.e.\ being a union of at least two clusters). A recent work of Czumaj, Peng, and Sohler (STOC'15) gave an ingenious sublinear time algorithm for testing $k$-clusterability in time $\tilde{O}(n^{1/2} \poly(k))$: their algorithm implicitly embeds a random sample of vertices of the graph into Euclidean space, and then clusters the samples based on estimates of  Euclidean distances between the points. This yields a very efficient testing algorithm, but only works if the cluster structure is very strong: it is necessary to assume that the gap between conductances of accepted and rejected graphs is at least  logarithmic in the size of the graph $G$. In this paper we show how one can leverage more refined geometric information, namely angles as opposed to distances, to obtain a  sublinear  time tester that works even when the gap is a sufficiently large constant. Our tester is based on the singular value decomposition of a natural matrix derived from random walk transition probabilities from a small sample of seed nodes.

We complement our algorithm with a matching lower bound on the query complexity of testing clusterability. Our lower bound is based on  a novel property testing problem, which we analyze using Fourier analytic tools. As a byproduct of our techniques, we also achieve new lower bounds for the problem of approximating MAX-CUT value in sublinear time.
\end{abstract}

\newpage

\input{intro.tex}

\input{prelims.tex}
\input{algorithm-nonuniform.tex}
\input{lowerbound.tex}

\input{decomposition.tex}
\bibliographystyle{alpha}
\bibliography{references}
\begin{appendix}
\input{app.tex}
\input{appc.tex}
\end{appendix}
\end{document}

%% file: intro.tex

\section{Introduction}\label{sec_intro}


Graph clustering is the problem of partitioning vertices of a graph based on the connectivity structure of the graph. It is a fundamental problem in many application domains where one wishes to identify groups of closely related objects, for instance, communities in a social network. The clustering problem is, thus, to partition a graph into vertex-disjoint subgraphs, namely clusters, such that each cluster contains vertices that are more similar to each other than the rest of the graph. 
There are many natural measures that have been proposed to assess the quality of a cluster; one particularly well-studied and well-motivated measure for graph clustering is conductance of a cluster~\cite{KannanVV_JACM04}. Roughly speaking, conductance of a graph measures the strength of connections across any partition of vertices relative to the strength of connections inside the smaller of the two parts. The higher the conductance inside a cluster, the harder it is to split it into non-trivial pieces. 
The conductance measure lends itself to a natural graph clustering objective, namely, partition the vertices of a graph into a small number of clusters such that each cluster has large conductance in the graph induced by it (the inner conductance of the cluster). Towards this objective, many efficient graph partitioning algorithms have been developed that partition vertices of a graph into a specified number of clusters with approximately high conductance (when possible). Any algorithm that outputs such a partition necessarily requires $\Omega(n)$ time -- simply to output the solution, and usually $\Omega(m)$ time, where $n$ and $m$ respectively denote the number of vertices and edges in the input graph. On very large-scale graphs, even linear-time algorithms may prove to be computationally prohibitive, and consequently, there has been considerable recent interest in understanding the cluster structure of a graph in sublinear time. Specifically, given a target number of clusters, say $k$, and a measure $\phi$ of desired cluster quality, how much exploration of the input graph is needed to distinguish between graphs that can be partitioned into at most $k$ clusters with inner conductance at least $\phi$ from graphs that are far from admitting such clustering? The focus of this paper is to understand the power of sublinear algorithms in discovering the cluster structure of a graph.

In our study, we use by now a standard model of graph exploration for sublinear algorithms, where at any step, the algorithm can either sample a uniformly at random vertex, query the degree $d(u)$ of a vertex $u$, or specify a pair $(u,i)$ and recover the $i^{{\rm th}}$ neighbor of $u$ for any $i \in [1..d(u)]$. 
For any positive $\epsilon > 0$, we say a pair of graphs is $\epsilon$-far if one needs to modify at least an $\epsilon$-fraction of edges to convert one graph into another.

The simplest form of the cluster structure problem is the case $k=1$: how many queries to the graph are needed to distinguish between graphs that are expanders (YES case) from graphs that are $\Omega(1)$-far from being expanders (NO case)?
A formal study of this basic question was initiated in the work of Goldreich and Ron~\cite{GoldreichR_Algorithmica02} where they showed that even on bounded degree graphs, $\Omega(\sqrt{n})$ queries to the input graph are necessary to distinguish between expanders and graphs that are far from expanders. On the positive side, it is known that a bounded degree expander graph with conductance at least $\phi$ can be distinguished from a graph that is $\Omega(1)$-far from a graph with conductance $\gamma \times \phi^2$ for some positive constant $\gamma$, using only $n^{\frac{1}{2} + O(\gamma)}$ queries~\cite{KaleS_SIAMJC11,NachmiasS_IC10}. Thus, even the simplest setting of the graph clustering problem is not completely understood -- the known algorithmic results require additional separation in the conductance requirements of YES and NO instances. Furthermore, even with this separation in conductance requirements, the best algorithmic result requires polynomially more queries than suggested by the lower bound. 

Lifting algorithmic results above for the case $k=1$ to larger values of $k$ turned out to be a challenging task. A breakthrough was made by Czumaj, Peng, and Sohler~\cite{CzumajPS_STOC15} who designed an algorithm that differentiates between bounded degree graphs that can be clustered into $k$ clusters with good inner conductance (YES case) from graphs that are far from such graphs (NO case), using only $\tilde{O}(n^{\frac{1}{2}} \poly(k))$ queries. This striking progress, however, required an even stronger separation between YES and NO instances of the problem. In particular, the algorithm requires that in the YES case, the graph can be partitioned into $k$ clusters with inner conductance at least $\phi$, while in the NO case, the graph is $\epsilon$-far from admitting $k$ clusters with conductance $\phi^2/\log n$. Thus the cluster quality in the NO case needs to be weakened by a factor that now depends on the size of the input graph.

The current state of the art raises several natural questions on both algorithmic and lower bound fronts. On the algorithmic front, does sublinear testing of cluster structure of a graph fundamentally require such strong separation between the cluster structures of YES and NO cases?
On the lower bound front, is there a stronger barrier than the current $\Omega(\sqrt{n})$ threshold for differentiating between the YES and NO cases? Even for the case of distinguishing an expander for a graph that is far from expander, the known algorithmic results require $n^{\frac{1}{2} + \Omega(1)}$ queries when the conductance guarantees of YES and NO cases are separated by only a constant factor. 

In this work, we make progress on both questions above. On the algorithmic side, we present a new sublinear testing algorithm that considerably weakens the separation required between the conductance of YES and NO instances. In particular, for any fixed $k$, our algorithm can distinguish between instances that can be partitioned into $k$ clusters with conductance at least $\phi$ from instances that are $\Omega(1)$-far from admitting $k$ clusters with conductance $\gamma \phi^2$, using $n^{\frac{1}{2} + O(\gamma)}$ queries. This generalizes the results of ~\cite{KaleS_SIAMJC11,NachmiasS_IC10} for $k=1$ to any fixed $k$ and arbitrary graphs. Similar to~\cite{CzumajPS_STOC15} our algorithm is based on sampling a small number of vertices and gathering information about the transition probabilities of suitably long random walks from the sampled points. However, instead of classifying points as pairwise similar or dissimilar based on $\ell_2$ similarity between the transition probability vectors, our approach is based on analyzing the structure of the Gram matrix of these transition probability vectors, which turns out to be a more robust mechanism for separating the YES and NO cases.

On the lower bound side, we show that arguably the simplest question in this setting, namely, differentiating a bounded degree expander graph with conductance $\Omega(1)$ from a graph that is $\Omega(1)$-far from a graph with conductance $\gamma$ for some positive constant $\gamma$, already requires $n^{\frac{1}{2} + \Omega(\gamma)}$ queries. This improves upon the long-standing previous lower bound of $\Omega(n^{\frac{1}{2}})$. Going past the $n^{\frac{1}{2}}$ threshold requires us to introduce new ideas to handle non-trivial dependencies that manifest due to unavoidable emergence of cycles once an $\omega(n^{\frac{1}{2}})$-sized component is uncovered in an expander. 
We use a Fourier analytic approach to handle emergence of cycles and create a distribution where $n^{\frac{1}{2} + \Omega(\gamma)}$ queries are necessary to distinguish between YES and NO cases.
We believe our lower bound techniques are of independent interest and will quite likely find applications to other problems. As one illustrative application, we show that our approach yields an $n^{\frac{1}{2} + \Omega(1)}$ query complexity lower bound for the problem of approximating the max-cut value in graph to within a factor better than $2$, improving the previous best lower bound of $\Omega(n^{\frac{1}{2}})$. 

In what follows, we formally define our clustering problem, present our main results, and give an overview of our techniques.

\subsection{Problem Statement}\label{sec_ourresults}

We start by introducing basic definitions, then proceed to define the problems that we design algorithms for (namely \partitiontesting~ and testing clusterability) in Section~\ref{sec:ub-intro}, and finally discuss the communication game that we use to derive query complexity lower bounds (namely the \noisyparities game) and state our results on lower bounds in Section~\ref{sec:lb-intro}.

\begin{definition}[\textbf{Internal and external conductance}]\label{def:conductance}
Let $G=(V_G,E_G)$ be a graph. Let $\text{deg}(v)$ be the degree of vertex $v$. For a set $S\subseteq V_G$, let $\text{vol}(S)=\sum_{v \in S} \text{deg}(v)$ denote the \textit{volume} of set $S$. For a set $S\subseteq C\subseteq V_G$, the \textit{conductance of $S$ within $C$}, denoted by $\phi^G_C(S)$, is the number of edges with one endpoint in $S$ and the other in $C\setminus S$ divided by $\text{vol}(S)$. Equivalently, $\phi^G_C(S)$ is the probability that a uniformly random neighbor, of a vertex in $S$ selected with probability proportional to degree, is in $C\setminus S$. The \textit{internal conductance} of $C$, denoted by $\phi^G(C)$, is defined to be  $\min_{S\subseteq C\text{,} 0<\text{vol}(S)\leq \frac{\text{vol}(C)}{2}}\phi^G_C(S)$ if $|C|>1$ and one otherwise. The \textit{external conductance} of $C$ is defined to be $\phi^G_{V_G}(C)$.
\end{definition}

Based on the conductance parameters, clusterability and unclusterability of graphs is defined as follows.

\begin{definition}[\textbf{Graph clusterability}]\label{def:clusterable}
Graph $G=(V_G,E_G)$ is defined to be \textit{$(k,\varphi)$-clusterable} if $V_G$ can be partitioned into $C_1,\ldots,C_h$ for some $h\leq k$ such that for all $i=1,\ldots,h$, $\phi^G(C_i)\geq\varphi$. Graph $G$ is defined to be \textit{$(k,\varphi,\beta)$-unclusterable} if $V_G$ contains $k+1$ pairwise disjoint subsets $C_1,\ldots,C_{k+1}$ such that for all 
$i=1,\ldots,k+1$, $\text{vol}(C_i)\geq\beta\cdot \frac{\text{vol}(V_G)}{k+1}$, and $\phi^G_{V_G}(C_i)\leq\varphi$.
\end{definition}

The following algorithmic problem was implicitly defined in~\cite{CzumajPS_STOC15}:
\begin{definition}\label{def_partitiontesting}
\partitiontesting$(k, \varphi_{\text{in}}, \varphi_{\text{out}},\beta)$ is the problem of distinguishing between the following two types of graphs.
\begin{enumerate}
\item The YES case: graphs which are $(k,\varphi_{\text{in}})$-clusterable
\item The NO case: graphs which are $(k,\varphi_{\text{out}},\beta)$-unclusterable
\end{enumerate}
\end{definition}

The ultimate problem that we would like to solve is the \clusterability~problem, defined below:
\begin{definition}\label{def_clusterability}
\clusterability$(k, \varphi,k',\varphi',\varepsilon)$ is the problem of distinguishing between the following two types of graphs.
\begin{enumerate}
\item The YES case: graphs which are $(k,\varphi)$-clusterable
\item The NO case: graphs which are $\varepsilon$-far from $(k',\varphi')$-clusterable.
\end{enumerate}
Here, a graph $G=(V,E)$ is $\varepsilon$-far from $(k',\varphi')$-clusterable if there does not exist a $(k',\varphi')$-clusterable graph $G'=(V,E')$ such that $|E\oplus E'|\leq\varepsilon\cdot|E|$ ($\oplus$ denotes the symmetric difference, or equivalently, the Hamming distance).
\end{definition}

Note that in the clusterability problem considered by Czumaj et al.~\cite{CzumajPS_STOC15}, the YES instances were required to have clusters with small outer conductance, whereas we have no such requirement.

\noindent\textbf{Queries and Complexity}. We assume that the algorithm has access to graph $G$ via the following queries.
\begin{enumerate}
\item Vertex query: returns a uniformly random vertex $v \in V_G$
\item Degree query: outputs degree $\text{deg}(v)$ of a given $v \in V_G$.
\item Neighbor query: given a vertex $v \in V_G$, and $i \in [n]$, returns the $i$-th neighbor of $v$ if $i \leq \text{deg}(v)$, and  returns \emph{fail} otherwise.

\end{enumerate}
The complexity of the algorithm is measured by number of access queries.

\subsection{Algorithmic Results}\label{sec:ub-intro}

\begin{theorem}\label{thm_alg_partitiontesting}
Suppose $\varphi_{\text{out}}\leq \frac{1}{480}\varphi_{\text{in}}^2$. Then there exists a randomized algorithm for \partitiontesting$(k, \varphi_{\text{in}}, \varphi_{\text{out}},\beta)$ which gives the correct answer with probability at least $2/3$, and which makes $\text{poly}(1/\varphi_{\text{in}})\cdot\text{poly}(k)\cdot\text{poly}(1/\beta)\cdot\text{poly}\log(m)\cdot m^{1/2+O(\varphi_{\text{out}}/\varphi_{\text{in}}^2)}$ queries on graphs with $m$ edges.
\end{theorem}

Observe that even when the \textit{average} degree of the vertices of the graph is constant, the dependence of query complexity on $n$, the number of vertices, is $\tilde{O}(n^{1/2+O(\varphi_{\text{out}}/\varphi_{\text{in}}^2)})$. We also note that our current analysis of the tester is probably somewhat loose: the tester likely requires no more than $\tilde{O}(n^{1/2+O(\varphi_{\text{out}}/\varphi_{\text{in}}^2)})$ for graphs of arbitrary volume (specifically, the variance bound provided by Lemma~\ref{lem_matrix_estimation} can probably be improved).

Theorem~\ref{thm_alg_partitiontesting} allows us to obtain the following result on testing clusterability, which removes the logarithmic gap assumption required for the results in~\cite{CzumajPS_arXiv15} in the property testing framework.
\begin{theorem}\label{thm:cluster-testing-boundedkk1}
Suppose $\varphi'\leq\alpha_{4.5}\varepsilon$, (for the constant $\alpha_{4.5}=\Theta(\min(d^{-1},k^{-1}))$ from Lemma 4.5 of \cite{CzumajPS_arXiv15}, where $d$ denotes the maximum degree), and $\varphi'\leq c'\varepsilon^2\varphi^2/k^2$ for some small constant $c'$. Then there exists a randomized algorithm for \clusterability$(k, \varphi,k,\varphi',\varepsilon)$ problem on degree $d$-bounded graphs that gives the correct answer with probability at least $2/3$, and which makes $\text{poly}(1/\varphi)\cdot\text{poly}(k)\cdot\text{poly}(1/\varepsilon)\cdot\text{poly}(d)\cdot\text{poly}\log(n)\cdot n^{1/2+O(\varepsilon^{-2}k^2\cdot\varphi'/\varphi^2)}$ queries on graphs with $n$ vertices.
\end{theorem}

The proof of the theorem follows by combining Theorem~\ref{thm_alg_partitiontesting} and Lemma 4.5 of \cite{CzumajPS_arXiv15}. The details of the proof are provided in Section~\ref{sec:decomp}.

Furthermore, we strengthen Lemma 4.5 of \cite{CzumajPS_arXiv15} to reduce the dependence of the gap between inner and outer conductance to logarithmic in $k$, albeit at the expense of a bicriteria approximation. This gives us the following theorem, whose proof is provided in Section~\ref{sec:decomp}.

\begin{theorem}\label{thm:cluster-testing-boundedk2k}
Let $0\leq \varepsilon \leq \frac{1}{2}$. 
Suppose $\varphi' \leq \alpha$, (for $\alpha = \min  \{\frac{c_{\text{exp}}}{150 d} ,  \frac{c_{\text{exp}} \cdot \varepsilon}{1400   \log \left( \frac{16k}{\varepsilon} \right)} \}$, where $d$ denotes  the maximum degree), and $\varphi'\leq c\cdot\varepsilon^2\varphi^2/\log(\frac{32k}{\varepsilon})$ for some small constant $c$. Then there exists a randomized algorithm for \clusterability$(k, \varphi,2k,\varphi',\varepsilon)$ problem on degree $d$-bounded graphs that gives the correct answer with probability at least $2/3$, and which makes $\text{poly}(1/\varphi)\cdot\text{poly}(k)\cdot\text{poly}(1/\varepsilon)\cdot\text{poly}(d)\cdot\text{poly}\log(n)\cdot n^{1/2+O(\varepsilon^{-2}\log (\frac{32k}{\varepsilon})\cdot\varphi'/\varphi^2)}$ queries on graphs with $n$ vertices.
\end{theorem}

\subsection{Lower bound Results}\label{sec:lb-intro}

Our lower bounds are based on the following communication problem that we refer to as the \noisyparities$(d, \varepsilon)$:

\begin{definition}\label{def_noisyparities}
\noisyparities$(d,\varepsilon)$ is the problem with parameters $d\geq 3$ and $\varepsilon\leq1/2$ defined as follows. An adversary samples a random $d$-regular graph $G=(V,E)$ from the distribution induced by the \textit{configuration model} of Bollob\'{a}s \cite{Bollobas80}.
The adversary chooses to be in the YES case or the NO case with probability $1/2$, and  generates a vector of binary edge labels $Y\in\{0,1\}^E$ as follows:
\begin{description}
\item[YES case:]  The vector $Y$ is chosen uniformly at random from $\{0,1\}^E$, that is, the labels $Y(e)$ for all edges $e\in E$ are independently $0$ or $1$ with probability $1/2$;
\item[NO case:] A vector $X\in\{0,1\}^V$ is sampled uniformly at random. Independently, a ``noise'' vector $Z\in\{0,1\}^E$ is sampled such that all the $Z(e)$'s are independent Bernoulli random variables which are $1$ with probability $\varepsilon$ and $0$ with probability $1-\varepsilon$. The label of an edge $e=(u,v)\in E$ is given by $Y(e)=X(u)+X(v)+Z(e)$.
\end{description}
The algorithm can query vertices $q\in V$ in an adaptive manner deterministically. Upon querying a vertex $q\in V$, the algorithm gets the edges incident on $q$ together with their labels as a response to the query, and must ultimately determine whether the adversary was in the YES or the NO case.
\end{definition}

Our main result is a tight lower bound on the query complexity of \noisyparities. Before stating our lower bound we note that it is easy to see that unless the set of edges that the algorithm has discovered contains a cycle, the algorithm cannot get any advantage over random guessing. Indeed, if the set of discovered edges were a path $P=(e_1,\ldots,e_T)$ where $e_i=(v_{i-1},v_i)$, the label $Y(e_i)=X(v_1)+X(v_{i-1})+Z(e_i)$ of $e_i$ in the NO case is uniform and independent of the labels of $e_1,\ldots,e_{i-1}$, because $X(v_i)$ is uniform and independent of $Y(e_1)\ldots,Y(e_{i-1})$. A similar argument holds when the set of discovered edges is a forest. Thus, the analysis must, at the very least, prove that $\Omega(\sqrt{n})$ queries are needed in our model for the algorithm to discover a cycle in the underlying graph $G$. In the noisy case (i.e.\ when $\varepsilon>0$) detecting a single cycle does not suffice. Indeed, a natural test would be to add up the labels over the edges of a cycle $C$, that is, consider $\sum_{e\in C}Y(e)$. In the YES case, this is uniformly $0$ or $1$, whereas in the NO case, it is equal to $\sum_{e\in C}Z(e)$, which is $0$ with probability $(1/2)\cdot(1+(1-2\varepsilon)^{|C|})$ and $1$ with probability $(1/2)\cdot(1-(1-2\varepsilon)^{|C|})$. Thus, the deviation of the distribution of $\sum_{e\in C}Y(e)$ from uniform is $n^{-\Theta(\varepsilon)}$, even in the NO case, if $|C|=\Theta(\log n)$.

\begin{theorem}\label{thm_main_aux_lb}
Any deterministic algorithm that solves the \noisyparities problem correctly with probability at least $2/3$ must make at least $n^{1/2+\Omega(\varepsilon)}$ queries on $n$-vertex graphs, for constant $d$.
\end{theorem}

We note that this lower bound is tight up to constant factors multiplying $\varepsilon$ in the exponent. For example, it suffices to find $n^{\Theta(\varepsilon)}$ disjoint cycles. This can be done as follows. Sample $n^{\Theta(\varepsilon)}$ vertices in $G$ uniformly at random, and run $\approx \sqrt{n}$ random walks from each of them. With at least constant probability, for most of the seed nodes the walks will intersect. Then for each cycle $C_i$, compute $\zeta_i=\sum_{e\in C_i}Y(e)$. In the YES case, this is uniformly $0$ and $1$, whereas in the NO case, it is $n^{-\Theta(\varepsilon)}$-far from uniform. Furthermore, since the cycles are disjoint, $\zeta_i$'s are independent. The Chernoff bound implies that, with a constant probability, less than $(1/2)\cdot(1+n^{-\Theta(\varepsilon)}/2)$ fraction of the $\zeta_i$'s will be zero in the YES case, and more than $(1/2)\cdot(1+n^{-\Theta(\varepsilon)}/2)$ fraction of the $\zeta_i$'s will be zero in the NO case.

As a consequence of Theorem \ref{thm_main_aux_lb} and appropriate reductions, we derive the following lower bounds.

\begin{theorem}\label{thm_main_lb}
Any algorithm that distinguishes between a $(1,\varphi_{\text{in}})$-clusterable graph (that is, a $\varphi_{\text{in}}$-expander) and a $(2,\varphi_{\text{out}},1)$-unclusterable graph on $n$ vertices (in other words, solves {\bf PartitionTesting}$(1, \varphi_{\text{in}}, \varphi_{\text{out}},1)$) correctly with probability at least $2/3$ must make at least $n^{1/2+\Omega(\varphi_{\text{out}})}$ queries, even when the input is restricted to regular graphs, for constant $\varphi_{\text{in}}$.
\end{theorem}

\begin{theorem}\label{thm_maxcut}
Any algorithm that approximates the maxcut of $n$-vertex graphs within a factor $2-\varepsilon'$ with probability at least $2/3$ must make at least $n^{1/2+\Omega(\varepsilon'/\log(1/\varepsilon'))}$ queries.
\end{theorem}

\begin{remark}
After posting our  paper on arXiv, we learnt that the above result was already known due to Yoshida (Theorem 1.2 of \cite{Yoshida_CCC11}; the proof appears in the full version \cite{Yoshida_arXiv10}). We note, however, that our proof is very different from Yoshida's proof, and may be of independent interest.
\end{remark}

\subsection{Our techniques}
In this section we give an overview of the new techniques involved in our algorithm and lower bounds.

\subsubsection{Algorithms} 

We start by giving an outline the approach of ~\cite{CzumajPS_STOC15}, outline the major challenges in designing robust tester of graph cluster structure, and then describe our approach. 

As ~\cite{CzumajPS_STOC15} show, the task of distinguishing between $(k, \varphi)$-clusterable graphs and graphs that are $\e$-far from $(k, \varphi')$-clusterable reduces to \partitiontesting$(k, \varphi_{\text{in}}, \varphi_{\text{out}}, \beta)$, where $\beta=\poly(\e)$.  In this problem we are given query access to a graph $G$, and would like to distinguish between two cases: either the graph can be partitioned into at most $k$ clusters with inner conductance at least $\varphi_{\text{in}}$ (the {\bf YES} case, or `clusterable' graphs) or there exists at least $k+1$ subsets $C_1,\ldots, C_{k+1}$ with outer conductance at most $\varphi_{\text{out}}$, and containing nontrivial (i.e.\ no smaller than $\beta n/(k+1)$) number of nodes (the {\bf NO} case, or `non-clusterable' graphs). Here $\varphi_{\text{in}}=\varphi$, and $\varphi_{\text{out}}$ is a function of the conductance $\varphi'$, the number of nodes $k$, and the precision parameter $\e$. 

A very natural approach to \partitiontesting$(k, \varphi_{\text{in}}, \varphi_{\text{out}}, \beta)$ is to sample $10 k$ nodes, say, run random walks of appropriate length from the sampled nodes, and compare the resulting distributions: if  a pair of nodes is in the same cluster, then the distributions of random walks should be `close', and if the nodes are in different clusters, the distributions of random walks should be `far'.  The work of~\cite{CzumajPS_STOC15} shows that this high level approach can indeed be made to work: if one compares distributions in $\ell_2$ norm, then for an appropriate separation between $\varphi_{\text{in}}$ and $\varphi_{\text{out}}$ random walks whose distributions are closer than a threshold $\theta$ in $\ell_2$ sense will indicate that the starting nodes are in the same cluster, and if the distributions are further than $2\theta$ apart in $\ell_2$, say, then the starting points must have been in different clusters. Using an ingenious analysis~\cite{CzumajPS_STOC15} show that one can construct a graph on the sampled nodes where `close' nodes are connected by an edge, and the original graph is clusterable if and only if the graph on the sampled nodes is a union of at most $k$ connected components. The question of estimating $\ell_2$ norm distance between distributions remains, but this can be done in about $\sqrt{n}$ time by estimating collision probabilities (by the birthday paradox), or by using existing results in the literature. The right threshold $\theta$ turns out to be $\approx 1/\sqrt{n}$.

\paragraph{The main challenge.} While very beautiful, the above approach unfortunately does not  work unless the cluster structure in our instances is very pronounced. Specifically, the analysis of~\cite{CzumajPS_STOC15} is based on arguing that random walks of $O(\log n)$ length from sampled nodes that come from the same cluster mostly don't leave the cluster, and this is true only if the outer conductance of the cluster is no larger than $1/\log n$. This makes the approach unsuitable for handling gaps between conductances that are smaller than $\log n$ (it is not hard to see that the walk length must be at least logarithmic in the size of the input graph, so shortening the walk will not help).  

One could think that this is a question of designing of a more refined analysis of the algorithm of~\cite{CzumajPS_STOC15}, but the problem is deeper: it is, in general, not possible to choose a threshold $\theta$ that will work even if the gap between conductances is constant, and even if we want to distinguish between $2$-clusterable and far from $2$-clusterable graphs (such a choice is, in fact, possible for $k=1$). The following simple example illustrates the issue. First consider a $d$-regular graph $G$ composed of two $\Omega(1)$-expanders $A$ and $B$, each of size $\frac{n}{2}$. Further, suppose that the outer conductance of both $A$ and $B$ is upper bounded by  $\frac{1}{4d}$, i.e.\ at most one quarter of the nodes in each of the clusters have connections to nodes on the other side.  Let $t=C\log n$ for a constant $C>0$, and let $\mathbf{p}_u^t$ denote the probability distribution of $t$-step random walk starting from vertex $u$. Lemma C.1 of \cite{CzumajPS_arXiv15} implies that for at least one of these two clusters (say $A$), $||\mathbf{p}_u^t-\mathbf{p}_v^t||_2 = \Omega(d^{-2} n^{-1/2})$ for all $u,v$ in some large subset $\tilde{A}$ of $A$. Thus, any tester that considers two sampled vertices close when their Euclidean distance is at most $\theta$ must use $\theta>d^{-2} n^{-1/2}$.  On the other hand, consider the following $3$-clusterable instance.

Fix $\e\in (0, 1)$, and let $C$ be regular graph with degree $\frac1{\e}-3$, inner conductance $\Omega(1)$ and size $\frac{n}{3}$. Let $G'=(V,E)$ be a $1/\e$-regular graph composed of three copies of $C$ (say $C_1, C_2, C_3$), where for each vertex $u\in C$, its three copies in $C_1, C_2, C_3$ are pairwise connected (i.e.\ form a triangle), and each vertex has a self-loop. We say that an edge is bad if it is a triangle edge or self-loop, otherwise we call it good. Notice that at any step the random walk takes a bad edge with probability $3\e$, and takes an edge inside one of the copies with probability $1-3\e$. We can think that at any step, the random walk first decides to take a good edge or a bad edge, and then takes a random edge accordingly. With probability $(1-3\e)^t$ the random walk never decides to take a bad edge and mixes inside the starting copy. On the other hand, if the random walk does decide to take a bad edge, it is thereafter equally likely to be in any of the three copies of any vertex of $C$.
Let $t=c\log n$ for a constant $c>0$, and let $\mathbf{p}_u^t$ denote the probability distribution of $t$-step random walk starting from vertex $u$. We are interested in bounding $||\mathbf{p}_u^t-\mathbf{p}_v^t||_2$ for a pair of nodes $u, v$ in different clusters (say $u\in C_1$ and $v\in C_2$).
For $u\in C_1$ and $a\in V$, let $\mathbf{q}_u^{t'}(a)$ denote the probability that a $t'$-step random walk in $C_1$ starting from $u$ ends up in the copy of $a$ in $C_1$. Notice that since $C_1$ is constant-expander, if $t'=\Theta(t)$, then for every $a$ we have $\mathbf{q}_u^{t'}(a)\simeq\frac{1}{n/3}$. Consider a $t$-step random walk from $u$ in $G$, and let $t'$ denote the number good edges taken. Notice that $t'$ is binomially distributed with parameters $(t,3\e)$, so that $t'\simeq t(1-3\e)$ with high probability.
Now, for $a \in C_1$, we have,
\[\mathbf{p}_u^t(a)=(1-3\e)^t \cdot \mathbf{q}_u^{t}(a) + \frac{1}{3}(1-(1-3\e)^t) \mathbb{E}_{t'}[\mathbf{q}_u^{t'}(a)|t'>0] \simeq 3n^{-1-3c\e}+\left(1-n^{-3c\e}\right)n^{-1}\text{,}\]
while for $b\notin C_1$, we have
$\mathbf{p}_u^t(b)= \frac{1}{3}(1-(1-3\e)^t) \mathbb{E}_{t'}[\mathbf{q}_u^{t'}(b)|t'>0] \simeq \left(1-n^{-3c\e}\right)n^{-1}$. A symmetric argument holds for $v$. Hence we have $||\mathbf{p}_u^t-\mathbf{p}_v^t||^2_2 \leq \Theta(n^{-1-6c\e})$. Therefore for a pair of nodes $u, v$ in different clusters one has $||p_u^t-p_v^t||_2\leq n^{-1/2-\Omega(\e)}\ll d^{-2} n^{-1/2}$ for constant $d$ and $\e$. Thus, in order to ensure that vertices in different clusters will be considered far, one must take the threshold $\theta$ to be smaller than $d^{-2} n^{-1/2}$. Thus, no tester that uses a fixed threshold can distinguish between the two cases correctly.
To summarize, euclidean distance between distributions is no longer a reliable metric if one would like to operate in a regime close to theoretical optimum, and a new proxy for clusterability is needed.

\paragraph{Our main algorithmic ideas.} Our main algorithmic contribution is a more geometric approach to analyzing the proximity of the sampled points: instead of comparing $\ell_2$ distances between points, our tester considers the Gram matrix of the random walk transition probabilities of the points, estimates this matrix entry-wise to a precision that depends on the gap between $\varphi_{\text{in}}$ and $\varphi_{\text{out}}$ in the instance of \partitiontesting$(k, \varphi_{\text{in}}, \varphi_{\text{out}}, \beta)$ that we would like to solve, and computes the $(k+1)$-st largest eigenvalue of the matrix. This quantity turns out to be a more robust metric, yielding a tester that operates close to the theoretical optimum, i.e.\ able to solve \partitiontesting$(k, \varphi_{\text{in}}, \varphi_{\text{out}}, \beta)$ as long as the gap $\varphi_{\text{out}}/\varphi_{\text{in}}^2$ is smaller than an absolute constant.\footnote{ Note that our runtime depends on $\varphi_{\text{out}}/\varphi_{\text{in}}^2$ as opposed to $\varphi_{\text{out}}/\varphi_{\text{in}}$ due to a loss in parameters incurred through Cheeger's inequality. This loss is quite common for spectral algorithms.} Specifically, our tester (see Algorithms~\ref{alg_estimate} and~\ref{disg_alg} in Section~\ref{sec_oracle} for the most basic version) samples a multiset $S$ of $s\approx \poly(k)\log n$ vertices of the graph $G$ independently and with probability proportional to the degree distribution (this can be achieved in $\approx \sqrt{n}$ time per sample using the result of Eden and Rosenbaum~\cite{EdenR_SOSA18}), and computes the matrix 
\begin{equation}\label{eq:matrix-a}
A:=(D^{-\frac{1}{2}}M^tS)^{\top}(D^{-\frac{1}{2}}M^tS),
\end{equation}
where $M$ is the random walk transition matrix of the graph $G$, and $D$ is the diagonal matrix of degrees. Note that this is the Gram matrix of the $t$-step distributions of random walks from the sampled nodes in $G$, for a logarithmic number of steps walk. Intuitively, the matrix $A$ captures pairwise collision probabilities of random walks from sampled nodes, weighted by inverse degree. The algorithm accepts the graph if the $(k+1)$-st largest eigenvalue of the matrix $A$ is below a threshold, and rejects otherwise. Specifically, the algorithm accepts if $\mu_{k+1}(A)\lesssim \text{vol}(V_G)^{-1-\Theta(\varphi_{\text{out}}/\varphi_{\text{in}}^2)}$ and rejects otherwise. Before outlining the proof of correctness for the tester, we note that, of course, the tester above cannot be directly implemented in sublinear time, as computing the matrix $A$ exactly is expensive. The actual sublinear time tester approximately computes the entries of the matrix $A$ to additive precision about  $\frac1{\poly(k)}\text{vol}(V_G)^{-1-\Theta(\varphi_{\text{out}}/\varphi_{\text{in}}^2)}$ and uses the eigenvalues of the approximately computed matrix to decide whether to accept or reject. Such an approximation can be computed in about $\text{vol}(V_G)^{\frac1{2}+\Theta(\varphi_{\text{out}}/\varphi_{\text{in}}^2)}$ queries by rather standard techniques (see Section~\ref{sec_liftingoracle}).

We now outline the proof of correctness of the tester above (the detailed proof is presented in Section~\ref{sec_oracle}). It turns out to be not too hard to show that the tester accepts graphs that are $(k, \varphi_{\text{in}})$-clusterable. One first observes that Cheeger's inequality together with the assumption that each of the $k$ clusters is a $\varphi_{\text{in}}$-expander implies that the $(k+1)$-st eigenvalue of the normalized Laplacian of $G$ is at least $\varphi_{\text{in}}^2/2$ (Lemma~\ref{lem_Cheeger}). It follows that the matrix $M^t$ of $t$-step random walk transition probabilities, for our choice of $t=(C/\varphi_{\text{in}}^2)\log n$, is very close to a matrix of rank at most $k$, and thus the $(k+1)$-st eigenvalue of the matrix $A$ above (see~\eqref{eq:matrix-a}) is smaller than $1/n^2$, say. The challenging part is to show that the tester rejects graphs that are $(k, \varphi_{\text{out}}, \beta)$-unclusterable, since in this case we do not have any assumptions on the inner structure of the clusters $C_1,\ldots, C_{k+1}$. The clusters $C_1,\ldots, C_{k+1}$  could either be good expanders, or, for instance, unions of small disconnected components. The random walks from nodes in those clusters behave very differently in these two cases, but the analysis needs to handle both. Our main idea is to consider a carefully defined $k+1$-dimensional subspace of the eigenspace of the normalized Laplacian of $G$ that corresponds to small (smaller than $O(\varphi_{\text{out}})$) eigenvalues, and show that our random sample of points is likely to have a well-concentrated projection onto this subspace. We then show that this fact implies that the matrix $A$ in~\eqref{eq:matrix-a} has a large $(k+1)$-st eigenvalue with high probability. The details of the argument are provided in Section~\ref{sec_no}: the definition of matrix $U$ and projection operator $P_h$ at the beginning of Section~\ref{sec_no} yield the $(k+1)$-dimensional subspace in question (this subspace is the span of the columns of $U$ projected onto the first $h$ eigenvectors of the normalized Laplacian of $G$), and a central claims  about the subspace in question are provided by Lemmas~\ref{lem_P-perpUz-small},~\ref{evenly_distributed} and~\ref{SVz}. The assumption that vertices in $S$ are sampled with probabilities proportional to their degrees is crucial to making the proof work for general (sparse) graphs.

One consequence of the fact that our algorithm for \partitiontesting$(k, \varphi_{\text{in}}, \varphi_{\text{out}}, \beta)$ estimates the entries of the Gram matrix referred to above to additive precision $\approx n^{-1-\Theta(\varphi_{\text{out}}/\varphi_{\text{in}}^2)}$ is that the runtime $\approx n^{1/2+\Theta(\varphi_{\text{out}}/\varphi_{\text{in}}^2)}$. If $\varphi_{\text{out}}\ll \varphi_{\text{in}}^2/\log n$, then we recover the $\approx \sqrt{n}$ runtime of~\cite{CzumajPS_STOC15}, but for any constant gap between $\varphi_{\text{out}}$ and $\varphi_{\text{in}}^2$ our runtime is polynomially larger than $\sqrt{n}$. Our main contribution on the lower bound side is to show that this dependence is necessary. We outline our main ideas in that part of the paper now.

\subsubsection{The lower bound} We show that the $n^{1+\Omega(\varphi_{\text{out}}/\varphi_{\text{in}}^2)}$ runtime is necessary for \partitiontesting$(k, \varphi_{\text{in}}, \varphi_{\text{out}}, \beta)$ problem, thereby proving that our runtime is essentially best possible for constant $k$.  More precisely, we show that even distinguishing between an expander and a graph that contains a cut of sparsity $\e$ for $\e \in (0,1/2)$ requires $n^{1+\Omega(\e)}$ adaptive queries, giving a lower bound for the query complexity (and hence runtime) of \partitiontesting$(1, \Omega(1), \e, 1)$ that matches our algorithm's performance.  

\paragraph{The \noisyparities problem.} Our main tool in proving the lower bound is a new communication complexity problem (the \noisyparities problem) that we define and analyze: an adversary chooses a regular graph $G=(V, E)$ and a hidden binary string $X\in \{0, 1\}^V$, which can be thought of as encoding a hidden bipartition of $G$. The algorithm can repeatedly (and adaptively) query vertices of $G$. Upon querying a vertex $v$, the algorithm receives the edges incident on $v$ and a binary label $Y(e)$ on each edge $e$. In the {\bf NO} case the labels $Y(e)$ satisfy $Y(e)=X(u)+X(v)+Z(e)$, where $Z(e)$ is an independent Bernoulli random variable with expectation $\e$ (i.e.\ the algorithm is told whether the edge crosses the hidden bipartition, but the answer is noisy). In the {\bf YES} case each label $Y(e)$ is uniformly random in $\{0, 1\}$. The task of the algorithm is to distinguish between the two cases using the smallest possible number of queries to the graph $G$. 

It is easy to see that if $\e=0$, then the algorithm can get a constant advantage over random guessing as long as it can query all edges along a cycle in $G$. If $G$ is a random $d$-regular graph unknown to the algorithm, one can show that this will take at least $\Omega(\sqrt{n})$ queries, recovering the lower bound for expansion testing due to Goldreich and Ron~\cite{GoldreichR_Algorithmica02}. In the noisy setting, however, detecting a single cycle is not enough, as cycles that the algorithm can locate in a random regular graph using few queries are generally of logarithmic length, and the noise added to each edge compounds over the length of the cycle, leading to only advantage of about $n^{-O(\e)}$ over random guessing that one can obtain from a single cycle. Intuitively. this suggests that the algorithm should find at least $n^{\Omega(\e)}$ cycles in order to get a constant advantage. Detecting a single cycle in an unknown sparse random graphs requires about $\sqrt{n}$ queries, which together leads to the $n^{1/2+\Omega(\e)}$ lower bound. Turning this intuition into a proof is challenging, however, as {\bf (a)} the algorithm may base its decisions on labels that it observes on its adaptively queried subgraph of $G$ and {\bf (b)} the algorithm does not have to base its decision on observed parities over cycles. We circumvent these difficulties by analyzing the distribution of labels on the edges of the subgraph that the algorithm queries in the {\bf NO} case and proving that this distribution is close to uniformly random in total variation distance, with high probability over the queries of the algorithm. We analyze this distribution using a Fourier analytic approach, which we outline now. 

Suppose that we are in the {\bf NO} case, i.e.\ the edge labels presented to the algorithm are an $\e$-noisy version of parities of the hidden boolean vector $X\in \{0, 1\}^n$, and suppose that the algorithm has discovered a subset $E_{\text{query}}\subseteq E_G$ of edges of the graph $G$ (recall that the graph $G$, crucially, is not known to the algorithm) together with their labels. The central question that our analysis needs to answer in this situation turns out to be the following: given the observed labels on edges in $E_{\text{query}}$ and an edge $e=(a, b)\in E_G$ what is the posterior distribution of $X(a)+X(b)$ given the information that the algorithm observed so far? For example, if $E_{\text{query}}$ does not contain any cycles (i.e.\ is a forest), then $X(a)+X(b)$ is a uniformly random Bernoulli variable with expectation $1/2$ if the edge $(a, b)$ does not close a cycle when added to $E_{\text{query}}$. If it does close a cycle but $E_{\text{query}}$ is still a forest, then one can show that if the distance in $E_{\text{query}}$ from $a$ to $b$ is large (at least $\Omega(\log n)$), then the posterior distribution of $X(a)+X(b)$ is still $n^{-\Omega(\e)}$ close, in total variation distance, to a Bernoulli random variable with expectation $1/2$. Our analysis needs to upper bound this distance to uniformity for a `typical' subset $E_{\text{query}}$ that arises throughout the interaction process of the algorithm with the adversary, and contains two main ideas. First, we show using Fourier analytic tools (see Theorem~\ref{thm_Fourier} in Section~\ref{sec:fourier}) that for `typical' subset of queried edges $E_{\text{query}}$ and any setting of observed labels, one has that the bias of $X(a)+X(b)$, i.e.\ the absolute deviation of the expectation of this Bernoulli random variable from $1/2$, satisfies 
\begin{equation}\label{eq:bias-ub}
\text{bias}(X(a)+X(b))\lesssim \sum_{E'\subseteq E_{\text{query}}\text{~s.t.~}E'\cup \{a, b\}\text{~is Eulerian}} (1-2\e)^{|E'|}.
\end{equation}
Note that for the special case of $E_{\text{query}}$ being a tree, the right hand side is exactly the $(1-2\e)^{\text{dist}(a, b)}$, where $\text{dist}(a, b)$ stands for the shortest path distance from $a$ to $b$ in $T$. Since `typical' cycles that the algorithm will discover will be of $\Omega(\log n)$ length due to the fact that $G$ is a constant degree random regular graph, this is $n^{-\Omega(\e)}$, as required. Of course, the main challenge in proving our lower bound is to analyze settings where the set of queried edges $E_{\text{query}}$ is quite far from being a tree, and generally contains many cycles, and control the sum in~\eqref{eq:bias-ub}. In other words, we need to bound the weight distribution of Eulerian subgraphs of $E_{\text{query}}$. The main insight here is the following structural claim about `typical' sets of queried edges $E_{\text{query}}$: we show that for typical interaction scenarios between the algorithm and the adversary one can decompose $E_{\text{query}}$ as $E_{\text{query}}=F\cup R$, where $F$ is a forest and $R$ is a small (about $n^{O(\e)}$ size) set of `off-forest' edges that further satisfies the property that the endpoints of edges in $R$ are $\Omega(\log n)$-far from each other in the shortest path metric induced by $F$. This analysis relies on basic properties of random graphs with constant degrees and is presented in Section~\ref{sec_analysis_interaction}. Once such a decomposition of $E_{\text{query}}=T\cup F$ is established, we get a convenient basis for the cycle space of $E_{\text{query}}$, which lets us control the right hand side in~\eqref{eq:bias-ub} as required (see Section~\ref{subsec_wrapping}). The details of the lower bound analysis are presented in Section~\ref{sec_lbproof}.

Finally, our lower bound on the query complexity of \noisyparities yields a lower bound for \partitiontesting$(1, \Omega(1), \e, 1)$ (Theorem~\ref{thm_main_lb}), as well as a lower bound for better than factor $2$ approximation to MAX-CUT value in sublinear time (Theorem~\ref{thm_maxcut}). Both reductions are presented in Section~\ref{sec_reductions}. The reduction to MAX-CUT follows using rather standard techniques (e.g. is very similar to~\cite{KapralovKSV_SODA17}; see Section~\ref{sec:reduction-maxcut}). The reduction to \partitiontesting$(1, \Omega(1), \e, 1)$ is more delicate and novel: the difficulty is that we need to ensure that the introduction of random noise $Z_e$ on the edge labels produces graphs that have the expansion property (in contrast, the MAX-CUT reduction produces graphs with a linear fraction of isolated nodes). This reduction is presented in Section~\ref{sec:reduction-partition}.

\subsection{Related Work}

Goldreich and Ron~\cite{GoldreichR_Algorithmica02} initiated the framework of testing graph properties via neighborhood queries. In this framework, the goal is to separate graphs having a certain property from graphs which are ``far'' from having that property, in the sense that they need many edge additions and deletions to satisfy the property. The line of work closest to this paper is the one on testing expansion of graphs~\cite{GoldreichR_ECCC00,NachmiasS_IC10,CzumajS_CPC10,KaleS_SIAMJC11} which proves that expansion testing can be done in about $\tilde{O}(\sqrt{n})$ queries, and $\Omega(\sqrt{n})$ queries are indeed necessary. Going beyond expansion (that is, $1$-clusterability), Kannan et al.~\cite{KannanVV_JACM04} introduced (internal) conductance as a measure of how well a set of vertices form a cluster. In order to measure the quality of a clustering, that is, a partition of vertices into clusters, Zhu et al.~\cite{ZhuLM_ICML13} and Oveis Gharan and Trevisan~\cite{GharanT_SODA14} proposed bi-criteria measures which take into account the (minimum) internal conductance and the (maximum) external conductance of the clusters. Considering this measure, Czumaj et al.~\cite{CzumajPS_STOC15} defined the notion of clusterable graphs parameterized requirements on the minimum internal expansion and by the maximum external expansion, and gave an algorithm for testing clusterability.

There has been an extensive work on testing many other graph properties in the framework of Goldreich and Ron. For instance, Czumaj et al.~\cite{CzumajGRSSS_RSA14} give algorithms for testing several properties including cycle-freeness, whereas Eden et al.~\cite{EdenLR_SODA18} design algorithms to test arboricity. Estimation of graph parameters such as degree distribution moments~\cite{EdenRS_ICALP17}, number of triangles~\cite{EdenLRS_SIAMJC17}, and more generally, number of $k$-cliques~\cite{EdenRS_arXiv17} has also received attention recently.

A closely related model of property testing is the one where the graph arrives as a random order stream and the property testing algorithm is required to use sublinear space. 
Although this appears to be a less powerful model because the algorithm no longer has the ability to execute whatever queries it wants, interestingly, Peng and Sohler~\cite{PengS_SODA18} show that sublinear property testing algorithms give rise to sublinear space algorithms for random order streams.

Other graph property testing models include extension to dense graphs~\cite{GonenR_Algorithmica10,GoldreichR_SIAMJC11} where the algorithm queries the entries of the adjacency matrix of the graph, and the non-deterministic property testing model~\cite{LovaszV_CPC13,GishbolinerS_ECCC13}, where the algorithm queries the graph and a certificate, and must decide whether the graph satisfies the property. We refer the reader to~\cite{CzumajPS_STOC15} for a more comprehensive survey of the related work.

%% file: algorithm-nonuniform.tex
\section{Algorithm for \partitiontesting}\label{sec:algo}

The goal of this section is to present an algorithm for the \partitiontesting problem, analyze it, and hence, prove Theorem \ref{thm_alg_partitiontesting}. We restate this theorem here for the reader's convenience, and its proof appears at the end of Section \ref{sec_liftingoracle}.

\noindent\textbf{Theorem \ref{thm_alg_partitiontesting} (restated).} Suppose $\varphi_{\text{out}}\leq \frac{1}{480}\varphi_{\text{in}}^2$. Then there exists a randomized algorithm for \partitiontesting$(k, \varphi_{\text{in}}, \varphi_{\text{out}},\beta)$ which gives the correct answer with probability at least $2/3$, and which makes $\text{poly}(1/\varphi_{\text{in}})\cdot\text{poly}(k)\cdot\text{poly}(1/\beta)\cdot\text{poly}\log(m)\cdot m^{1/2+O(\varphi_{\text{out}}/\varphi_{\text{in}}^2)}$ queries on graphs with $m$ edges.

Towards proving this theorem, we first make the following simplifying assumption. We assume that we have the following oracle at our disposal: the oracle takes a vertex $a$ as input, and returns $D^{-\frac{1}{2}}M^t\mathds{1}_a$, where $D$ is the diagonal matrix of the vertex degrees, $M$ is the transition matrix of the lazy random walk associated with the input graph, and $\mathds{1}_a$ is the indicator vector of $a$. We first present and analyze, in Section \ref{sec_oracle}, an algorithm for \partitiontesting which makes use of this oracle. Following this, in Section \ref{sec_liftingoracle}, we show how the oracle can be (approximately) simulated, and thereby, get an algorithm for {\bf PartitionTesting}.

We remark that our algorithms use the value of $\text{vol}(V_G)$, which is not available directly through the access model described in Section \ref{sec_ourresults}. However, by the result of \cite{seshadhri2015simpler}, it is possible to approximate the value of $\text{vol}(V_G)$ with an arbitrarily small multiplicative error using $\tilde{O}(\sqrt{|V_G|})$ queries.

\subsection{The algorithm under an oracle assumption}\label{sec_oracle}

Let $G=(V_G,E_G)$ be a graph and let $A$ be its adjacency matrix. Recall that in Definition \ref{def_randwalk} we associated with such a graph the random walk given by the transition matrix $M=\frac{I+AD^{-1}}{2}$. That is, from any vertex, the walk takes each edge incident on the vertex with probability $\frac{1}{2\cdot\text{deg}(v)}$, and stays at the same vertex with probability $\frac{1}{2}$. Fix $t$, the length of the random walk.  For this section, we assume that we have the following oracle at our disposal: the oracle takes a vertex $a \in V_G$ as input, and returns $D^{-\frac{1}{2}}M^t\mathds{1}_a$. Our algorithm for \partitiontesting is given by Algorithm \ref{disg_alg} called \textsc{PartitionTest}. The goal of this section is to prove guarantees about this algorithm, as stated in the following theorem.

\begin{theorem}\label{thm_main}
Suppose $\varphi_{\text{in}}^2>480\varphi_{\text{out}}$. For every graph $G$, integer $k\geq 1$, and $\beta \in (0,1)$, 
\begin{enumerate}
\item If $G$ is $(k,\varphi_{\text{in}})$-clusterable (YES case), then \textsc{PartitionTest}($G,k,\varphi_{\text{in}},\varphi_{\text{out}},\beta$) accepts.
\item If $G$ is $(k,\varphi_{\text{out}},\beta)$-unclusterable (NO case), then \textsc{PartitionTest}($G,k,\varphi_{\text{in}},\varphi_{\text{out}},\beta$) rejects with probability at least $\frac{2}{3}$.
\end{enumerate}
\end{theorem}


\begin{algorithm}
\caption{\textsc{Estimate}($G,k,s,t,\eta$)}\label{alg_estimate}
\begin{algorithmic}[1]
\Procedure{\textsc{Estimate}($G,k,s,t,\eta$)}{}
	\State Sample $s$ vertices from $V_G$ independently and with probability proportional to the degree of the vertices  at random with replacement using \textbf{sampler}($G,\eta$) (See Lemma \ref{lem_vertexsampling}). Let $S$ be the multiset of sampled vertices.
	\State Compute $D^{-\frac{1}{2}}M^tS$ using the oracle.
	\State Return $\mu_{k+1}((D^{-\frac{1}{2}}M^tS)^{\top}(D^{-\frac{1}{2}}M^tS))$.
\EndProcedure
\end{algorithmic}
\end{algorithm}

\begin{algorithm}
\caption{\textsc{PartitionTest}($G,k,\varphi_{\text{in}},\varphi_{\text{out}},\beta$)}\label{disg_alg}
\begin{algorithmic}[1]
\Procedure{\textsc{PartitionTest}($G,k,\varphi_{\text{in}},\varphi_{\text{out}},\beta$)}{}
	\Comment Need: $\varphi_{\text{in}}^2>480\varphi_{\text{out}}$
	\State $\eta:=0.5$.
	\State $s:=1600(k+1)^2\cdot \ln(12(k+1))\cdot \ln (\text{vol}(V_G))/(\beta(1-\eta))$. 
	\State $c:= \frac{20}{\varphi_{\text{in}}^2 }$, $t:=c\ln (\text{vol}(V_G))$ \Comment Observe: $c>0$.
	\State $\mu_{\text{thres}}:=\frac{1}{2}\cdot\frac{8(k+1)\ln(12(k+1))}{\beta\cdot(1-\eta)}\times \text{vol}(V_G)^{-1-120c\varphi_{\text{out}}}$.
	\If {\textsc{Estimate}($G,k,s,t,\eta$) $\leq \mu_{\text{thres}}$} 
		\State Accept $G$.
	\Else 
		\State Reject $G$.
	\EndIf
\EndProcedure
\end{algorithmic}
\end{algorithm}

Algorithm \textsc{PartitionTest} calls the procedure \textsc{Estimate} given by Algorithm \ref{alg_estimate}, compares the value returned with a threshold, and then decides whether to accept or reject. Procedure \textsc{Estimate} needs to draw several samples of vertices, where each vertex of the input graph is sampled with probability proportional to its degree. This, by itself, is not allowed in the query model under consideration defined in Section \ref{sec_ourresults}. Therefore, procedure \textsc{Estimate} makes use of the following result by Eden and Rosenbaum to (approximately) sample vertices with probabilities proportional to degree.

\begin{lemma}[Corollary 1.5 of \cite{EdenR_SOSA18}]\label{lem_vertexsampling}
Let $G = (V_G,E_G)$ be an arbitrary graph, and $\eta> 0$. Let $\mathcal{D}$ denote the degree distribution of $G$ (i.e., $\mathcal{D}(v) = \frac{\deg(v)}{\text{vol}(G)} $). Then there exists an algorithm, denoted by \textbf{sampler}($G,\eta$),
that with probability at least $\frac{2}{3}$ produces a vertex $v$ sampled from a distribution $\mathcal{P}$ over $V_G$, and outputs ``Fail'' otherwise. The distribution $\mathcal{P}$ is such that for all $v \in V_G$,
\[|\mathcal{P}(v)-\mathcal{D}(v)| \leq \eta\cdot \mathcal{D}(v)\text{.}\]
The algorithm uses $\tilde{O}\left(\frac{ |V_G|}{\sqrt{\eta\cdot \text{vol}(V_G)}}\right)$ vertex, degree and neighbor queries.
\end{lemma}



The proof of Theorem \ref{thm_main} relies on the following guarantees about the behavior of the algorithm in the YES case, and the NO case respectively, whose proofs are given in Section \ref{sec_yes} and Section \ref{sec_no} respectively.

\begin{theorem}\label{thm_yes}
Let $\varphi_{\text{in}}>0$ and integer $k \geq 1$. Then for every $(k,\varphi_{\text{in}})$-clusterable graph $G=(V_G, E_G)$ (see definition \ref{def:clusterable}), with $\min_{v \in V_G}{\text{deg}(v)}\geq 1$ the following holds:
\[\textsc{Estimate}(G,k,s,t,\eta) \leq  s \cdot \left(1-\frac{\varphi_{\text{in}}^2}{4}\right)^{2t} \text{.}\]
\end{theorem}

\begin{theorem}\label{thm_no}
Let $\varphi_{\text{out}}>0$, $\beta \in (0,1)$, and integer $k\geq 1$. Let  $$s=1600(k+1)^2\cdot \ln(12(k+1))\cdot \ln (\text{vol}(V_G))/(\beta\cdot(1-\eta)) \text{.}$$ Then for every $(k,\varphi_{\text{out}},\beta)$-unclusterable graph $G=(V_G, E_G)$ (see definition \ref{def:clusterable}), with $\min_{v \in V_G}{\text{deg}(v)}\geq 1$, the following holds  with probability at least $\frac{2}{3}$.
\[\textsc{Estimate}(G,k,s,t,\eta) \geq  
\frac{8(k+1)\ln(12(k+1))}{\beta \cdot(1-\eta)\cdot \text{vol}(V_G)}\times(1-30\varphi_{out})^{2t} \text{.}\]
\end{theorem}

\begin{proof}[Proof of Theorem \ref{thm_main}]
Let $t=c\ln (\text{vol}(V_G))$ for $c=\frac{20}{\varphi_{\text{in}}^2}$. We call the procedure \textsc{Estimate} with 
$$
s={1600(k+1)^2\cdot \ln(12(k+1))\cdot \ln (\text{vol}(V_G))}/ (\beta\cdot(1-\eta)),
$$ and $t=c\ln (\text{vol}(V_G))$. In the YES case, by Theorem \ref{thm_yes}, \textsc{Estimate} returns a value at most 
\begin{align*}
s \cdot \left(1-\frac{\varphi_{\text{in}}^2}{4}\right)^{2t}  
&\leq s\cdot \exp\left(-\frac{\varphi_{\text{in}}^2 t}{2}\right)
=s\cdot \exp\left(-\frac{\varphi_{\text{in}}^2c\ln (\text{vol}(V_G))}{2}\right) \\
&=\frac{1600(k+1)^2\cdot \ln(12(k+1))\cdot \ln (\text{vol}(V_G))}{\beta\cdot(1-\eta)}\times \text{vol}(V_G)^{-c\frac{\varphi_{\text{in}}^2}{2}} \\
&\leq \frac{1}{2}\cdot\frac{8(k+1)\ln(12(k+1))}{\beta\cdot(1-\eta)}\times \text{vol}(V_G)^{2-c\frac{\varphi_{\text{in}}^2}{2}} \text{.}
\end{align*}
In the last inequality we use the fact that $k+1\leq \text{vol}(V_G)$, and $|V_G|$ is large enough to insure that $200\ln (\text{vol}(V_G)) \leq \text{vol}(V_G)$. In the NO case, by Theorem \ref{thm_no}, with probability at least $\frac{2}{3}$, \textsc{Estimate} returns a value at least
\begin{align*}
\frac{8(k+1)\ln(12(k+1))}{\beta \cdot (1-\eta)\cdot \text{vol}(V_G)}\times(1-30\varphi_{out})^{2t} 
&\geq  \frac{8(k+1)\ln(12(k+1))}{\beta \cdot (1-\eta) \cdot \text{vol}(V_G)}\times\exp\left(-120\varphi_{\text{out}}c\ln (\text{vol}(V_G)) \right) \\
&\geq \frac{1}{2}\cdot\frac{8(k+1)\ln(12(k+1))}{\beta \cdot (1-\eta)}\times \text{vol}(V_G)^{-1-120c\varphi_{\text{out}}}\text{.}
\end{align*}
Since $\varphi_{\text{in}}^2>480\varphi_{\text{out}}$, the value of $c=\frac{20}{\varphi_{\text{in}}^2}$, chosen in \textsc{PartitionTest} is such that $2-c\frac{\varphi_{\text{in}}^2}{2}<-1-120c\varphi_{\text{out}}$. Therefore for $|V_G|$ large enough, the upper bound on the value returned by \textsc{Estimate} in the YES case is less than $\mu_{\text{thres}}=\frac{1}{2}\cdot\frac{8(k+1)\ln(12(k+1))}{\beta \cdot (1-\eta)}\times \text{vol}(V_G)^{-1-120c\varphi_{\text{out}}}$, which is less than the lower bound on the value returned by \textsc{Estimate} in the NO case. 
\end{proof}


\subsubsection{Proof of Theorem~\ref{thm_yes} (the YES case)}\label{sec_yes}

The main result of this section is a proof of Theorem~\ref{thm_yes}, restated below for convenience of the reader:

\noindent{\em {\bf Theorem~\ref{thm_yes}} (Restated)
Let $\varphi_{\text{in}}>0$ and integer $k \geq 1$. Then for every $(k,\varphi_{\text{in}})$-clusterable graph $G=(V_G, E_G)$ (see definition \ref{def:clusterable}), with $\min_{v \in V_G}{\text{deg}(v)}\geq 1$ the following holds:
\[\textsc{Estimate}(G,k,s,t,\eta) \leq  s \cdot \left(1-\frac{\varphi_{\text{in}}^2}{4}\right)^{2t} \text{.}\]
}

Consider the YES case, where the vertices of $G$ can be partitioned into $h$ subsets with $C_1,\ldots,C_h$ for some $h\leq k$, such that for each $i$, $\phi^G(C_i)\geq\varphi_{\text{in}}$. We are interested in bounding $\mu_{k+1}((D^{-\frac{1}{2}}M^tS)^{\top}(D^{-\frac{1}{2}}M^tS))$ from above.

\begin{lemma}\label{lem_yes_algo}
Let $\varphi_{\text{in}}>0$, integer $k\geq 1$, and $G=(V_G, E_G)$ be a $(k,\varphi_{\text{in}})$-clusterable graph (see definition \ref{def:clusterable}), with $\min_{v \in V_G}{\text{deg}(v)}\geq 1$. Let $L$ be its normalized Laplacian matrix, and $M$ be the transition matrix of the associated random walk. Let $S$ be a (multi)set of $s$ vertices of $G$. Then \[\mu_{k+1}((D^{-\frac{1}{2}}M^tS)^{\top}(D^{-\frac{1}{2}}M^tS)) \leq s\cdot  \left(1-\frac{\lambda_{k+1}}{2}\right)^{2t},\] where $\lambda_{k+1}$ is the $(k+1)$-st smallest eigenvalue of $L$.
\end{lemma}

\begin{proof}
Recall from Section \ref{sec_prelim} that $M=D^{\frac{1}{2}}\overline{M}D^{-\frac{1}{2}}$, hence, $M^t=D^{\frac{1}{2}}{\overline{M}}^tD^{-\frac{1}{2}}$. Thus we can write
\begin{align*}
\mu_{k+1}((D^{-\frac{1}{2}}M^tS)^{\top}(D^{-\frac{1}{2}}M^tS)) &= \mu_{k+1}((D^{-\frac{1}{2}}D^{\frac{1}{2}}\overline{M}D^{-\frac{1}{2}}S)^{\top}(D^{-\frac{1}{2}}D^{\frac{1}{2}}\overline{M}D^{-\frac{1}{2}}S)) \\
&= \mu_{k+1}(S^{\top}D^{-\frac{1}{2}}{\overline{M}}^{2t}D^{-\frac{1}{2}}S).
\end{align*}

Recall from Section \ref{sec_prelim} that $1-\frac{\lambda_1}{2}\geq \cdots\geq 1-\frac{\lambda_n}{2}$ are the eigenvalues of ${\overline{M}}$, $\Sigma$ is the diagonal matrix of these eigenvalues in descending order, and $V$ is the matrix whose columns are orthonormal eigenvectors arranged in descending order of their eigenvalues. We have $\overline{M}^{2t}=V\Sigma^{2t}V^{\top}$. Let $\Sigma_{1:k}$ be $n \times n$ diagonal matrix with first $k$ entries  $1-\frac{\lambda_1}{2}\geq \cdots\geq 1-\frac{\lambda_k}{2}$ and the rest zero and let $\Sigma_{k+1:n}$ denote  $n \times n$ diagonal matrix with first $k$ entries  zero and rest $1-\frac{\lambda_{k+1}}{2}\geq \cdots\geq 1-\frac{\lambda_n}{2}$. We have $\Sigma^{2t}=\Sigma_{1:k}^{2t}+\Sigma_{k+1:n}^{2t}$, thus we get
\begin{align*}
\mu_{k+1}((D^{-\frac{1}{2}}M^tS)^{\top}(D^{-\frac{1}{2}}M^tS))&=\mu_{k+1}(S^{\top}D^{-\frac{1}{2}}{\overline{M}}^{2t} D^{-\frac{1}{2}}S) \\
&=\mu_{k+1}(S^{\top}D^{-\frac{1}{2}}(V\Sigma^{2t}V^{\top})D^{-\frac{1}{2}}S) \\
&=\mu_{k+1}(S^{\top}D^{-\frac{1}{2}}V(\Sigma_{1:k}^{2t}+\Sigma_{k+1:n}^{2t})V^{\top}D^{-\frac{1}{2}}S) \\
&\leq \mu_{k+1}(S^{\top}D^{-\frac{1}{2}}V\Sigma_{1:k}^{2t}V^{\top}D^{-\frac{1}{2}}S) + \mu_{\max}(S^{\top}D^{-\frac{1}{2}}V\Sigma_{k+1:n}^{2t}V^{\top}D^{-\frac{1}{2}}S) 
\end{align*}

The last inequality follows from Lemma \ref{lem_Weyl}. Here $\mu_{k+1}(S^{\top}D^{-\frac{1}{2}}V\Sigma_{1:k}^{2t}V^{\top}D^{-\frac{1}{2}}S)=0$, because the rank of $\Sigma_{1:k}^{2t}$ is $k$. We are left to bound $\mu_{\max}(S^{\top}D^{-\frac{1}{2}}V\Sigma_{k+1:n}^{2t}V^{\top}D^{-\frac{1}{2}}S)$. By Lemmas \ref{lem_commute} and \ref{lem_submult}, we have,
\begin{align*}
&\mu_{\max}(S^{\top}D^{-\frac{1}{2}}V\Sigma_{k+1:n}^{2t}V^{\top}D^{-\frac{1}{2}}S) \\
&=\mu_{\max}(D^{-\frac{1}{2}}V\Sigma_{k+1:n}^{2t}V^{\top}D^{-\frac{1}{2}}SS^\top) && {(\text{By Lemma \ref{lem_commute}})}\\
&\leq \mu_{\max}(D^{-\frac{1}{2}}V\Sigma_{k+1:n}^{2t}V^{\top}D^{-\frac{1}{2}}) \cdot \mu_{\max}(SS^\top) && {(\text{By Lemma \ref{lem_submult}})} \\
&=\mu_{\max}(V\Sigma_{k+1:n}^{2t}V^{\top}D^{-1}) \cdot \mu_{\max}(SS^\top) && {(\text{By Lemma \ref{lem_commute}})} \\
&\leq\mu_{\max}(V\Sigma_{k+1:n}^{2t}V^{\top}) \cdot \mu_{\max}(SS^\top) \cdot\mu_{\max}(D^{-1}) && {(\text{By Lemma \ref{lem_submult}})} \\
&=\mu_{\max}(\Sigma_{k+1:n}^{2t}V^{\top}V) \cdot \mu_{\max}(SS^\top) \cdot\mu_{\max}(D^{-1}) && {(\text{By Lemma \ref{lem_commute}})} \\
&=\mu_{\max}(\Sigma_{k+1:n}^{2t}) \cdot \mu_{\max}(SS^\top) \cdot\mu_{\max}(D^{-1}) && {(\text{Since } V^\top V=I)} \\
\end{align*}
Next, observe that $SS^{\top}\in N\times N$ is a diagonal matrix whose $(a,a)^{\text{\tiny{th}}}$ entry is the multiplicity of vertex $a$ in $S$. Thus, $\mu_{\max}(SS^{\top})$ is the maximum multiplicity over all vertices, which is at most $s$. Also notice that $\mu_{\max}(D^{-1})=\max_{v \in V_G}\frac{1}{\text{ deg}(v)}\leq 1$, and $\mu_{\max}(\Sigma_{k+1:n}^{2t}) = (1-\frac{\lambda_{k+1}}{2})^{2t}$. Thus we get,
\[\mu_{k+1}((D^{-\frac{1}{2}}M^tS)^{\top}(D^{-\frac{1}{2}}M^tS)) \leq \mu_{\max}(S^{\top}D^{-\frac{1}{2}}V\Sigma_{k+1:n}^{2t}V^{\top}D^{-\frac{1}{2}}S) \leq s\cdot \left(1-\frac{\lambda_{k+1}}{2}\right)^{2t} \text{.}\]

\end{proof}

Next, we bound $\lambda_{k+1}$ from below. For this, we prove a lemma that can be seen as a strengthening of Lemma 5.2 of \cite{CzumajPS_arXiv15}.  Let us first recall Cheeger's inequality that we use later in the proof of Lemma \ref{lem_Cheeger}. 

\begin{lemma}[Cheeger's inequality]
For a general graph $G$, let $L$ denote the normalized Laplacian of $G$, and $\lambda_2$ be the second smallest eigenvalue of $L$. Then
\[\frac{\phi(G)^2}{2} \leq \lambda_2 \leq 2\phi(G) \text{.}\] 
\end{lemma}

\begin{lemma}\label{lem_Cheeger}
Let $G$ be any graph which is $(k,\varphi_{\text{in}})$-clusterable. Let $L$ be its normalized Laplacian matrix, and $\lambda_{k+1}$ be the $(k+1)$st smallest eigenvalue of $L$. Then $\lambda_{k+1}\geq\frac{\varphi_{\text{in}}^2}{2}$.
\end{lemma}

\begin{proof}
Let $C_1,\ldots,C_h$ be a partition of $V_G$ which achieves $\phi^G(C_i)\geq\varphi$ for all $i$, and $h\leq k$. Let $G_{\text{in}}$ be the graph consisting of edges of $G$ with endpoints in the same cluster $C_i$ for some $i$. Let $G_{\text{out}}$ be the graph consisting of edges of $G$ with endpoints in different clusters. Let $D$, $D_{\text{in}}$, and $D_{\text{out}}$ be the diagonal matrices of the degrees of the vertices in $G$, $G_{\text{in}}$, and $G_{\text{out}}$ respectively, so that $D=D_{\text{in}}+D_{\text{out}}$. Let $A$, $A_{\text{in}}$, and $A_{\text{out}}$ be the adjacency matrices of $G$, $G_{\text{in}}$, and $G_{\text{out}}$ respectively, so that $A=A_{\text{in}}+A_{\text{out}}$. Recall that $\lambda_{k+1}$ is the $(k+1)$st eigenvalue of the the normalized Laplacian $L$. Observe that, 
\[L=I-\overline{A} = D^{-\frac{1}{2}}(D-A)D^{-\frac{1}{2}} = D^{-\frac{1}{2}}(D_{\text{in}}-A_{\text{in}})D^{-\frac{1}{2}} + D^{-\frac{1}{2}}(D_{\text{out}}-A_{\text{out}})D^{-\frac{1}{2}}\]

Let $\lambda^{\text{in}}_{k+1}$ be the $(k+1)$st smallest eigenvalue of $D^{-\frac{1}{2}}(D_{\text{in}}-A_{\text{in}})D^{-\frac{1}{2}}$ and $\lambda^{\text{out}}_1$ be the minimum eigenvalue of $D^{-\frac{1}{2}}(D_{\text{out}}-A_{\text{out}})D^{-\frac{1}{2}}$. Then by Lemma \ref{lem_Weyl}, $\lambda_{k+1}\geq\lambda^{\text{in}}_{k+1}+\lambda^{\text{out}}_1$. Observe that $\lambda^{\text{out}}_1\geq0$, since $D^{-\frac{1}{2}}(D_{\text{out}}-A_{\text{out}})D^{-\frac{1}{2}}$ is  positive semi-definite. Therefore, it is sufficient to lower bound $\lambda^{\text{in}}_{k+1}$.

Let graph $G_{\text{in}}'$ is obtained from $G_{\text{in}}$ by increasing the degree of every $a\in V_G$ by $D(aa)-D_{\text{in}}(aa)$, by adding self-loops. Let $A_{\text{in}}'$, and $L_{\text{in}}'$ be the adjacency matrix and the normalized Laplacian of $G_{\text{in}}$ respectively. Observe that $D^{-\frac{1}{2}}(D_{\text{in}}-A_{\text{in}})D^{-\frac{1}{2}}=D^{-\frac{1}{2}}(D-A_{\text{in}}')D^{-\frac{1}{2}}=L_{\text{in}}'$.

Consider the graph $G_{\text{in}}'$. It is composed of $h$ disconnected components, each of which has internal expansion $\varphi_{\text{in}}$. Thus, by applying Cheeger's inequality to each component, we get that, the second smallest eigenvalue of the normalized Laplacian of each component is at least $\frac{\varphi_{\text{in}}^2}{2}$. 
Now the set of eigenvalues of $L_{\text{in}}'$ is the multi-union of the sets of eigenvalues of the components. Thus, we have $\lambda^{\text{in}}_1=\cdots=\lambda^{\text{in}}_h=0$ and $\frac{\varphi_{\text{in}}^2}{2}\leq\lambda^{\text{in}}_{h+1}\leq\ldots\leq\lambda^{\text{in}}_{k+1}$. This implies $\lambda_{k+1}\geq\frac{\varphi_{\text{in}}^2}{2}$, as required.
\end{proof}

\begin{proof}[Proof of Theorem \ref{thm_yes}]
Follows from Lemma \ref{lem_yes_algo} and Lemma \ref{lem_Cheeger}.
\end{proof}

\subsubsection{Proof of Theorem~\ref{thm_no} (the NO case)}\label{sec_no}

The main result of this section is a proof of Theorem~\ref{thm_no}, restated below for convenience of the reader:

\noindent{\em {\bf Theorem~\ref{thm_no}} }(Restated) Let $\varphi_{\text{out}}>0$, $\beta \in (0,1)$, and integer $k\geq 1$. Let  $$s=1600(k+1)^2\cdot \ln(12(k+1))\cdot \ln (\text{vol}(V_G))/(\beta \cdot (1-\eta)) \text{.}$$ Then for every $(k,\varphi_{\text{out}},\beta)$-unclusterable graph $G=(V_G, E_G)$ (see definition \ref{def:clusterable}), with $\min_{v \in V_G}{\text{deg}(v)}\geq 1$, the following holds  with probability at least $\frac{2}{3}$.
\[\textsc{Estimate}(G,k,s,t,\eta) \geq  
\frac{8(k+1)\ln(12(k+1))}{\beta  \cdot (1-\eta)\cdot \text{vol}(V_G)}\times(1-30\varphi_{out})^{2t} \text{.}\]

Consider the NO case, where the vertex set of $G$ contains $k+1$ subsets $C_1,\ldots,C_{k+1}$ of volume at least $\frac{\beta}{k+1}\text{vol}(V_G)$ each, such that for each $i$, $\phi^G_{V_G}(C_i)\leq\varphi_{\text{out}}$. We are interested in bounding the quantity $\mu_{k+1}((D^{-\frac{1}{2}}M^tS)^{\top}(D^{-\frac{1}{2}}M^tS))$ from below. 
Let $\theta$ be a large enough absolute constant (say $\theta=60$). Let $h$ be the largest index such that $\lambda_h<\theta\varphi_{\text{out}}$. 

Recall that $(v_1,\ldots,v_n)$ is an orthonormal basis of eigenvectors of $L$. $V\in\mathbb{R}^{V_G\times[n]}$ is the matrix whose columns are the orthonormal eigenvectors of $L$ arranged in increasing order of eigenvalues. Let $P_h=V_{1:h}V_{1:h}^{\top}$ and $P_h^{\perp}=V_{h+1:n}V_{h+1:n}^{\top}$, so that for any vector $v\in\mathbb{R}^{V_G}$, $P_hv$ is the projection of $v$ onto the span of $\{v_1,\ldots,v_h\}$, and $P_h^{\perp}v$ is its projection on the span of $\{v_{h+1},\ldots,v_n\}$, that is, the orthogonal complement of the span of $\{v_1,\ldots,v_h\}$. Also, observe that $P_h+P_h^{\perp}=I$, $P_h^2=P_h$, and $(P_h^{\perp})^2=P_h^{\perp}$. Let $P=D^{-\frac{1}{2}}P_h$ and $P^{\perp}=D^{-\frac{1}{2}}P_h^{\perp}$. Let $U \in \mathbb{R}^{V_G\times[k+1]}$ be the matrix with orthonormal columns, where for $a\in V_G$, and $1\leq i \leq k+1$,   the $(a,i)$-th entry of $U$ has $\sqrt{\frac{\text{deg}(a)}{\text{vol}(C_i)}}$ if $a\in C_i$, and zero otherwise.


\begin{lemma}\label{lem_WMW}
Let $G=(V_G,E_G)$ be a graph with $\min_{v \in V_G}{\text{deg}(v)}\geq 1$, and with normalized Laplacian $L$ (Definition \ref{def_Laplacian}), and $M$ be the transition matrix of the random walk associated with $G$ (Definition \ref{def_randwalk}). Let $C_1\ldots,C_{k+1}$ be pairwise disjoint subsets of vertices of $G$. Let $U \in \mathbb{R}^{V_G\times[k+1]}$ be the matrix with orthonormal columns, where for $a\in V_G$, and $1\leq i \leq k+1$,   the $(a,i)$-th entry of $U$ has $\sqrt{\frac{\text{deg}(a)}{\text{vol}(C_i)}}$ if $a\in C_i$, and zero otherwise. For $\theta>0$ and $\varphi_{\text{out}}\geq0$, let $h$ be the largest index such that $\lambda_h$, the $h^{\text{\tiny{th}}}$ smallest eigenvalue of $L$, is less than $\theta\varphi_{\text{out}}$. Let $P=D^{-\frac{1}{2}}P_h$.  Let $S$ be any multiset of vertices. (Recall our abuse of notation from Section \ref{sec_prelim}.) Then
\[\mu_{k+1}((D^{-\frac{1}{2}}M^tS)^{\top}(D^{-\frac{1}{2}}M^tS))\geq\left(1-\frac{\theta\varphi_{\text{out}}}{2}\right)^{2t} \cdot\min_{z\in\mathbb{R}^{k+1}\text{, }\lVert z\rVert_2=1}\lVert S^{\top}PUz\rVert_2^2\text{.}\]
\end{lemma}

\begin{proof}
We can write $\mu_{k+1}((D^{-\frac{1}{2}}M^tS)^{\top}(D^{-\frac{1}{2}}M^tS))$ as 
\begin{align*}
\mu_{k+1}((D^{-\frac{1}{2}}M^tS)^{\top}(D^{-\frac{1}{2}}M^tS)) &=\mu_{k+1}((M^tS)^{\top} D^{-1} (M^tS)) \\
&\geq \mu_{k+1}((M^tS)^{\top} D^{-\frac{1}{2}} V_{1:h}V_{1:h}^{\top}D^{-\frac{1}{2}}(M^tS)) && {\text{By Lemma \ref{lem_projection}}} \\
&= \mu_{k+1}((M^tS)^{\top} D^{-\frac{1}{2}} V_{1:h}V_{1:h}^{\top}V_{1:h}V_{1:h}^{\top}D^{-\frac{1}{2}}(M^tS)) && {\text{Since } V_{1:h}^{\top}V_{1:h}=I } \\
&=\mu_{k+1}(S^{\top}{M^t}^{\top}P P ^\top M^tS).
\end{align*}
Recall from Section \ref{sec_prelim} that $ v_i^\top D^{-\frac{1}{2}}$ is a left eigenvector of $M$, and $ D^{-\frac{1}{2}}v_i$ is a right eigenvector of $M^\top$ with eigenvalue $1-\frac{\lambda_i}{2}$.  Thus we can write $V_{1:h}^\top  D^{-\frac{1}{2}} M^t=\Sigma_{1:h}^t V_{1:h}^\top  D^{-\frac{1}{2}}$ and ${M^t}^\top D^{-\frac{1}{2}}V_{1:h}=D^{-\frac{1}{2}}V_{1:h}\Sigma_{1:h}^t$, where $\Sigma_{1:h}^t$ is a $h\times h$ diagonal matrix with entries ${(1-\frac{\lambda_1}{2}})^t,\ldots,({{1-\frac{\lambda_h}{2}}})^t$. Observe that
\[(M^t)^\top P = (M^t)^\top D^{-\frac{1}{2}}V_{1:h}V_{1:h}^{\top} =D^{-\frac{1}{2}}V_{1:h}\Sigma_{1:h}^t V_{1:h}^{\top} \]
Thus we have,
\begin{align*}
\mu_{k+1}((D^{-\frac{1}{2}}M^tS)^{\top}(D^{-\frac{1}{2}}M^tS)) &\geq \mu_{k+1}(S^{\top}{M^t}^{\top}PP^\top M^tS) \\
&=\mu_{k+1}(S^{\top}D^{-\frac{1}{2}}V_{1:h}\Sigma_{1:h}^t V_{1:h}^{\top}V_{1:h}\Sigma_{1:h}^tV_{1:h}^{\top}D^{-\frac{1}{2}} S) \\
&=\max_U\{\quad\min_y \{\lVert V_{1:h}\Sigma_{1:h}^tV_{1:h}^{\top}D^{-\frac{1}{2}} Sy\rVert_2^2 \quad | y \in U,  \lVert y\rVert_2=1   \} \quad| \text{dim}(U)=k+1\}\text{,}
\end{align*}
where the last equality follows from Courant-Fischer min-max principle (Lemma \ref{lem_Courant-Fischer}). Observe that $\Sigma_{1:h}^t$ is a $h\times h$ diagonal matrix with entries ${(1-\frac{\lambda_1}{2}})^t,\ldots,({{1-\frac{\lambda_h}{2}}})^t$, hence  

\[ \lVert V_{1:h}\Sigma_{1:h}^tV_{1:h}^{\top}D^{-\frac{1}{2}} Sy\rVert_2^2 \geq {\left({1-\frac{\lambda_h}{2}}\right)}^{2t} \cdot \lVert V_{1:h}V_{1:h}^{\top}D^{-\frac{1}{2}} Sy\rVert_2^2 = {\left({1-\frac{\lambda_h}{2}}\right)}^{2t} \cdot \lVert P^\top Sy\rVert_2^2\]
Notice that by Courant-Fischer min-max principle (Lemma \ref{lem_Courant-Fischer}) we have 
\[\mu_{k+1}(S^\top P P^\top S)= \max_U\{\quad\min_y \{\lVert P^\top Sy\rVert_2^2 \quad | y \in U,  \lVert y\rVert_2=1   \} \quad| \text{dim}(U)=k+1\}\text{.}\]
Let $U^{*}$ be the subspace with $\text{dim}(U^{*})=k+1$ which maximizes $\min_y \{\lVert P^\top Sy\rVert_2^2 \quad | y \in U,  \lVert y\rVert_2=1\}$. Thus we get,
\begin{align*}
\mu_{k+1}((D^{-\frac{1}{2}}M^tS)^{\top}(D^{-\frac{1}{2}}M^tS))&=  \max_U\{\quad\min_y \{\lVert V_{1:h}\Sigma_{1:h}^tV_{1:h}^{\top}D^{-\frac{1}{2}} Sy\rVert_2^2 \quad | y \in U,  \lVert y\rVert_2=1   \} \quad| \text{dim}(U)=k+1\} \\
&\geq  \min_y \{\lVert V_{1:h}\Sigma_{1:h}^tV_{1:h}^{\top}D^{-\frac{1}{2}} Sy\rVert_2^2 \quad | y \in U^*,  \lVert y\rVert_2=1   \} \\
&\geq  \min_y \left\{{\left({1-\frac{\lambda_h}{2}}\right)}^{2t} \cdot \lVert P^\top  Sy\rVert_2^2 \quad | y \in U^*,  \lVert y\rVert_2=1   \right\} \\
&=  {\left({1-\frac{\lambda_h}{2}}\right)}^{2t} \cdot \min_y \{ \lVert P^\top Sy\rVert_2^2 \quad | y \in U^*,  \lVert y\rVert_2=1   \} \\
&={\left({1-\frac{\lambda_h}{2}}\right)}^{2t} \cdot \mu_{k+1}(S^\top P P^\top S)\text{.}
\end{align*}
Therefore we have
\begin{align*}
\mu_{k+1}(\left(D^{-\frac{1}{2}}M^tS)^{\top}(D^{-\frac{1}{2}}M^tS)\right) &\geq{\left({1-\frac{\lambda_h}{2}}\right)}^{2t} \cdot \mu_{k+1}(S^\top P P^\top S) \\ 
&\geq {\left({1-\frac{\lambda_h}{2}}\right)}^{2t}\cdot\mu_{k+1}(S^{\top}PUU^{\top}P^{\top}S) && \text{By Lemma \ref{lem_projection}} \\
&= {\left({1-\frac{\lambda_h}{2}}\right)}^{2t}\cdot\mu_{k+1}(U^{\top}P^{\top}SS^{\top}PU) && \text{By Lemma \ref{lem_commute}} \\
&={\left({1-\frac{\lambda_h}{2}}\right)}^{2t}\cdot\mu_{\text{min}}(U^{\top}P^{\top}SS^{\top}PU) && \text{Since $U^{\top}P^{\top}SS^{\top}PU$ is $k+1 \times k+1$ matrix}\\
&\geq {\left(1-\frac{\lambda_h}{2}\right)}^{2t}\cdot \min_{z\in\mathbb{R}^{k+1}\text{, }\lVert z\rVert_2=1}\lVert S^{\top}PUz\rVert_2^2 \\
&\geq {\left(1-\frac{\theta\varphi_{\text{out}}}{2}\right)}^{2t}\cdot \min_{z\in\mathbb{R}^{k+1}\text{, }\lVert z\rVert_2=1}\lVert S^{\top}PUz\rVert_2^2.
\end{align*}
\end{proof}

Our goal is to prove that by selecting a random (multi)set $S$ of vertices of a ``reasonable'' size, with at least a constant probability (say $2/3$), for all $z\in\mathbb{R}^{k+1}$ with $\left\|z\right\|_2=1$, we have $\left\|S^{\top}PUz\right\|_2^2 $ is ``large''. 

\begin{lemma} \label{lem_xLx small}
Let $G=(V_G,E_G)$ be a graph with $\min_{v \in V_G}{\text{deg}(v)}\geq 1$, and with normalized Laplacian $L$ (Definition \ref{def_Laplacian}). Let $C_1\ldots,C_{k+1}$ be pairwise disjoint subsets of vertices of $G$ such that $\phi^G_{V_G}(C_i)\leq\varphi_{\text{out}}$ for all $i$. Let $U \in \mathbb{R}^{V_G\times[k+1]}$ be the matrix with orthonormal columns, where for $a\in V_G$, and $1\leq i \leq k+1$,   the $(a,i)$-th entry of $U$ has $\sqrt{\frac{\text{deg}(a)}{\text{vol}(C_i)}}$ if $a\in C_i$, and zero otherwise. Then for every $z\in\mathbb{R}^{k+1}$ with $\lVert z\rVert_2^2=1$, \[z^{\top}U^{\top}LUz \leq 2\varphi_{\text{out}}\].
\end{lemma}

\begin{proof}
We have, $z^{\top}U^{\top}LUz= z^{\top}U^{\top}Uz  - z^{\top}U^{\top}\overline{A}Uz =1-z^{\top}U^{\top}\overline{A}Uz$, where $\overline{A}$ is the normalized adjacency matrix of $G$, since $U^{\top}U=I$. We will prove that every eigenvalue of $U^{\top}\overline{A}U$ lies in $[1-2\varphi_{\text{out}},1]$, and this implies the claim.

Let $m_{ij}$ denote the number of edges between $C_i$ and $C_j$, and $m'_i$ denote the number of edges between $C_i$ and $V_G\setminus\bigcup_{j=1}^{k+1}C_j$. Then observe that
\[(U^{\top}\overline{A}U)_{ij}=u_i^{\top}Au_j=\frac{m_{ij}}{\sqrt{\text{vol}(C_i)\cdot \text{vol}(C_j)}}\text{.}\]

Thus, $U^{\top}\overline{A}U=W^{-1/2}HW^{-1/2}$, where $W=\text{diag}(\text{vol}(C_1),\ldots,\text{vol}(C_{k+1}))$, and $H$ is given by $H_{ij}=m_{ij}$. By Lemma \ref{lem_commute}, the eigenvalues of $W^{-1/2}HW^{-1/2}$ are same as the eigenvalues of $W^{-1}H$. Therefore, it is sufficient to prove that the eigenvalues of $W^{-1}H$ lie in $[1-2\varphi_{\text{out}},1]$.

We know that for all $i$, $\phi^G_{V_G}(C_i) \leq \varphi_{\text{out}}$, and thus, for all $i$, $m'_i+\sum_{j \neq i} m_{ij}\leq  \varphi_{\text{out}}\cdot\text{vol}(C_i)$, and $m_{ii} \geq (1-\varphi_{\text{out}})\text{vol}(C_i)$. Therefore $W^{-1}H$ is the ${(k+1)\times (k+1)}$ matrix such that for all $i$,
\[(W^{-1}H)_{ii} = \frac{m_{ii}}{\text{vol}(C_i)} \geq 1-\varphi_{\text{out}}\]
and
\[\sum_{j \neq i} (W^{-1}H)_{ij} = \frac{1}{\text{vol}(C_i)} \sum_{j \neq i} m_{ij} \leq\varphi_{\text{out}}\text{.}\]
Thus, for every $i$,
\begin{equation} \label{eq-diag-val}
(W^{-1}H)_{ii}+\sum_{j \neq i} (W^{-1}H)_{ij}=\frac{1}{\text{vol}(C_i)}\sum_{j=1}^{k+1}m_{ij}\leq 1\text{,}
\end{equation}
and
\begin{equation} \label{eq-row-val}
(W^{-1}H)_{ii}-\sum_{j \neq i} (W^{-1}H)_{ij}\geq1-2\varphi_{\text{out}}\text{.}
\end{equation}
From \eqref{eq-diag-val} and \eqref{eq-row-val}, and by using the Gershgorin circle theorem (Lemma \ref{lem_Gershgorin}), we conclude that every eigenvalue of $W^{-1}H$ lies within $[1-2\varphi_{\text{out}}, 1]$, as required.
\end{proof}

\begin{lemma} \label{lem_P-perpUz-small}
Let $G=(V_G,E_G)$ be a graph with $\min_{v \in V_G}{\text{deg}(v)}\geq 1$, and  with normalized Laplacian $L$ (Definition \ref{def_Laplacian}). Let $C_1\ldots,C_{k+1}$ be pairwise disjoint subsets of vertices of $G$ such that $\phi^G_{V_G}(C_i)\leq\varphi_{\text{out}}$ for all $i$. Let $U \in \mathbb{R}^{V_G\times[k+1]}$ be the matrix with orthonormal columns, where for $a\in V_G$, and $1\leq i \leq k+1$,   the $(a,i)$-th entry of $U$ has $\sqrt{\frac{\text{deg}(a)}{\text{vol}(C_i)}}$ if $a\in C_i$, and zero otherwise. Let $z\in\mathbb{R}^{k+1}$ with $\lVert z\rVert_2^2=1$. For a constant $\theta>0$, let $h$ be the largest index such that $\lambda_h$, the $h$-th smallest eigenvalue of $L$, is less than $\theta\varphi_{\text{out}}$.  Then \[\left\| P_h^{\perp}Uz \right\|_2^2\leq \frac{2}{\theta}\].
\end{lemma}

\begin{proof}
Recall that $0=\lambda_1\leq\cdots\leq\lambda_n$ are the eigenvalues of $L$ and $v_1,\ldots,v_n$ are the corresponding orthonormal eigenvectors forming a basis of $\mathbb{R}^{V_G}$. Write $Uz\in\mathbb{R}^{V_G}$ in the eignebasis as $Uz=\sum_{i=1}^n \alpha_i v_i$. Then we have, 
\begin{equation} \label{lw_bound_xLx}
(Uz)^{\top}LUz =\left(\sum_{i=1}^n \alpha_i v_i^{\top}\right)L\left(\sum_{i=1}^n \alpha_i v_i\right)=\sum_{i=1}^n \lambda_i\alpha_i^2 \geq \sum_{i=h+1}^n \lambda_i\alpha_i^2 \geq \theta\varphi_{\text{out}}\sum_{i=h+1}^n \alpha_i^2.
\end{equation} 
On the other hand, by Lemma \ref{lem_xLx small}, we have
\begin{equation} \label{up_bound_xLx}
z^{\top}U^{\top}LUz \leq 2\varphi_{\text{out}}.
\end{equation} 
Putting \eqref{lw_bound_xLx} and \eqref{up_bound_xLx} together, we get
$\theta\varphi_{\text{out}}\sum_{i=h+1}^n \alpha_i^2 \leq 2\varphi_{\text{out}}$, and thus  $\sum_{i=h+1}^n \alpha_i^2  \leq \frac{2}{\theta}$.
Recall that $P_h^\perp=V_{h+1:n}^{\top}V_{h+1:n}$, and therefore, 
\[\left\| P_h^{\perp}Uz \right\|_2^2 = \left\| \sum_{i=1}^n \alpha_i P_h^{\perp} v_i \right\|_2^2 = \left\| \sum_{i=h+1}^n \alpha_i  v_i \right\|_2^2 = \sum_{i=h+1}^n \alpha_i^2 \leq \frac{2}{\theta}. \]
\end{proof}

The next two lemmas concern random samples of vertices $S'$, and prove, for any fixed $z$, a lower bound on $\left\| S'^{\top}PUz \right\|_2$ as a function of the size of $S'$. 


\begin{lemma}\label{evenly_distributed}
Let $C_1,\ldots,C_{k+1}$ be subsets of some universe $V_G$, such that for all $j$, $\text{vol}(C_j) \geq \frac{\beta}{k+1} \text{vol}(V_G)$ for some $\beta>0$. Let $\eta \in (0,1)$, and \[s'=\frac{200(k+1)\ln(12(k+1))}{\beta\cdot(1-\eta)} \text{.}\]
Let $S'$ be a multiset of $s'$ independent random vertices in $V_G$, sampled from  distribution $\mathcal{P}$ over $V_G$ such that for all $v \in V_G$, $\left|\mathcal{P}(v)-\frac{\deg(v)}{\text{vol}(G)}\right| \leq \eta\cdot \frac{\deg(v)}{\text{vol}(G)}$. Then with probability at least $\frac{11}{12}$, for every $1\leq j \leq k+1$,
\[|S'\cap C_j| \geq \frac{9}{10} \cdot \frac{\text{vol}(C_j)}{\text{vol}(V_G)}  s'(1-\eta) \text{.}\]
\end{lemma}

\begin{proof}
For $v\in V_G$, and $1\leq r \leq s'$, let $X_{v}^r$ be a random variable which is $1$ if the $r$-th sampled vertex is $v$, and $0$ otherwise. Thus $\mathbb{E}[X_{v}^r]=\mathcal{P}(v)\geq (1-\eta)\frac{\text{deg}(v)}{\text{vol}(V_G)}$. Observe that $|S'\cap C_j|$ is a random variable defined as $\sum_{r=1}^{s'}\sum_{v \in C_j} X_v^r$, where its expectation is given by
\[\mathbb{E}[|S'\cap C_j|]=\sum_{r=1}^{s'}\sum_{v \in C_j} \mathbb{E}[X_v^r]\geq s'(1-\eta)\frac{\text{vol}(C_j)}{\text{vol}(V_G)} \geq s'(1-\eta)\frac{\beta}{k+1}\text{.}\]
Notice that the random variables $X_{v}^r$ are negatively associated, since for each $r$, $\sum_{v \in V_G} X_{v}^r=1$. Therefore, by Chernoff bound,
\[\Pr\left[|S'\cap C_j|<\frac{9s'(1-\eta)}{10}\cdot\frac{\text{vol}(C_j)}{\text{vol}(V_G)}\right] \leq \exp\left(-\frac{s'(1-\eta)}{200}\cdot\frac{\beta}{k+1}\right)\text{.}\]
By union bound,
\[\Pr\left[\exists j\text{: }|S'\cap C_j|<\frac{9s'(1-\eta)}{10}\cdot\frac{\text{vol}(C_j)}{\text{vol}(V_G)}\right]\leq (k+1)\cdot\exp\left(-\frac{s'(1-\eta)}{200}\cdot\frac{\beta}{k+1}\right)\leq\frac{1}{12}\text{,}\]
by our choice of $s'$.
\end{proof}

\begin{lemma}\label{SVz}
Let $G=(V_G,E_G)$ be a graph with $\min_{v \in V_G}{\text{deg}(v)}\geq 1$, and with normalized Laplacian $L$ (Definition \ref{def_Laplacian}). Let $C_1\ldots,C_{k+1}$ be pairwise disjoint subsets of vertices of $G$ of volume at least $\frac{\beta}{k+1}\text{vol}(V_G)$ each, such that $\phi^G_{V_G}(C_i)\leq\varphi_{\text{out}}$ for all $i$. Let $U \in \mathbb{R}^{V_G\times[k+1]}$ be the matrix with orthonormal columns, where for $a\in V_G$, and $1\leq i \leq k+1$,   the $(a,i)$-th entry of $U$ has $\sqrt{\frac{\text{deg}(a)}{\text{vol}(C_i)}}$ if $a\in C_i$, and zero otherwise. Let $z\in\mathbb{R}^{k+1}$ with $\lVert z\rVert_2^2=1$. For $\theta=60$, let $h$ be the largest index such that $\lambda_h$, the $h^{\text{\tiny{th}}}$ smallest eigenvalue of $L$, is less than $\theta\varphi_{\text{out}}$. Let $P=D^{-\frac{1}{2}}P_h$ and $P^{\perp}=D^{-\frac{1}{2}}P_h^{\perp}$. Let $0<\eta\leq\frac{1}{2} $, and \[s'=\frac{200(k+1)\ln(12(k+1))}{\beta\cdot(1-\eta)} \text{.}\]
Let $S'$ be a multiset of $s'$ independent random vertices in $V_G$, sampled from  distribution $\mathcal{P}$ over $V_G$ such that for all $v \in V_G$, $\left|\mathcal{P}(v)-\frac{\deg(v)}{\text{vol}(G)}\right| \leq \eta\cdot \frac{\deg(v)}{\text{vol}(G)}$.
Then $\left\| S'^{\top}PUz \right\|_2 \geq \frac{1}{2}\sqrt{\frac{s'}{\text{vol}(V_G)}}$, with probability at least $\frac{7}{12}$.
\end{lemma}

\begin{proof}
By triangle inequality $\left\| S'^{\top}PUz \right\|_2 \geq \left\| S'^{\top}D^{-\frac{1}{2}}Uz \right\|_2 - \left\| S'^{\top}P^{\perp}Uz \right\|_2$. Observe that the vector $D^{-\frac{1}{2}}Uz$ takes a uniform value $\mu_j=\frac{z_j}{\sqrt{\text{vol}(C_j)}}$ on each set $C_j$. By Lemma \ref{evenly_distributed}, with probability at least $\frac{11}{12}$, we have $|S'\cap C_j| \geq \frac{9s'(1-\eta)}{10} \cdot \frac{\text{vol}(C_j)}{\text{vol}(V_G)}$, for all $j$. In this event, we have,
\begin{align}
\left\| S'^{\top}D^{-\frac{1}{2}}Uz \right\|_2^2 &\geq\sum_{j=1}^{k+1}\frac{9s'(1-\eta)}{10}\cdot\frac{\text{vol}(C_j)}{\text{vol}(V_G)}\cdot\mu_j^2 \nonumber \\
&=\frac{9}{10}\cdot\frac{s'(1-\eta)}{{\text{vol}(V_G)}}\sum_{j=1}^{k+1}\text{vol}(C_j)\cdot\frac{z_j^2}{\text{vol}(C_j)} \nonumber \\
&=\frac{9}{10}\cdot\frac{s'(1-\eta)}{{\text{vol}(V_G)}}\cdot\left\|z\right\|_2^2 \nonumber \\
&=\frac{9}{10}\cdot\frac{s'(1-\eta)}{{\text{vol}(V_G)}} \text{,} \label{eq:1}
\end{align}
where the last equality follows because  $z$ is a unit vector.

Let $y=P_h^{\perp}Uz$. By Lemma \ref{lem_P-perpUz-small}, we have $\left\| y\right\|_2^2\leq \frac{2}{\theta}$. Note that $P^{\perp}Uz=D^{-\frac{1}{2}}P_h^{\perp}Uz=D^{-\frac{1}{2}}y$. Thus we have
\begin{align*}
\mathbb{E}_{S'}\left[\left\| S'^{\top}P^{\perp}Uz\right\|_2^2\right] 
&= \mathbb{E}_{S'}\left[\left\| S'^{\top}D^{-\frac{1}{2}}P_h^{\perp}Uz\right\|_2^2\right]  \\
&= s'\cdot\sum_{v \in V_G} \mathcal{P}(v)\cdot\frac{y(v)^2}{\text{deg}(v)}\\
&\leq s'\cdot\sum_{v \in V_G} (1+\eta)\cdot\frac{\text{deg}(v)}{\text{vol}(V_G)}\cdot\frac{y(v)^2}{\text{deg}(v)}\\
&=\frac{s'(1+\eta)}{{\text{vol}(V_G)}} \left\| y\right\|_2^2 \\
&\leq \frac{2}{\theta}\cdot \frac{s'(1+\eta)}{{\text{vol}(V_G)}} \text{.}
\end{align*}
Thus, by Markov's inequality, with probability at least $\frac{2}{3}$, 
\begin{equation} \label{eq:2} 
\left\| S'^{\top}P^{\perp}Uz\right\|_2^2 \leq 3\cdot \mathbb{E}[\left\| S'^{\top}P^{\perp}Uz\right\|_2^2] \leq \frac{6}{\theta}\cdot \frac{s'}{{\text{vol}(V_G)}}=\frac{1}{10}\cdot \frac{s'(1+\eta)}{{\text{vol}(V_G)}}\text{,}
\end{equation}
where we get the last equality by recalling that $\theta=60$. Putting \eqref{eq:1} and \eqref{eq:2} together, we get with probability at least $1-\frac{1}{3}-\frac{1}{12}=\frac{7}{12}$,
\[ \left\| S'^{\top}PUz \right\|_2 \geq \left\| S'^{\top}D^{-\frac{1}{2}}Uz \right\|_2 - \left\| S'^{\top}P^{\perp}Uz \right\|_2 \geq \left(\sqrt{\frac{9(1-\eta)}{10}}-\sqrt{\frac{(1+\eta)}{10}}\right)\sqrt{\frac{s'}{{\text{vol}(V_G)}}} \geq \frac{1}{4} \sqrt{\frac{s'}{{\text{vol}(V_G)}}}\text{.} \]
\end{proof}

The above lemma gives a lower bound on $\left\| S'^{\top}PUz \right\|_2$ which holds with a constant probability. We next show how we can trade off the lower bound to get a guarantee that holds with probability arbitrarily close to one.

\begin{lemma} \label{SVz-fix-z}
Let $G=(V_G,E_G)$ be a graph with $\min_{v \in V_G}{\text{deg}(v)}\geq 1$, and with normalized Laplacian $L$ (Definition \ref{def_Laplacian}). Let $C_1\ldots,C_{k+1}$ be pairwise disjoint subsets of vertices of $G$ of volume at least $\frac{\beta}{k+1}\text{vol}(V_G)$ each, such that $\phi^G_{V_G}(C_i)\leq\varphi_{\text{out}}$ for all $i$. Let $U \in \mathbb{R}^{V_G\times[k+1]}$ be the matrix with orthonormal columns, where for $a\in V_G$, and $1\leq i \leq k+1$,   the $(a,i)$-th entry of $U$ has $\sqrt{\frac{\text{deg}(a)}{\text{vol}(C_i)}}$ if $a\in C_i$, and zero otherwise. Let $z\in\mathbb{R}^{k+1}$ with $\lVert z\rVert_2^2=1$. For $\theta=60$, let $h$ be the largest index such that $\lambda_h$, the $h^{\text{\tiny{th}}}$ smallest eigenvalue of $L$, is less than $\theta\varphi_{\text{out}}$. Let $P=D^{-\frac{1}{2}}P_h$ and $P^{\perp}=D^{-\frac{1}{2}}P_h^{\perp}$. Let $0<\eta\leq\frac{1}{2} $, and \[s=\frac{1600(k+1)^2\cdot \ln(12(k+1))\cdot \ln (\text{vol}(V_G))}{\beta\cdot(1-\eta)} \text{.}\]
Let $S'$ be a multiset of $s'$ independent random vertices in $V_G$, sampled from  distribution $\mathcal{P}$ over $V_G$ such that for all $v \in V_G$, $\left|\mathcal{P}(v)-\frac{\deg(v)}{\text{vol}(G)}\right| \leq \eta\cdot \frac{\deg(v)}{\text{vol}(G)}$.
Then for $\tau=8(k+1)\ln \text{vol}(V_G)$, we have $\left\| S^{\top}PUz \right\|_2 \geq \frac{1}{4}\sqrt{\frac{s}{\tau \cdot \text{vol}(V_G)}}$, with probability at least $1-\left(\frac{5}{12}\right)^\tau$.
\end{lemma}

\begin{proof}
We represent $S$ as a multi-union $S=S_1 \cup S_2 \cup \ldots S_{\tau}$ of independently drawn sets containing $s'=\frac{s}{\tau}$ independent samples each, where
\[s'=\frac{s}{\tau}=\frac{200(k+1)\ln(12(k+1))}{\beta\cdot(1-\eta)}\text{.}\]
Using Lemma \ref{SVz}, for any $i$, $\Pr\left[\left\| S_i^{\top}PUz \right\|_2 \geq \frac{1}{4}\sqrt{\frac{s}{\tau \cdot \text{vol}(V_G)}}\right] \geq \frac{7}{12}$. Since the $S_i$'s are independent sets of samples, $\Pr\left[\exists i\text{ }\left\| S_i^{\top}PUz \right\|_2 \geq\frac{1}{4}\sqrt{\frac{s}{\tau \cdot \text{vol}(V_G)}}\right] \geq 1-(\frac{5}{12})^\tau$. Therefore with probability at least $1-\left(\frac{5}{12}\right)^\tau$, we have
\[\left\|S^{\top}PUz\right\|_2^2=\sum_i \left\|S_i^{\top}PUz\right\|_2^2 \geq \left(\frac{1}{4}\sqrt{\frac{s}{\tau \cdot \text{vol}(V_G)}}\right)^2 = \frac{s}{16\tau  \text{vol}(V_G)} \]
Thus, $\Pr\left[\left\|S^{\top}PUz\right\|_2 \geq \frac{1}{4} \sqrt\frac{s}{\tau \cdot \text{vol}(V_G)}\right]\geq 1-(\frac{5}{12})^\tau $.
\end{proof}

In the next lemma, we switch the order of quantification and prove that with a constant probability, a random $S$ achieves a large value for $\left\|S^{\top}PUz\right\|_2$ for all $z$ of unit norm simultaneously, with constant probability. 

\begin{lemma} \label{eps-net}
Let $G=(V_G,E_G)$ be a graph with $\min_{v \in V_G}{\text{deg}(v)}\geq 1$, and with normalized Laplacian $L$ (Definition \ref{def_Laplacian}). Let $C_1\ldots,C_{k+1}$ be pairwise disjoint subsets of vertices of $G$ of volume at least $\frac{\beta}{k+1}\text{vol}(V_G)$ each, such that $\phi^G_{V_G}(C_i)\leq\varphi_{\text{out}}$ for all $i$. Let $U \in \mathbb{R}^{V_G\times[k+1]}$ be the matrix with orthonormal columns, where for $a\in V_G$, and $1\leq i \leq k+1$,   the $(a,i)$-th entry of $U$ has $\sqrt{\frac{\text{deg}(a)}{\text{vol}(C_i)}}$ if $a\in C_i$, and zero otherwise. Let $z\in\mathbb{R}^{k+1}$ with $\lVert z\rVert_2^2=1$. For $\theta=60$, let $h$ be the largest index such that $\lambda_h$, the $h^{\text{\tiny{th}}}$ smallest eigenvalue of $L$, is less than $\theta\varphi_{\text{out}}$. Let $P=D^{-\frac{1}{2}}P_h$ and $P^{\perp}=D^{-\frac{1}{2}}P_h^{\perp}$. Let $0<\eta\leq\frac{1}{2} $, and \[s=\frac{1600(k+1)^2\cdot \ln(12(k+1))\cdot \ln (\text{vol}(V_G))}{\beta\cdot(1-\eta)} \text{.}\]
Let $S'$ be a multiset of $s'$ independent random vertices in $V_G$, sampled from  distribution $\mathcal{P}$ over $V_G$ such that for all $v \in V_G$, $\left|\mathcal{P}(v)-\frac{\deg(v)}{\text{vol}(G)}\right| \leq \eta\cdot \frac{\deg(v)}{\text{vol}(G)}$.
Then with probability at least $\frac{2}{3}$, for all $z\in\mathbb{R}^{k+1}$ with $\left\|z\right\|_2^2=1$,  we have $\left\|S^{\top}PUz\right\|_2 \geq \frac{1}{5}\sqrt{\frac{s}{\tau \cdot \text{vol}(V_G)}}$, where $\tau=8(k+1)\ln \text{vol}(V_G)$.
\end{lemma}

\begin{proof}
Let $\mathcal{N}$ be a $k+1$ dimensional $\delta$-net of size $(\frac{4}{\delta})^{k+1}$ on the Euclidean sphere of radius $1$, for some small enough $\delta$. Let us first explain how to construct such $\delta$-net. Pick $y_1$ of unit norm in $\mathbb{R}^{k+1}$, and then for every $t\geq2$ pick $y_t$ of unit norm such that $\left\|y_t-y_j\right\|_2 \geq \delta$ for all $j = 1,\ldots,t-1$, until no such $y$ can be picked. Note that balls of radius $\frac{\delta}{2}$ centered at the $y_t$’s are disjoint, and their union belongs to the
ball of radius $1 + \frac{\delta}{2}$ centered at zero. Thus $|\mathcal{N}|\leq \frac{(1+\frac{\delta}{2})^{k+1}}{(\frac{\delta}{2})^{k+1}} \leq (\frac{4}{\delta})^{k+1}$.

For $z\in\mathbb{R}^{k+1}$ with $\left\|z\right\|_2=1$, let $\mathcal{N}(z)$, be the closest point to $z$ from $\mathcal{N}$. 
Using Lemma \ref{SVz-fix-z}, by union bound over all points in $\mathcal{N}$ we have that with probability at least $1-(\frac{5}{12})^\tau(\frac{4}{\delta})^{k+1}$, for all $y\in\mathcal{N}$, $\left\|S^{\top}PU y\right\|_2 \geq \frac{1}{4}\sqrt{\frac{s}{\tau \cdot \text{vol}(V_G)}}$. Therefore, with probability at least $1-(\frac{5}{12})^\tau(\frac{4}{\delta})^{k+1}$, we have for every $z\in\mathbb{R}^{k+1}$ with $\left\|z\right\|_2=1$,
$ \left\|S^{\top}PU \mathcal{N}(z)\right\|_2 \geq \frac{1}{4}\sqrt{\frac{s}{\tau \cdot \text{vol}(V_G)}} $. 

Observe that 
\begin{align*}
 \left\|S^{\top}PU (z-\mathcal{N}(z)) \right\|_2^2  &=  \left\|S^{\top}PU \right\|_2^2 \cdot  \left\|(z-\mathcal{N}(z)) \right\|_2^2  \\
&\leq \mu_{\max}(U^\top P^\top S S^{\top}PU) \cdot \delta\\ &\leq \mu_{\max}(UU^\top)\cdot \mu_{\max}(PP^\top)\cdot  \mu_{\max}(SS^\top) \cdot \delta && \text{(By Lemma \ref{lem_commute}, and Lemma \ref{lem_submult})} \\
&\leq \mu_{\max}(PP^\top)\cdot  \mu_{\max}(SS^\top) \cdot \delta && (\text{Since } UU^\top=I) \\
&\leq \mu_{\max}(V_hV_h^\top)\cdot \mu_{\max}(D^{-1})\cdot  \mu_{\max}(SS^\top) \cdot \delta && (\text{Since } PP^\top=D^{-\frac{1}{2}}V_hV_h^\top D^{-\frac{1}{2}})
\end{align*}

Next, observe that $SS^{\top}\in N\times N$ is a diagonal matrix whose $(a,a)^{\text{\tiny{th}}}$ entry is the multiplicity of vertex $a$ in $S$. Thus, $\mu_{\max}(SS^{\top})$ is the maximum multiplicity over all vertices, which is at most $s$. Also notice that $\mu_{\max}(D^{-1})=\max_{v \in V_G}\frac{1}{\text{ deg}(v)}\leq 1$, and $\mu_{\max}(V_hV_h^\top) = 1$, since $P_h$ is a projection matrix. Thus we get
\[ \left\|S^{\top}PU (z-\mathcal{N}(z)) \right\|_2^2  \leq s \cdot \delta \]
Therefore, with probability at least $1-(\frac{5}{12})^\tau(\frac{4}{\delta})^{k+1}$, for every $z\in\mathbb{R}^h$ with $\left\|z\right\|_2=1$, we have
\[ \left\|S^{\top}PU z \right\|_2 \geq \left\|S^{\top}PU \mathcal{N}(z) \right\|_2-\left\|S^{\top}PU (z-\mathcal{N}(z)) \right\|_2 \geq \frac{1}{4}\sqrt{\frac{s}{\tau \cdot \text{vol}(V_G)}} - s\cdot\delta. \]
By setting $\delta=\frac{1}{20s} \sqrt{\frac{s}{\tau \cdot \text{vol}(V_G)}}$, we get
$\left\|S^{\top}PU z \right\|_2 \geq  \frac{1}{5}\sqrt{\frac{s}{\tau \cdot \text{vol}(V_G)}}$ with probability at least 
\[1 -\left(\frac{5}{12}\right)^\tau\left(\frac{4}{\delta}\right)^{k+1}=1-\left(\frac{5}{12}\right)^\tau\left(80\sqrt{s\cdot \tau \cdot \text{vol}(V_G)}\right)^{k+1}\text{.}\]
Observe that $\tau=8(k+1)\ln  \text{vol}(V_G)$ is large enough to ensure that the above probability is at least $\frac{2}{3}$.
\end{proof}

\begin{proof}[Proof of Theorem \ref{thm_no}]
Follows from Lemma \ref{lem_WMW} and Lemma \ref{eps-net}.
\end{proof}


\subsection{Lifting the Oracle Assumption}\label{sec_liftingoracle}

The goal of this section is to show how we can remove the oracle assumption that we made in Section \ref{sec_oracle}, and get an algorithm for the \partitiontesting problem that fits into the query complexity model, defined in Section \ref{sec_ourresults}, that only allows (uniformly random) vertex, degree, and neighbor queries. This will then establish Theorem \ref{thm_main}. The algorithm is presented as a main procedure \textsc{PartitionTestWithoutOracle} (Algorithm \ref{disg_alg_wo_oracle}) that calls the subroutine \textsc{EstimateWithoutOracle} (Algorithm \ref{alg_estimate_wo_oracle}). These two procedures can be seen as a analogs of the procedures \textsc{PartitionTest} (Algorithm \ref{disg_alg}) and \textsc{Estimate} (Algorithm \ref{alg_estimate}) respectively, from Section \ref{sec_oracle}.

\begin{algorithm}
\caption{\textbf{\textsc{EstimateWithoutOracle}($G,k,s,t,\sigma,R,\eta$)}}\label{alg_estimate_wo_oracle}
\begin{algorithmic}[1]
\Procedure{\textsc{EstimateWithoutOracle}($G,k,s,t,\sigma,R,\eta$)}{}
	\State Sample $s$ vertices from $N$ independently and with probability proportional to the degree of the vertices  at random with replacement using \textbf{sampler}($G,\eta$).
Let $S$ be the multiset of sampled vertices.
	\State $r= 192s\sqrt{\text{vol}(V_G)}$.
	\For{Each sample $a\in S$}
		\If{\textbf{$\ell_2^2$-norm tester}($G,a,\sigma,r$) rejects} 
			\Return $\infty$. \Comment High collision probability
		\EndIf
	\EndFor
	\For{Each sample $a\in S$}
		\State Run $R$ random walks starting from $a$, and let $\mathbf{q}_a$ be the distribution of the $t$-step random walk started at $a$.
	\EndFor
	\State Let $Q$ be the matrix whose columns are $\{D^{-\frac{1}{2}}\mathbf{q}_a:a\in S\}$.
	\State Return $\mu_{k+1}(Q^{\top}Q)$.
\EndProcedure
\end{algorithmic}
\end{algorithm}

\begin{algorithm}
\caption{\textsc{PartitionTestWithoutOracle}($G,k,\varphi_{\text{in}},\varphi_{\text{out}},\beta$)}\label{disg_alg_wo_oracle}
\begin{algorithmic}[1]
\Procedure{\textsc{PartitionTestWithoutOracle}($G,k,\varphi_{\text{in}},\varphi_{\text{out}},\beta$)}{}
	\Comment Need: $\varphi_{\text{in}}^2>480\varphi_{\text{out}}$
	\State $\eta:=0.5$
	\State $s:=1600(k+1)^2\cdot \ln(12(k+1))\cdot \ln (\text{vol}(V_G)/(\beta(1-\eta))$.
	\State $c:= \frac{20}{\varphi_{\text{in}}^2}$, $t:=c\ln (\text{vol}(V_G))$  \Comment Observe: $c>0$.
	\State $\sigma:=\frac{192sk(1+\eta)}{\text{vol}(V_G)}$.
	
	\State $\mu_{\text{thres}}:=\frac{1}{2}\cdot\frac{8(k+1)\ln(12(k+1))}{\beta\cdot(1-\eta)}\times \text{vol}(V_G)^{-1-120c\varphi_{\text{out}}}$.
	
	\State $\mu_{\text{err}}=\frac{1}{3}\cdot\frac{8(k+1)\ln(12(k+1))}{\beta\cdot(1-\eta)}\times \text{vol}(V_G)^{-1-120c\varphi_{\text{out}}}$
	
	\State $R:=\max\left(\frac{100s^2\sigma^{1/2}}{\mu_{\text{err}}},\frac{200s^4\sigma^{3/2}}{\mu_{\text{err}}^2}\right)$.

	\If {\textsc{EstimateWithoutOracle}($G,k,s,t,\sigma,R,\eta$) $\leq \mu_{\text{thres}}$} 
		\State Accept $G$.
	\Else 
		\State Reject $G$.
	\EndIf
\EndProcedure
\end{algorithmic}
\end{algorithm}

Recall from Definition \ref{def_randwalk} that with the graph $G$ we associated a random walk, and let $M$ be the transition matrix of that random walk. For a vertex $a$ of $G$, denote by $\mathbf{p}^t_a=M^t\mathds{1}_a$ the probability distribution of of a $t$ step random walk starting from $a$. 
Recall that \textsc{Estimate} assumed the existence of an oracle that takes a vertex $a$ of $G$ as input, and returns $D^{-\frac{1}{2}}M^t\mathds{1}_a$. \textsc{EstimateWithoutOracle} simulates the behavior of the oracle by running several $t$-step random walks from $a$. For any vertex $b$, the fraction of the random walks ending in $b$ is taken as an estimate of $\mathbf{p}^t_a(b)=\mathds{1}_b^{\top}M^t\mathds{1}_a$, the probability that the $t$-step random walk started from $a$ ends in $b$. However, for this estimate to have sufficiently small variance, the quantity $\lVert D^{-\frac{1}{2}} \mathbf{p}^t_a\rVert_2^2$ needs to be small enough. To check this, \textsc{EstimateWithoutOracle} uses the procedure \textbf{$\ell_2^2$-norm tester}, whose guarantees are formally specified in the following lemma.

\begin{lemma}
\label{lem:l2norm-tester}
Let $G=(V_G,E_G)$. Let $a\in V_G$, $\sigma>0$, $0<\delta<1$, and $R\geq \frac{16\sqrt{\text{vol}(G)}}{\delta} $. Let $t\geq1$, and $\mathbf{p}^t_a$ be the probability distribution of the endpoints of a $t$-step random walk starting from $a$. There exists an algorithm, denoted by \textbf{$\ell_2^2$-norm tester}($G,a,\sigma,R$), that outputs accept if $\lVert D^{-\frac{1}{2}} \mathbf{p}^t_a\rVert_2^2\leq \frac{\sigma}{4}$, and outputs reject if $\lVert D^{-\frac{1}{2}} \mathbf{p}^t_a\rVert_2^2>\sigma$, with probability at least $1-\delta$. The running time of the tester is $O(R\cdot t)$.
\end{lemma}

Our \textbf{$\ell_2^2$-norm tester} is a modification of \textbf{$\ell_2^2$-norm tester} in \cite{CzumajPS_arXiv15}. We defer the proof of this lemma to Appendix \ref{app:matrix-estimation}. The running time of \textbf{$\ell_2^2$-norm tester} is independent of $\sigma$, since $\lVert D^{-\frac{1}{2}} \mathbf{p}^t_a\rVert_2^2\geq \frac{1}{\text{vol}(G)}$ for all $a \in V_G$. We will use the following definition of a $(\sigma,t)$-good vertex for the rest of the section.

\begin{definition}\label{lem_l2tester}
We say that a vertex $a\in V_G$, is \textit{$(\sigma,t)$-good} if $\lVert D^{-\frac{1}{2}}\mathbf{p}^t_a\rVert_2^2\leq\sigma$.
\end{definition}

We first claim that for all multisets $S$ containing only $(\sigma,t)$-good vertices, with a good probability over the $R$ random walks, the quantity $Q^{\top}Q$ that Algorithm \ref{alg_estimate_wo_oracle} returns is a good approximation to $(D^{-\frac{1}{2}}M^tS)^{\top}(D^{-\frac{1}{2}}M^tS)$ in Frobenius norm.

\begin{lemma}\label{lem_matrix_estimation}
Let $G=(V_G,E_G)$ be a graph. Let $0<\sigma\leq1$ $t>0$, $\mu_{\text{err}}>0$, $k$ be an integer, and let $S$ be a multiset of $s$ vertices, all whose elements are $(\sigma,t)$-good. Let
\[R=\max\left(\frac{100s^2\sigma^{1/2}}{\mu_{\text{err}}},\frac{200s^4\sigma^{3/2}}{\mu_{\text{err}}^2}\right)\text{.}\]
For each $a\in S$ and each $b\in V_G$, let $\mathbf{q}_a(b)$ be the random variable which denotes the fraction out of the $R$ random walks starting from $a$, which end in $b$. Let $Q$ be the matrix whose columns are $(D^{-\frac{1}{2}} \mathbf{q}_a)_{a\in S}$. Then with probability at least $49/50$,
$|\mu_{k+1}(Q^{\top}Q)-\mu_{k+1}((D^{-\frac{1}{2}}M^tS)^{\top}(D^{-\frac{1}{2}}M^tS))|\leq\mu_{\text{err}}$.
\end{lemma}

The proof of the lemma is given in Appendix~\ref{app:matrix-estimation}. We now prove that Algorithm \ref{disg_alg_wo_oracle} indeed outputs a YES with good probability on a YES instance. For this, we need the following lemma which is a modification of Lemma 4.3 of  \cite{CzumajPS_arXiv15}, and the proof of this lemma is deferred to Appendix~\ref{app:matrix-estimation}.

\begin{lemma}\label{lem_CPS4.3}
For all $0<\alpha<1$, and all $G=(V_G,E_G)$ which is $(k,\varphi_{\text{in}})$-clusterable, there exists $V'_G\subseteq V_G$ with $\text{vol}(V'_G)\geq (1-\alpha) \text{vol}(V_G)$ such that for any $t\geq \frac{2\ln (\text{vol}(V_G))}{\varphi_{\text{in}}^2}$, every $u\in V'_G$ is $\left(\frac{2k}{\alpha \cdot \text{vol}(V_G)},t\right)$-good.
\end{lemma}

\begin{theorem} \label{thm:yes_wo_oracle}
Let $\varphi_{\text{in}}>0$, and integer $k \geq 1$. Then for every $(k,\varphi_{\text{in}})$-clusterable graph $G=(V_G, E_G)$ (see definition \ref{def:clusterable}), with $\min_{v \in V_G}{\text{deg}(v)}\geq 1$, Algorithm \ref{disg_alg_wo_oracle} accepts $G$ with probability at least $\frac{5}{6}$.
\end{theorem}

\begin{proof}
If Algorithm \ref{disg_alg_wo_oracle} outputs a NO one of the following events must happen.
\begin{itemize}
\item $E_1$: Some vertex in $S$ is not $(\frac{\sigma}{4} ,t)$-good.
\item $E_2$: All vertices in $S$ are $(\frac{\sigma}{4},t)$-good, but \textbf{$\ell_2^2$-norm tester} fails on some vertex.
\item $E_3$: All vertices in $S$ are $(\frac{\sigma}{4},t)$-good, and \textbf{$\ell_2^2$-norm tester} succeeds on all vertices, but $|\mu_{k+1}(Q^{\top}Q)-\mu_{k+1}((D^{-\frac{1}{2}}M^tS)^{\top}(D^{-\frac{1}{2}}M^tS))|>\mu_{\text{err}}$.
\end{itemize}
If none of the above happen then Algorithm \ref{alg_estimate_wo_oracle} returns
\[\mu_{k+1}(Q^{\top}Q)\leq\mu_{k+1}((D^{-\frac{1}{2}}M^tS)^{\top}(D^{-\frac{1}{2}}M^tS))+\mu_{\text{err}}\leq\mu_{\text{yes}}+\mu_{\text{err}}<\mu_{\text{thres}}\text{,}\]
and Algorithm \ref{disg_alg_wo_oracle} accepts.

Recall that we use \textbf{sampler}($G,\eta$) to sample vertices, where $\eta=\frac{1}{2}$.
Apply Lemma \ref{lem_CPS4.3} with $\alpha=\frac{1}{24s(1+\eta)}$. Then by the union bound, with probability at least $1-\alpha\cdot(1+\eta)=1-\frac{1}{24}$ all the vertices in $S$ are $\left(\frac{48sk(1+\eta)}{\text{vol}(V_G)},t\right)$-good, that is, $(\frac{\sigma}{4},t)$-good, where $\sigma=\frac{192sk}{\text{vol}(V_G)}$, as chosen in Algorithm \ref{alg_estimate_wo_oracle}. Thus, $\Pr[E_1]\leq\frac{1}{24}$. Given that $E_1$ doesn't happen, by Lemma \ref{lem_l2tester}, on any sample, \textbf{$\ell_2^2$-norm tester} fails with probability at most $\frac{16\sqrt{\text{vol}(V_G)}}{r}<\frac{1}{12s}$ for $r=192s\sqrt{\text{vol}(V_G)}$, as chosen in Algorithm \ref{alg_estimate_wo_oracle}. Thus, with probability at least $1- \frac{1}{12}$, \textbf{$\ell_2^2$-norm tester} succeeds on all the sampled vertices, which implies $\Pr[E_2]\leq\frac{1}{12}$. Given that both $E_1$ and $E_2$ don't happen, by Lemma \ref{lem_matrix_estimation}, with probability at least $\frac{49}{50}$, Algorithm \ref{alg_estimate_wo_oracle} returns a value that is at most $\mu_{\text{err}}$ away from $\mu_{k+1}((D^{-\frac{1}{2}}M^tS)^{\top}(D^{-\frac{1}{2}}M^tS))$. Thus, $\Pr[E_3]\leq \frac{1}{50}$. By the union bound, the probability that Algorithm \ref{disg_alg_wo_oracle} rejects is at most $ \frac{1}{24}+\frac{1}{12}+\frac{1}{50}<\frac{1}{6}$.
\end{proof}

Next, we prove that Algorithm \ref{disg_alg_wo_oracle} indeed returns a NO with good probability on a NO instance.

\begin{theorem} \label{thm:no_wo_oracle}
Let $\varphi_{\text{out}}>0$, $\beta \in (0,1)$, and integer $k\geq 1$. Then for every $(k,\varphi_{\text{out}},\beta)$-unclusterable graph $G=(V_G, E_G)$ (see definition \ref{def:clusterable}), with $\min_{v \in V_G}{\text{deg}(v)}\geq 1$, Algorithm \ref{disg_alg_wo_oracle} rejects $G$ with probability at least $\frac{4}{7}$.
\end{theorem}

\begin{proof}
If the algorithm outputs a YES, then one of the following events must happen.
\begin{itemize}
\item $E_1$: Some vertices in $S$ are not $(\sigma,t)$-good, but \textbf{$\ell_2^2$-norm tester} misses these and passes all vertices.
\item $E_2$: All vertices in $S$ are $(\sigma,t)$-good, but $|\mu_{k+1}(Q^{\top}Q)-\mu_{k+1}((D^{-\frac{1}{2}}M^tS)^{\top}(D^{-\frac{1}{2}}M^tS))|>\mu_{\text{err}}$.
\item $E_3$: $\mu_{k+1}((D^{-\frac{1}{2}}M^tS)^{\top}(D^{-\frac{1}{2}}M^tS))<\mu_{\text{no}}=\frac{8(k+1)\ln(12(k+1))}{\beta\cdot(1-\eta)}\times \text{vol}(V_G)^{-1-120c\varphi_{\text{out}}}$.
\end{itemize}
If none of the above happen then Algorithm \ref{alg_estimate_wo_oracle} returns
\[\mu_{k+1}(Q^{\top}Q)\geq\mu_{k+1}((D^{-\frac{1}{2}}M^tS)^{\top}(D^{-\frac{1}{2}}M^tS))-\mu_{\text{err}}\geq\mu_{\text{no}}-\mu_{\text{err}}>\mu_{\text{thres}}\text{,}\]
and Algorithm \ref{disg_alg_wo_oracle} rejects.

By Lemma \ref{lem_l2tester}, the probability that \textbf{$\ell_2^2$-norm tester} passes a bad vertex is at most $\frac{16\sqrt{\text{vol}(V_G)}}{r}<\frac{1}{12s}$ for $r=192s\sqrt{\text{vol}(V_G)}$, as chosen in Algorithm \ref{alg_estimate_wo_oracle}. Thus, $\Pr[E_1]\leq\frac{1}{12}$. If all vertices in $S$ are $(\sigma,t)$-good, by Lemma \ref{lem_matrix_estimation}, with probability at least $\frac{49}{50}$, Algorithm \ref{alg_estimate_wo_oracle} returns a value that is at most $\mu_{\text{err}}$ away from $\mu_{k+1}((D^{-\frac{1}{2}}M^tS)^{\top}(D^{-\frac{1}{2}}M^tS))$. Thus, $\Pr[E_2]\leq \frac{1}{50}$. Finally, by Theorem \ref{thm_no}, $\Pr[E_3]\leq \frac{1}{3}$. By the union bound, the probability that Algorithm \ref{disg_alg_wo_oracle} accepts is at most $\frac{1}{12} + \frac{1}{50}+\frac{1}{3}<\frac{3}{7}$
\end{proof}


Now we are set to prove Theorem \ref{thm_alg_partitiontesting}.


\begin{proof}[Proof of Theorem \ref{thm_alg_partitiontesting}]
The correctness of the algorithm is guaranteed by Theorems \ref{thm:yes_wo_oracle} and \ref{thm:no_wo_oracle}. Since these theorems give correctness probability that is a constant larger than $1/2$, it can be boosted up to $2/3$ using standard techniques (majority of the answers of a sufficiently
large constant number of independent runs). It remains to analyze the query complexity. The running time of the \textbf{sampler} algorithm to sample each vertex is $\tilde{O}(\frac{|V_G|}{\text{vol}(G)})$. Hence in total the query complexity of sampling is $\tilde{O}(s\cdot\sqrt{\text{vol}(G)})$. For each of the $s$ sampled vertices, we run \textbf{$\ell_2^2$-norm tester} once, followed by $R$ random walks of $t$ steps each. Each call to the \textbf{$\ell_2^2$-norm tester} takes $O(rt)=O(st\sqrt{\text{vol}(V_G)})=O(st\sqrt{m})$ queries, as guaranteed by Lemma \ref{lem:l2norm-tester}. The random walks from each vertex take $O(Rt)$ time. Thus, the overall query complexity is $O(srt+sRt+s\sqrt{m})$. Substituting the values of $s$, $r$, $R$, and $t$ as defined in Algorithm \ref{disg_alg_wo_oracle}, and noting that $m=\text{vol}(V_G)/2$, we get the required bound.
\end{proof}

\input{app.tex}
\end{appendix}

\bibliographystyle{alpha}
\bibliography{references}

%% file: lowerbound.tex
\section{Lower Bound for the \noisyparities Problem with Applications}

Recall the definition of \noisyparities from Section \ref{sec_intro}, and consider a deterministic algorithm querying $T$ out of $n$ vertices in an instance. We wish to prove the following lower bound on the number of queries $T$ needed to get a nontrivial advantage over a random guess.

\noindent\textbf{Theorem \ref{thm_main_aux_lb} (restated).} Consider a deterministic algorithm ALG for the \noisyparities problem with parameters $d$ and $\varepsilon$. Let $b=1/(8\ln d)$. Suppose ALG makes at most $n^{1/2+\delta}$ queries on $n$ vertex graphs, where $\delta<\min(1/16,b\varepsilon)$. Then ALG gives the correct answer with probability at most $1/2+o(1)$.

We present the proof of this theorem in Section \ref{sec_lbproof}, but first we setup some preliminaries here, and then use this theorem to establish query complexity lower bounds in Section \ref{sec_reductions}.

\subsection{Preliminaries and notation}

\subsubsection{Random $d$-regular Graphs and the Configuration Model}

Recall that the \noisyparities problem (Definition \ref{def_noisyparities}) has a random $d$-regular graph generated according to the configuration model as its underlying graph. The configuration model of Bollob\'{a}s generates a random $d$ regular graph $G=(V,E)$ over a set $V$ of $n$ vertices (provided $dn$ is even) as follows. It first generates $d$ half-edges on each vertex and identifies the set of half-edges with $V\times[d]$. Then in each round, an arbitrary unpaired half-edge $(u,i)$ of some arbitrary vertex $u$ is picked, and it is paired up with a uniformly random unpaired half-edge $(v,j)$. This results in the addition of an edge $(u,v)$ to $E$. This continues until all the half-edges are paired up. (This might result in self-loops and parallel edges, so $G$ is not necessarily simple.) The following is known about the expansion of random $d$-regular graphs generated by the configuration model \cite{Bollobas88}.

\begin{fact}\label{fact_Bollobas}
For $d\geq 3$ let $\eta(d)\in(0,1)$ be such that $(1-\eta(d))\log_2(1-\eta(d))+(1+\eta(d))\log_2(1+\eta(d))>4/d$. Then with probability $1-o(1)$, a random $d$-regular graph on $n$ vertices generated from the configuration model has expansion at least $(1-\eta(d))/2$.
\end{fact}

\begin{definition}
Given a graph $G=(V,E)$ and a vertex $v\in V$, $B_G(v,r)$ denotes the ball centered at $v$ with radius $r$, that is, the set of vertices which are at a distance at most $r$ from $v$ in $G$.
\end{definition}

The following bound on the size of a ball follows from a simple calculation.

\begin{proposition}\label{prop_ball}
If $G$ is a (subgraph of a) $d$-regular graph for $d\geq 2$, then for any vertex $v$, $|B_G(v,0)|=1$, $|B_G(v,1)|\leq d+1$, $|B_G(v,2)|\leq d^2+1$, and for $r>2$, $|B_G(v,r)|\leq d^r$.
\end{proposition}

\subsubsection{Fourier Transform of Boolean Functions}

Fix a finite set $E$ with $|E|=m$, and identify each subset $S$ of $E$ with a boolean vector in $\{0,1\}^E$ in the natural way. The set of functions $f:\{0,1\}^E\longrightarrow\mathbb{R}$ form a $2^m$ dimensional vector space. Define an inner product $\langle\cdot,\cdot\rangle$ on this vector space as $\langle f,g\rangle=2^{-m}\sum_{x\in\{0,1\}^E}f(x)g(x)$. For each $\alpha\in\{0,1\}^n$, define its \textit{characteristic function} $\chi_{\alpha}:\{0,1\}^E\longrightarrow\mathbb{R}$ as $\chi_{\alpha}(x)=(-1)^{\alpha\cdot x}$. Then the set of functions $\{\chi_{\alpha}:\alpha\in\{0,1\}^E\}$ form an orthonormal basis with respect to the inner product $\langle\cdot,\cdot\rangle$. Thus, any function $f:\{0,1\}^E\longrightarrow\mathbb{R}$ can be resolved in this basis as $f=\sum_{\alpha\in\{0,1\}^E}\widehat{f}(\alpha)\chi_{\alpha}$, where
\[\widehat{f}(\alpha)=\langle f,\chi_{\alpha}\rangle=2^{-m}\sum_{x\in\{0,1\}^E}f(x)\chi_{\alpha}(x)\text{.}\]

We will need the following properties of the Fourier transform, whose proofs can be found in \cite{ODonnell}.

\begin{proposition}\label{prop_noise}
Let $f:\{0,1\}^E\rightarrow\mathbb{R}$ be given by $f(z)=\varepsilon^{|z|}(1-\varepsilon)^{|E|-|z|}$, where $|z|$ denotes the number of ones in the the vector $z$. Then $\widehat{f}$ is given by $\widehat{f}(\alpha)=2^{-|E|}(1-2\varepsilon)^{|\alpha|}$ for all $\alpha\in\{0,1\}^{E}$.
\end{proposition}

\begin{proposition}[Fourier transforms of affine subspaces]\label{prop_affine}
Let $S$ be a subspace of $\{0,1\}^{E}$ of dimension $r$, and $b\in\{0,1\}^{E}$. Let $f:\{0,1\}^E\rightarrow\mathbb{R}$ be given by $f(z)=1$ if $\gamma\cdot z=\gamma\cdot b$ for all $\gamma\in S$, and $f(z)=0$ otherwise. Then $\widehat{f}$ is given by $\widehat{f}(\alpha)=2^{-(m-r)}(-1)^{\alpha\cdot b}$ if $\alpha\in S$, and $\widehat{f}(\alpha)=0$ otherwise.
\end{proposition}

\begin{proposition}[Convolution Theorem]\label{prop_convolution}
Let $f,g:\{0,1\}^E\rightarrow\mathbb{R}$. Then $\widehat{fg}$ is given by
\[\widehat{fg}(\alpha)=\sum_{\beta\in\{0,1\}^n}f(\beta)g(\alpha+\beta)\text{.}\]
\end{proposition}

\subsubsection{Incidence Matrices, Eulerian Subgraphs, Spanning Forests}

Given a graph $G=(V,E)$, its \textit{incidence matrix} is the binary $V\times E$ matrix whose $(v,e)$-entry is $1$ if $v$ is an endpoint of $e$, and $0$ otherwise. A graph is \textit{Eulerian} if and only if each of its vertices has even degree, or equivalently, the mod-2 nullspace of its incidence matrix contains the all ones vector $\mathds{1}_E$. The set of subgraphs of $G$ is in the natural one-to-one correspondence with $\{0,1\}^E$, where the set of Eulerian subgraphs of $G$ corresponds to the nullspace of the incidence matrix of $G$.

We define the \textit{rank} of a graph to be the cardinality of its spanning forest, which is also equal to the rank of its incidence matrix. Fix a spanning forest $F$ of a graph $G=(V,E)$. We use this forest to construct a basis for $\{0,1\}^E$, the vector space of subgraphs of $G$, as follows. For each $e\in F$, let $v_e(e)=1$ and $v_e(e')=0$ for $e'\neq e$. For each $e\in E\setminus F$, define the vector $v_e\in\{0,1\}^E$ as $v_e(e')=1$ if $e'$ belongs to the unique cycle in $F\cup\{e\}$, and $v_e(e')=0$ otherwise. Then the collection of vectors $\{v_e:e\in F\}\cup\{v_e:e\in E\setminus F\}$ forms a basis of $\{0,1\}^E$. Here, the set $\{v_e:e\in E\setminus F\}$ spans the subspace of Eulerian subgraphs of $G$, whereas $\{v_e:e\in F\}$ spans a complementary subspace: the space of sub-forests of $F$. As a consequence, we have that every subgraph of $G$ can be written uniquely as a symmetric difference of an Eulerian subgraph of $G$ and a sub-forest of $F$.

\begin{lemma}\label{lem_Eulerian}
Let $F$ be a spanning forest of a graph $G=(V,E)$. Then the Eulerian subgraphs of $G$ are in one-to-one correspondence with subsets of $E\setminus F$, where the bijection is given by $E^*\leftrightarrow E^*\setminus F$. In other words, for every $S\subseteq E\setminus F$, there exists a unique Eulerian subgraph $E^*$ of $G$ such that $E^*\setminus F=S$.
\end{lemma}

\begin{proof}
For each $e\in E\setminus F$, let $C(e)$ denote the unique cycle in $F\cup\{e\}$. First we prove that $E^*\rightarrow E^*\setminus F$ is surjective. Let $S\subseteq E\setminus F$. Then $E^*=\bigoplus_{e\in S}C(e)$ is Eulerian and $E^*\setminus F=S$. Next, we prove that $E^*\rightarrow E^*\setminus F$ is injective. Let $S\subseteq E\setminus F$ and let $E^*$ and $E'$ be Eulerian subgraphs of $G$ such that $E^*\setminus F=E'\setminus F=S$. Then $(E^*\oplus E')\setminus F=\emptyset$, which means $E^*\oplus E'\subseteq F$. But $F$ is a forest and $E^*\oplus E'$ is Eulerian. Therefore, $E^*\oplus E'=\emptyset$, which means $E^*=E'$.
\end{proof}

\begin{lemma}\label{lem_sep}
Let $F$ be a spanning forest of a graph $G=(V,E)$ such that the endpoints of the edges in $E\setminus F$ are pairwise distance $\Delta$ apart in $F$. Let $G^*=(V,E^*)$ be an Eulerian subgraph of $G$. Then $|E^*|\geq\Delta\cdot|E^*\setminus F|$.
\end{lemma}

\begin{proof}
Partition the edges of $G^*$ into cycles, and consider each cycle one by one. Between any two occurrences of non-forest edges in the cycle, we have a path consisting of edges from $F$. By the separation condition on the endpoints of edges not in $F$, each such path must have length at least $\Delta$. Thus, we have at least $\Delta$ forest edges per non-forest edge in $G^*$.
\end{proof}

\subsubsection{Total Variation Distance}

\begin{definition}
Let $\Omega$ be a finite set. The \textit{Total Variation Distance (TVD)} between two probability distributions $p$ and $p'$ over $\Omega$, denoted by $\text{TVD}(p,p')$ (resp.\ two random variables $X$ and $X'$ taking values from $\Omega$, denoted by $\text{TVD}(X,X')$) is defined as $(1/2)\cdot\sum_{s\in\Omega}|p(s)-p'(s)|$ (resp.\ $(1/2)\cdot\sum_{s\in\Omega}|\Pr[X=s]-\Pr[X'=s]|$). 
\end{definition}

\begin{lemma}
Let $X_1$, $X_1'$ be random variables taking values in $\Omega_1$, and let $X_2$, $X_2'$ be random variables taking values in $\Omega_2$. Then
\[\text{TVD}((X_1,X_2),(X_1',X_2'))\leq\text{TVD}(X_1,X_1')+\sum_{s\in\Omega_1}\Pr[X_1=s]\cdot\text{TVD}((X_2|X_1=s),(X_2'|X_1'=s))\text{.}\]
\end{lemma}

\begin{proof}
For $s_1\in\Omega_1$, let $p(s_1)=\Pr[X_1=s_1]$, and $p'(s_1)=\Pr[X_1'=s_1]$. For $s_1\in\Omega_1$ and $s_2\in\Omega_2$, let $q_{s_1}(s_2)=\Pr[X_2=s_2|X_1=s_1]$, and $q'_{s_1}(s_2)=\Pr[X'_2=s_2|X'_1=s_1]$. Then we have, $\Pr[(X_1,X_2)=(s_1,s_2)]=p(s_1)\cdot q_{s_1}(s_2)$ and $\Pr[(X_1',X_2')=(s_1,s_2)]=p'(s_1)\cdot q'_{s_1}(s_2)$.
\begin{eqnarray*}
\text{TVD}((X_1,X_2),(X_1',X_2')) & = & \frac{1}{2}\sum_{(s_1,s_2)\in\Omega_1\times\Omega_2}\left|p(s_1)\cdot q_{s_1}(s_2)-p'(s_1)\cdot q'_{s_1}(s_2)\right| \\
 & \leq & \frac{1}{2}\sum_{(s_1,s_2)\in\Omega_1\times\Omega_2}\left[\left|p(s_1)\cdot q_{s_1}(s_2)-p(s_1)\cdot q'_{s_1}(s_2)\right|+\left|p(s_1)\cdot q'_{s_1}(s_2)-p'(s_1)\cdot q'_{s_1}(s_2)\right|\right]\text{.}
\end{eqnarray*}
The first term above is
\begin{eqnarray*}
\frac{1}{2}\sum_{(s_1,s_2)\in\Omega_1\times\Omega_2}\left|p(s_1)\cdot q_{s_1}(s_2)-p(s_1)\cdot q'_{s_1}(s_2)\right| & = & \sum_{s_1\in\Omega_1}p(s_1)\cdot\frac{1}{2}\sum_{s_2\in\Omega_2}\left|q_{s_1}(s_2)-q'_{s_1}(s_2)\right|\\
 & = & \sum_{s\in\Omega_1}\Pr[X_1=s]\cdot\text{TVD}((X_2|X_1=s),(X_2'|X_1'=x))\text{,}
\end{eqnarray*}
while the second term is
\[\frac{1}{2}\sum_{(s_1,s_2)\in\Omega_1\times\Omega_2}\left|p(s_1)\cdot q'_{s_1}(s_2)-p'(s_1)\cdot q'_{s_1}(s_2)\right|=\frac{1}{2}\sum_{s_1\in\Omega_1}\left|p(s_1)-p'(s_1)\right|\cdot\sum_{s_2\in\Omega_2}q'_{s_1}(s_2)=\text{TVD}(X_1,X_1')\text{,}\]
since $\sum_{s_2\in\Omega_2}q'_{s_1}(s_2)=\sum_{s_2\in\Omega_2}\Pr[X_2'=s_2|X_1'=s_1]=1$ for all $s_1\in\Omega_1$.
\end{proof}

\begin{corollary}\label{cor_tvd_error}
Let $X_1$, $X_1'$ be random variables taking values in $\Omega_1$, and let $X_2$, $X_2'$ be random variables taking values in $\Omega_2$. Let $\mathcal{E}_1\subseteq\Omega_1$. Then
\[\text{TVD}((X_1,X_2),(X_1',X_2'))\leq\text{TVD}(X_1,X_1')+\Pr[X_1\notin\mathcal{E}]+\sum_{s\in\mathcal{E}}\Pr[X_1=s]\cdot\text{TVD}((X_2|X_1=s),(X_2'|X_1'=s))\text{.}\]
\end{corollary}

\begin{proof}
Follows since for all $s\in\Omega_1$ (and in particular, for $s\notin\mathcal{E}$), $\text{TVD}((X_2|X_1=s),(X_2'|X_1'=s))\leq1$ by definition.
\end{proof}


\begin{corollary}\label{cor_coupling}
For $i=1$ to $T$, let $X_i$ and $X_i'$ be random variables taking values in $\Omega_i$. For $i=1$ to $T-1$, let $\mathcal{E}_i\subseteq\Omega_1\times\cdots\times\Omega_i$, such that $(s_1,\ldots,s_i)\in\mathcal{E}_i$ implies $(s_1,\ldots,s_{i-1})\in\mathcal{E}_{i-1}$. Then
\[\text{TVD}((X_1,\ldots,X_T),(X_1',\ldots,X_T'))\leq\Pr[(X_1,\ldots,X_T)\notin\mathcal{E}_T]+\]
\[\sum_{i=1}^T\sum_{(s_1,\ldots,s_{i-1})\in\mathcal{E}_{i-1}}\Pr\left[\bigwedge_{j=1}^{i-1}X_j=s_j\right]\cdot\text{TVD}\left(\left(X_i\mid\bigwedge_{j=1}^{i-1}X_j=s_j\right),\left(X_i'\mid\bigwedge_{j=1}^{i-1}X_j'=s_j\right)\right)\text{.}\]
\end{corollary}

\begin{proof}
Follows by repeated application of Corollary \ref{cor_tvd_error}.
\end{proof}

\subsection{Reductions to \partitiontesting and MAX-CUT}\label{sec_reductions}

\subsubsection{Reduction to {\bf PartitionTesting}}\label{sec:reduction-partition}

In this section, we show how the problem \noisyparities reduces to testing {\bf PartitionTesting}$(k,\varphi_{\text{in}},\varphi_{\text{out}},\beta)$, even for $k=1$ and any $\beta\leq1$. By this reduction, we establish a lower bound of $n^{1/2+\Omega(\varphi_{\text{out}})}$ on the number of queries required to test whether a graph is $(1,\varphi_{\text{in}})$-clusterable for some constant $\varphi_{\text{in}}$ (the YES case), or it is $(2,\varphi_{\text{out}},\beta)$-unclusterable for any constant $\beta\leq 1$ (the NO case). 

\noindent\textbf{Theorem \ref{thm_main_lb} (restated).}
There exist positive constants $\varphi_{\text{in}}$ and $b$ such that for all $\varphi_{\text{out}}\leq 1$, any algorithm that distinguishes between a $(1,\varphi_{\text{in}})$-clusterable graph (that is, a $\varphi_{\text{in}}$-expander) and a $(2,\varphi_{\text{out}},1)$-unclusterable graph on $n$ vertices with success probability at least $2/3$ must make at least $(n/2)^{1/2+b\varphi_{\text{out}}/2}$ queries, even when the input is restricted to $d$-regular graphs for a large enough constant $d$.

The reduction is given by Algorithm \ref{alg_reduction}.

\begin{algorithm}
\caption{\textsc{ReductionToPartitionTesting} ($G=(V,E)$, $y:E\longrightarrow\{0,1\}$)}\label{alg_reduction}
\begin{algorithmic}[1]
\State Input: $G=(V,E)$,  labeling $y:E\longrightarrow\{0,1\}$
\State $V':=V\times\{0,1\}$. \Comment{We denote the vertex $(v,b)\in V\times\{0,1\}$ by $v^b$ for readability.}
\State $E_0':=\bigcup_{e=(u,v)\in E\text{: }y(e)=0}\{(u^0,v^0),(u^1,v^1)\}$.
\State $E_1':=\bigcup_{e=(u,v)\in E\text{: }y(e)=1}\{(u^0,v^1),(u^1,v^0)\}$.
\State $E'=E_0'\cup E_1'$.
\State \Return $G'=(V',E')$.
\end{algorithmic}
\end{algorithm}

Observe that the reduction is ``query complexity preserving'' in the sense that any query from a {\bf PartitionTesting} algorithm asking the neighbors of a vertex $v^b\in V'$ can be answered by making (at most) one query, asking the yet undisclosed edges incident on $v$ in $G$ and their labels. To establish the correctness of the reduction, it is sufficient to prove:
\begin{enumerate}
\item The YES case: If the edges of $G$ are labeled independently and uniformly at random, then $G'$ is an expander with high probability.
\item The NO case: If each edge $e=(u,v)$ of $G$ is labeled $X(u)+X(v)+Z(u,v)$, where $Z(u,v)$ is $1$ with probability $\varepsilon$, then with high probability $G'$ contains a cut with $n$ vertices on each side whose expansion is $O(\varepsilon)$.
\end{enumerate}

\begin{lemma}\label{lem_yes}
Let $G=(V,E)$ be a $d$-regular $\varphi$-expander with $|V|=n$. Suppose each edge $(u,v)\in E$ independently and uniformly given label $Y(u,v)\in\{0,1\}$. Suppose \textsf{ReductionToPartitionTesting} on input $(G,Y)$ returns the graph $G'=(V',E')$. Then $G'$ is a $\min(\varphi/4,1/32)$-expander with probability at least $1-2^{2n}\cdot\exp(-dn/256)$.
\end{lemma}

\begin{proof}
We need to prove that every $C\subseteq V'$ with $|C|\leq|V'|/2=n$ expands well. Let $C_0=\{v\in V\text{ }:\text{ }v^0\in C\}$ and $C_1=\{v\in V\text{ }:\text{ }v^0\in C\}$ be the ``projections'' of $C$ on the two halves of $V'=V\times\{0,1\}$, each half identified with $V$, so that $|C|=|C_0|+|C_1|$. Then at least one of the following must hold.
\begin{enumerate}
\item $|C_0\cup C_1|\leq n/2$.
\item $|C_0\cap C_1|\geq n/4$.
\item $|C_0\oplus C_1|\geq n/4$.
\end{enumerate}

In the first case, consider the set of edges which cross the set $C_0\cup C_1$ in $G$. Since $G$ is a $\varphi$-expander and $|C_0\cup C_1|\leq n/2$, the number of such edges is at least $\varphi d|C_0\cup C_1|$. For each such edge, one of its two copies in $G'$ must cross the set $C$. Therefore, the expansion of $C$ is at least
\[\frac{\varphi d|C_0\cup C_1|}{d|C|}\geq\frac{\varphi}{2}\text{.}\]

In the second case, we cannot have $|C_0\cap C_1|>n/2$, otherwise we contradict the assumption that $|C|\leq n$. Therefore, $|C_0\cap C_1|\leq n/2$. Consider the set of edges which cross $|C_0\cap C_1|$ in $G$. Again, since $G$ is a $\varphi$-expander and $|C_0\cap C_1|\leq n/2$, the number of such edges is at least $\varphi d|C_0\cap C_1|$. As before, for each such edge $e$, at least one of its two copies in $E'$ must cross the set $C$. Therefore, the expansion of $C$ is at least
\[\frac{\varphi d|C_0\cap C_1|}{d|C|}\geq\frac{\varphi n/4}{n}=\frac{\varphi}{4}\text{.}\]

Finally, consider the third case. Let $m$ be the number of edges in $G$, both of whose endpoints are in $C_0\oplus C_1$. Therefore, the number of edges in $G$ with exactly one endpoint in $C_0\oplus C_1$ is $d|C_0\oplus C_1|-2m$. We split into two sub-cases depending on whether $m\leq d|C_0\oplus C_1|/4$, or $m>d|C_0\oplus C_1|/4$.

In the first sub-case, consider the $d|C_0\oplus C_1|-2m$ edges of $G$ with exactly one endpoint in $C_0\oplus C_1$. For each of such edge, (exactly) one of its copies in $G'$ crosses the set $C$. Therefore, the expansion of $C$ is at least
\[\frac{d|C_0\oplus C_1|-2m}{d|C|}\geq\frac{d|C_0\oplus C_1|-d|C_0\oplus C_1|/2}{d|C|}=\frac{|C_0\oplus C_1|/2}{|C|}\geq\frac{n/8}{n}=\frac{1}{8}\text{.}\]

In the second sub-case, consider each edge $(u,v)\in E$ with $u,v\in C_0\oplus C_1$. Suppose both $u$ and $v$ belong to the same $C_i$. If $Y(u,v)=0$, none of the copies of the edge $(u,v)$ in $E'$ crosses the set $C$, and if $Y(u,v)=1$, then both copies cross. Similarly, suppose one of $u$ and $v$ belongs to $C_0$ and the other belongs to $C_1$. If $Y(u,v)=1$, none of the copies of the edge $(u,v)$ in $E'$ crosses the set $C$, and if $Y(u,v)=0$, then both copies cross. Thus for each of the $m$ edges $(u,v)\in E$ with $u,v\in C_0\oplus C_1$, both of its copies cross $C$ with probability $1/2$, and none crosses with probability $1/2$. This happens independently for all the $m$ edges. Therefore, by Chernoff bound, the number of edges both of whose copies cross the set $C$ is at least $m/4\geq d|C_0\oplus C_1|/16$ with probability at least
\[1-\exp(-m/16)\geq1-\exp(-d|C_0\oplus C_1|/64)\geq1-\exp(-dn/256)\text{.}\]
Assuming this happens, the expansion of $C$ is at least
\[\frac{2\times d|C_0\oplus C_1|/16}{d|C|}=\frac{|C_0\oplus C_1|/8}{|C|}\geq\frac{n/32}{n}=\frac{1}{32}\text{.}\]
This holds for each set $C$ falling into this sub-case. Applying the union bound over all the at most $2^{2n}$ sets falling into this sub-case, we have that with probability at least $1-2^{2n}\cdot\exp(-dn/256)$, all sets falling into this sub-case have expansion at least $1/32$. 
\end{proof}

\begin{lemma}\label{lem_no}
Let $G=(V,E)$ be a $d$-regular graph with $|V|=n$. For each $v\in V$, let $X(v)$ be an independent uniformly random bit. For each edge $(u,v)\in E$, let $Z(u,v)$ be an independent random bit which is $1$ with probability $\varepsilon$ and $0$ otherwise. Suppose each edge $(u,v)\in E$ is labeled $Y(u,v)=X(u)+X(v)+Z(u,v)$. Suppose \textsf{ReductionToPartitionTesting} on input $(G,Y)$ returns the graph $G'=(V',E')$. Then with probability at least $1-\exp(-\varepsilon nd/6)$, there exists a set $V^*\subseteq V'$ with $|V^*|=n=|V'|/2$ whose expansion is at most $2\varepsilon$.
\end{lemma}

\begin{proof}
Define $V^*$ as
\[V^*=\{v^0\text{ }:\text{ }v\in V\text{, }X(v)=0\}\cup\{v^1\text{ }:\text{ }v\in V\text{, }X(v)=1\}\text{,}\]
so that $|V^*|=n$. By a case-by-case consideration, it is easy to verify that if for $e=(u,v)\in E$ we have $Z(e)=1$, then the two edges between $u^0$, $v^0$, $u^1$, $v^1$ cross the cut $(V^*,V'\setminus V^*)$. Conversely, if $Z(e)=0$, then one of the edges between $u^0$, $v^0$, $u^1$, $v^1$ lies within $V^*$ and the other lies outside $V^*$. Thus, the number of edges crossing the cut is twice the size of the set $\{e\in E\text{ }:\text{ }Z(e)=1\}$. The expectation of the size of this set is $\varepsilon\cdot|E|=\varepsilon nd/2$. Since $\{Z(e)\}_{e\in E}$ are independent and identically distributed, application of the Chernoff bound gives us that the size of the set $\{e\in E\text{ }:\text{ }Z(e)=1\}$ is at most $\varepsilon nd$ with probability at least $1-\exp(-\varepsilon nd/6)$. Thus, the number of edges crossing the cut $(V^*,V'\setminus V^*)$ is at most $2\varepsilon nd$ with high probability. Dividing by $nd$, the volume of $V^*$, we have that the expansion of $V^*$ is at most $2\varepsilon$ with probability at least $1-\exp(-\varepsilon nd/6)$.
\end{proof}


\begin{proof}[Proof of Theorem \ref{thm_main_lb}]
Let $d=512$, $\varphi=(1-\eta(d))/2\approx0.45$, and $\varphi_{\text{in}}=\min(\varphi/4,1/32)$. As before, let $b=1/(8\ln d)$ (with $d=512$ now). Given $\varphi_{\text{out}}\leq1$, let $\varepsilon=\varphi_{\text{out}}/2\leq1/2$. Suppose there is an algorithm for {\bf PartitionTesting} which makes $(n/2)^{1/2+\delta}$ queries on $n$ vertex graphs and outputs the correct answer with probability at least $2/3$, for some $\delta<b\varphi_{\text{out}}/2=b\varepsilon=\min(1/16,b\varepsilon)$ (note that $b\varepsilon\leq1/(16\ln 512)\leq1/16$). Then for any probability distribution $\mathcal{D}$ over $n$-vertex {\bf PartitionTesting} instances, there exists a deterministic algorithm ALG$(\mathcal{D})$ making $O(n^{1/2+\delta})$ queries which outputs the correct answer with probability at least $2/3$, on a random instance of {\bf PartitionTesting} drawn from the distribution.

Let $(G,y)$ be a random instance of the \noisyparities problem with parameters $d$ and $\varepsilon$, where $G=(V,E)$ is a graph and $y:E\longrightarrow\{0,1\}$ is an edge lebeling. Apply \textsf{ReductionToPartitionTesting} to $(G,y)$, and thus, get a random instance $G'$ of {\bf PartitionTesting} from the appropriate probability distribution $\mathcal{D}$. Run ALG$(\mathcal{D})$ on this instance and return the answer. Note that to answer one query of ALG$(\mathcal{D})$, we make at most one query into $G$. Thus, this reduction gives an algorithm $\text{ALG}'$ for \noisyparities making at most $n^{1/2+\delta}$ queries.

The underlying graph $G$ is a random $d$-regular graph on $n$ vertices. By Fact \ref{fact_Bollobas}, with high probability, $G$ is a $\varphi$-expander. Hence, by Lemma \ref{lem_yes}, if $(G,y)$ is a YES instance, then with high probability the reduced graph $G'$ is a $\varphi_{\text{in}}$-expander for $\varphi_{\text{in}}=\min(\varphi/4,1/32)$ (we chose $d$ large enough so that the failure probability $2^{2n}\cdot\exp(-dn/256)$ in Lemma \ref{lem_yes} becomes $o(1)$). On the other hand, if $(G,y)$ is a NO instance, then by Lemma \ref{lem_no}, the reduced graph $G'$ is a graph on $2n$ vertices containing with high probability a subset of $n$ vertices whose expansion is at most $2\varepsilon=\varphi_{\text{out}}$. Thus, $G'$ is $(2,\varphi_{\text{out}},1)$-unclusterable. Hence, the reduction succeeds with probability $1-o(1)$. Since ALG answers correctly with probability at least $2/3$, $\text{ALG}'$ answers correctly with probability at least $2/3-o(1)$.

However, by Theorem \ref{thm_main_aux_lb}, since $\text{ALG}'$ makes at most $n^{1/2+\delta}$ queries, $\text{ALG}'$ can be correct with probability at most $1/2+o(1)$. This is a contradiction.
\end{proof}

\subsubsection{Reduction to Approximating MAX-CUT Value}\label{sec:reduction-maxcut}

In this section, we show how the problem \noisyparities reduces to estimating maxcut. By this reduction, we establish the following theorem.

\noindent\textbf{Theorem \ref{thm_maxcut} (restated).}
There exists a constant $\beta$ such that for any $\varepsilon'>0$ and any $d\geq3$, any algorithm that distinguishes correctly with probability $2/3$ between the following two types of $n$-vertex $d$-degree bounded graphs must make at least $n^{1/2+\min(1/16,\varepsilon'/(24\ln d))}$ queries.
\begin{itemize}
\item The YES instances: Graphs which have a cut of size at least $(nd/4)\cdot(1-\varepsilon')$.
\item The NO instances: Graphs which do not have a cut of size more than $(1+\beta d^{-1/2})\cdot(nd/8)$.
\end{itemize}

The reduction from \noisyparities to MAX-CUT works as follows.

\begin{algorithm}
\begin{algorithmic}[1]
\Procedure{\textsf{ReductionToMAXCUT}}{Graph $G=(V,E)$, Edge labeling $y:E\longrightarrow\{0,1\}$}
\State $E'=\{e\in E\text{ }:\text{ }y(e)=1\}$.
\State \Return{$G'=(V,E')$}.
\EndProcedure
\end{algorithmic}
\end{algorithm}

We claim that a YES instance of \noisyparities is reduced with high probability to a NO instance of MAX-CUT, and vice versa. To prove that the reduction correctly converts a YES instance of \noisyparities to a NO instance of MAX-CUT, we need the following fact which is implied by Theorem 1.6 of \cite{DemboMS_arXiv15}.

\begin{fact}\label{fact_maxcut}
There exists an absolute constant $\alpha$ such that the following holds for all all $d$ large enough. Let $G$ be a random $d$-regular $n$-vertex graph generated from the configuration model. Then with probability $1-o(1)$, the maximum cut in $G$ cuts at most $1/2+\alpha d^{-1/2}$ fraction of the edges.
\end{fact}

\begin{lemma}\label{lem_yesno}
There exists a constant $\beta$ such that for all $d$ the following holds. Let $G=(V,E)$ be a $d$-regular $\varphi$-expander with $|V|=n$. Suppose each edge $(u,v)\in E$ independently and uniformly given label $Y(u,v)\in\{0,1\}$. Suppose \textsf{ReductionToMAXCUT} on input $(G,Y)$ returns the graph $G'=(V,E')$. Then with probability $1-o(1)$, the maxcut in $G'$ is at most $(1+\beta d^{-1/2})\cdot nd/8$.
\end{lemma}

\begin{proof}
By Fact \ref{fact_maxcut}, with probability $1-o(1)$ every cut in $G$ has size at most $(1/2+\alpha d^{-1/2})\cdot nd/2$. Given that every cut in $G$ has size at most $(1/2+\alpha d^{-1/2})\cdot nd/2$, we have the following. Consider an arbitrary cut in $G'$. The expected number of edges in this cut with label $1$ is at most $(1/2+\alpha d^{-1/2})\cdot nd/4=(1+2\alpha d^{-1/2})\cdot nd/8$. By Chernoff bound, the probability that more than $(1+\varepsilon')\cdot(1+2\alpha d^{-1/2})\cdot nd/8$ edges in $G'$ lie in this cut is at most
\[\exp\left(-\frac{\varepsilon'^2\cdot(1+2\alpha d^{-1/2})\cdot nd}{24}\right)\leq\exp\left(-\frac{\varepsilon'^2\cdot nd}{24}\right)\leq\exp(-n)\text{,}\]
for $\varepsilon'=24d^{-1/2}$. By union bound over all the $2^n$ cuts in $G'$, we have that the probability that some cut value exceeds $(1+24d^{-1/2})\cdot(1+2\alpha d^{-1/2})\cdot nd/8\leq(1+(24+50\alpha)d^{-1/2})\cdot nd/8$ is at most $(2/e)^n=o(1)$. Adding to this the $o(1)$ probability that $G$ itself has a large cut, and setting $\beta=24+50\alpha$, we get the claim.
\end{proof}

Next, we prove that the reduction correctly converts a NO instance of \noisyparities to a YES instance of MAX-CUT, we need the following claim.

\begin{lemma}\label{lem_var}
Let $G=(V,E)$ be an arbitrary $d$-regular graph, and let $\{X(v)\text{ }:\text{ }v\in V\}$ be a set of independent binary random variables, each of which is $0$ and $1$ with probability $1/2$. Let $V_0=\{v\in V\text{ }:\text{ }X(v)=0\}$ $V_1=\{v\in V\text{ }:\text{ }X(v)=1\}$. Let $C$ be the random variable whose value is the number of edges in the $(V_0,V_1)$ cut. Then $\text{Var}[C]\leq2d\cdot\mathbb{E}[C]$.
\end{lemma}

\begin{proof}
For each $e\in E$, let $C(e)$ be the indicator random variable that is $1$ if $e$ lies in the $(V_0,V_1)$ cut, and $0$ otherwise. Since $X(v)$'s are independent, for any two edges $e$ and $e'$ which do not share an endpoint, $C(e)$ and $C(e')$ are independent. $C=\sum_{e\in E}C(e)$, and therefore,
\[(\mathbb{E}[C])^2=\sum_{e,e'\in E}\mathbb{E}[C(e)]\mathbb{E}[C(e')]\geq\sum_{e,e'\in E\text{:}e\cap e'=\emptyset}\mathbb{E}[C(e)]\mathbb{E}[C(e')]=\sum_{e,e'\in E\text{:}e\cap e'=\emptyset}\mathbb{E}[C(e)C(e')]\text{.}\]
Now, we have
\[\mathbb{E}[C^2]=\sum_{e,e'\in E}\mathbb{E}[C(e)C(e')]=\sum_{e,e'\in E\text{:}e\cap e'=\emptyset}\mathbb{E}[C(e)C(e')]+\sum_{e,e'\in E\text{:}e\cap e'\neq\emptyset}\mathbb{E}[C(e)C(e')]\text{.}\]
Using the lower bound on $(\mathbb{E}[C])^2$, we have
\[\mathbb{E}[C^2]\leq(\mathbb{E}[C])^2+\sum_{e,e'\in E\text{:}e\cap e'\neq\emptyset}\mathbb{E}[C(e)C(e')]\text{,}\]
which implies,
\[\text{Var}[C]=\mathbb{E}[C^2]-(\mathbb{E}[C])^2\leq\sum_{e,e'\in E\text{:}e\cap e'\neq\emptyset}\mathbb{E}[C(e)C(e')]\leq\sum_{e\in E}\mathbb{E}[C(e)]\cdot|\{e'\in E\text{ }:\text{ }e\cap e'\neq\emptyset\}|\]
where we used in the last inequality that $C(e')\leq 1$ for any $e'$. Using the fact that the graph is $d$-regular, we have that $|\{e'\in E\text{ }:\text{ }e\cap e'\neq\emptyset\}|\leq2d$ for any $e$. Therefore,
\[\text{Var}[C]\leq2d\cdot\sum_{e\in E}\mathbb{E}[C(e)]=2d\cdot\mathbb{E}[C]\text{,}\]
as required.
\end{proof}

\begin{lemma}\label{lem_noyes}
Let $G=(V,E)$ be a $d$-regular graph with $|V|=n$. For each $v\in V$, let $X(v)$ be an independent uniformly random bit. For each edge $(u,v)\in E$, let $Z(u,v)$ be an independent random bit which is $1$ with probability $\varepsilon$ and $0$ otherwise. Suppose each edge $(u,v)\in E$ is labeled $Y(u,v)=X(u)+X(v)+Z(u,v)$. Suppose \textsf{ReductionToMAXCUT} on input $(G,Y)$ returns the graph $G'=(V,E')$. Then with probability at least $1-o(1)$, there exists a cut $(V_0,V_1)$ in $G'$ of value at least $(nd/4)\cdot(1-2\varepsilon)\cdot(1-o(1))$.
\end{lemma}

\begin{proof}
Let $V_0=\{v\in V\text{ }:\text{ }X(v)=0\}$ and $V_1=\{v\in V\text{ }:\text{ }X(v)=1\}$, as in the statement of Lemma \ref{lem_var}, and let $C$ be the random variable whose value is the number of edges of $G$ in the $(V_0,V_1)$ cut. Then $\mathbb{E}[C]=nd/4$, since $|E|=nd/2$ and each edge is cut with probability $1/2$. By Chebyshev's inequality, we have,
\[\Pr\left[|C-\mathbb{E}[C]|>\frac{\mathbb{E}[C]}{n^{1/4}}\right]\leq\frac{\text{Var}[C]\cdot n^{1/2}}{(\mathbb{E}[C])^2}\leq\frac{2d\cdot\mathbb{E}[C]\cdot n^{1/2}}{(\mathbb{E}[C])^2}=\frac{8dn^{1/2}}{nd}=\frac{8}{n^{1/2}}\text{,}\]
where the second inequality follows from Lemma \ref{lem_var}. Therefore, with probability at least $1-8/n^{1/2}$, we have $C\geq\mathbb{E}[C](1-n^{-1/4})=(nd/4)\cdot(1-n^{-1/4})$.

Now, let us condition on the values of $X(v)$'s which ensure $C\geq\mathbb{E}[C](1-n^{-1/4})$. Then the cut $(V_0,V_1)$ is fixed, and the labels on the edges become independent. Each edge $(u,v)$ of $G$ in the $(V_0,V_1)$ cut has label $0$ with probability $\varepsilon$ and $1$ with probability $1-\varepsilon$. By the Chernoff bound, with probability $1-\exp(-\varepsilon C/3)\geq1-\exp(-(\varepsilon nd/4)\cdot(1-n^{-1/2}))$, at most $2\varepsilon C$ out of the $C$ edges of $G$ in the $(V_0,V_1)$ cut have label $0$, and therefore, at least $(1-2\varepsilon)C$ edges have label $1$. All these edges with label $1$ appear in the $(V_0,V_1)$ cut of $G'$. Thus, with probability at least $1-8/n^{1/2}-\exp(-(\varepsilon nd/4)\cdot(1-n^{-1/2}))=1-o(1)$, $G'$ contains a cut of size at least $(nd/4)\cdot(1-n^{-1/4})\cdot(1-2\varepsilon)=(nd/4)\cdot(1-2\varepsilon)\cdot(1-o(1))$.
\end{proof}


\begin{proof}[Proof of Theorem \ref{thm_maxcut}]
Consider an algorithm that approximates maxcut within a factor $2-\varepsilon'$ with probability at least $2/3$, and assume it makes $n^{1/2+\delta}$ queries. Then for any distribution over the instances, there exists a deterministic algorithm ALG making $n^{1/2+\delta}$ queries and having the same approximation guarantee on a random instance drawn from the distribution.

Let $(G,y)$ be a random instance of the \noisyparities problem with parameters $\varepsilon=\varepsilon'/24$ and $d=(4\beta/\varepsilon')^2$, where $\beta$ is the constant from Lemma \ref{lem_yesno}. Here, $G=(V,E)$ is a graph and $y:E\longrightarrow\{0,1\}$ is an edge lebeling. Apply \textsf{ReductionToMAXCUT} to $(G,y)$, and thus, get a random instance $G'$ of MAX-CUT from the appropriate probability distribution. Run ALG on this instance and obtain an estimate $z$ of the maxcut. Return YES if $z>(1+\varepsilon'/4)nd/8$, otherwise return NO. Note that to answer one query of ALG, we make one query into $G$. Thus, this reduction gives an algorithm $\text{ALG}'$ for \noisyparities making at most $n^{1/2+\delta}$ queries.

If $(G,y)$ is a YES instance of the \noisyparities problem, then by Lemma \ref{lem_yesno}, with probability $1-o(1)$, $G'$ has maxcut at most $(1+\beta d^{-1/2})\cdot(nd/8)=(1+\varepsilon/4)\cdot(nd/8)$. Thus, the estimate of maxcut given by ALG is at most $(1+\varepsilon/4)\cdot(nd/8)$ with probability at least $2/3$, and we return YES. On the other hand, if  $(G,y)$ is a NO instance of the \noisyparities problem, then by Lemma \ref{lem_noyes}, with probability $1-o(1)$, $G'$ has maxcut at least
\[\frac{nd}{4}\cdot(1-2\varepsilon)\cdot(1-o(1))\geq(1-3\varepsilon)\cdot\frac{nd}{4}=\left(1-\frac{\varepsilon'}{8}\right)\cdot\frac{nd}{4}\text{.}\]
Therefore, the estimate of maxcut given by ALG, with probability at least $2/3$, is at least
\[\frac{1-\varepsilon'/8}{2-\varepsilon'}\frac{nd}{4}=\frac{1-\varepsilon'/8}{1-\varepsilon'/2}\frac{nd}{8}>\left(1-\frac{\varepsilon'}{8}\right)\left(1+\frac{\varepsilon'}{2}\right)\cdot\frac{nd}{8}>\left(1+\frac{\varepsilon'}{4}\right)\cdot\frac{nd}{8}\text{,}\]
and we return NO. Thus, ALG' is correct with probability at least $2/3-o(1)$. Therefore, by Theorem \ref{thm_main_aux_lb}, $\delta\geq\min(1/16,b\varepsilon)$, where $b=1/(8\ln d)$. Thus, $\delta=\Omega(\varepsilon'/\log(1/\varepsilon'))$.
\end{proof}

\subsection{Query Lower Bound for the \noisyparities Problem}\label{sec_lbproof}

Recall the \noisyparities problem (Definition \ref{def_noisyparities}). In this section, we prove Theorem \ref{thm_main_aux_lb}, which gives a lower bound on the query complexity of the \noisyparities problem. We start out by formalizing the the execution of the algorithm's query as a process which generates the instance incrementally. 

\subsubsection{Interaction and Closure}\label{subsec_interaction}

Formally, the interaction that takes place between the algorithm and the adversary is given by the procedure \textsf{Interaction}. Here \textsf{NextQuery} is the function which simulates the behavior of the algorithm: it takes as input the uncovered edge-labeled graph and the set of vertices already queried, and returns an unqueried vertex. It is helpful to make the following observations.
\begin{enumerate}
\item The random graph is generated according to the configuration model. As soon as a vertex $q$ is queried, the unpaired half-edges on $q$ are paired up one-by-one to random unpaired half-edges incident on the yet unqueried vertices. 
\item In the YES case, the label generated for any edge is uniformly random, as per the problem definition in Section \ref{sec_intro}. In the NO case, although the label is generated without referring to the parities $X$ of the vertices in the problem specification, the parities are built-in in an implicit manner, and the distribution of the labels is still consistent with the problem specification. Also, this is the only place where the behavior of \textsf{Interaction} differs depending on whether it is executing the YES case or the NO case.
\end{enumerate}

\begin{algorithm}
\begin{algorithmic}[1]
\Procedure{\textsf{Interaction}(Answer,$V$,$\varepsilon$)}{}
\State $Q_0:=\emptyset$, $H_0:=\emptyset$, $F_0:=0$, $T=n^{1/2+\delta}$.
\For {$t$ $=$ $1$ to $T$}
\State \Comment{Invariant: All half-edges incident on vertices in $Q_{t-1}$ are paired.}
\State \Comment{Invariant: $F_{t-1}$ is a spanning forest of $H_{t-1}$.}
\State \Comment{Invariant: If Answer $=$ NO then for any cycle $C\subseteq H_{t-1}$, $\sum_{e\in C}(Y(e)+Z(e))=0$.}
\State $q_t$ $:=$ \textsf{NextQuery}$(Q_{t-1},H_{t-1})$. \Comment{Assumption: $q_t\notin Q_{t-1}$.}
\State $Q_t:=Q_{t-1}\cup\{q_t\}$.
\State $H_t:=H_{t-1}$, $F_t:=F_{t-1}$.
\While {$q$ has an unpaired half-edge $(q,i)$}
\State Pair up $(q,i)$ with a random unpaired half-edge, say $(v,j)$. Call the resulting edge between $q$ and $v$ as $e$. \Comment{$v\notin Q_{t-1}$ unless $v=q$.}
\If {Answer $=$ YES}
\If {$F_t\cup\{e\}$ is acyclic}
\State $F_t:=F_t\cup\{e\}$.
\EndIf
\State Generate label $Y(e)$ $:=$ $0$ or $1$ with probability $1/2$ each.
\Else \Comment{Answer $=$ NO.}
\State Generate noise $Z(e)$ $:=$ $1$ with probability $\varepsilon$ and $0$ with probability $1-\varepsilon$.
\If {$F_t\cup\{e\}$ is acyclic}
\State $F_t:=F_t\cup\{e\}$.
\State Generate label $Y(e)$ $:=$ $0$ or $1$ with probability $1/2$ each.
\Else \Comment{$F_t\cup\{e\}$ contains a single cycle.}
\State Let $P$ be the unique path from $q$ to $v$ in $F_t$.
\State Generate label $Y(e)$ $:=$ $Z(e)+\sum_{e'\in P}(Y(e')+Z(e'))$.
\EndIf
\EndIf
\State $H_t:=H_t\cup\{(e,Y(e))\}$.
\EndWhile
\EndFor
\EndProcedure
\end{algorithmic}
\end{algorithm}

Recall that our goal is to prove that the algorithm does not gain sufficient information with $n^{1/2+\delta}$ queries to distinguish between the YES case and the NO case. In order to facilitate our analysis, we give the following additional information to the algorithm for free, and refer to  this modified version of \textsf{Interaction} as \textsf{InteractionWithClosure}, and then argue that the algorithm fails nonetheless.
\begin{enumerate}
\item \textsf{InteractionWithClosure} ensures that the pairwise distance in $F$ between vertices on which a non-forest edge is incident is at least $b\ln n$, where $b=1/(8\ln d)$, as defined in Theorem \ref{thm_main_aux_lb}. As soon as the algorithm manages to uncover a new edge resulting in violation of this invariant, \textsf{InteractionWithClosure} throws an error and conservatively assumes that the algorithm found the correct answer already. (In particular, this includes the scenarios where self-loops and parallel edges are discovered.) We say that event $\text{Err}_t$ happened if \textsf{InteractionWithClosure} throws an error in round $t$.
\item As soon as a new edge incident on the queried vertex $q_t$ that cannot be added to $F_{t-1}$ is discovered, \textsf{InteractionWithClosure} generates the whole ball of radius $b\ln n$ around $q_t$. This might already result in the event $\text{Err}_t$ as defined above. If not, \textsf{InteractionWithClosure} adds the BFS tree around $q$ to $F_t$, labels its edges uniformly at random, samples and records the noise for its edges, adds these labeled edges to $H_t$, and gives $H_t$ back to the algorithm.
\end{enumerate}

\subsubsection{Analysis of \textsf{InteractionWithClosure}}\label{sec_analysis_interaction}

\begin{definition}
After any round $t$ of \textsf{InteractionWithClosure}, we call a vertex $v\in V$ \textit{discovered} if it was queried (that is, $v\in Q_t$) or if one of its neighbors in $G$ was queried (that is, it had degree at least one in $H_t$). We denote by $D_t$ the set of vertices discovered after round $t$.
\end{definition}

We now define certain ``good'' events $\mathcal{E}_t$ which are sufficient to ensure that our analysis works and gets us the query lower bound. Moreover, we will also show that these events are extremely likely to happen.

\begin{definition}\label{def_E}
Define $R_t=H_t\setminus F_t$ to be the set of non-forest edges seen by the end of round $t$. Then $R_t\supseteq R_{t-1}$ for all $t$. For any round $t$, we say that event $\mathcal{E}_t$ happened if the following conditions hold.
\begin{enumerate}
\item \textsf{InteractionWithClosure} completes the $t^{\text{\tiny{th}}}$ round without throwing an error, that is, none of the events $\text{Err}_j$ for $j\leq t$ happen.
\item $|D_j|\leq dn^{1/2+\delta}\ln n$ for all $j\leq t$.
\end{enumerate}
\end{definition}

First, let us prove a bound on the probability that a non-forest edge is found in round $t$.

\begin{lemma}\label{lem_prob_nonforest}
If event $\mathcal{E}_{t-1}$ happens then the probability that \textsf{InteractionWithClosure} encounters an edge in round $t$ which forms a cycle with edges in $F_{t-1}$ is at most $(2d^2\ln n)/n^{1/2-\delta}$.
\end{lemma}

\begin{proof}
The number of undiscovered vertices is at least $n-|D_{t-1}|\geq n-dn^{1/2+\delta}\ln n\geq n/2+2$, and therefore, there are at least $d(n/2+2)$ free half-edges incident on undiscovered vertices. Therefore, the probability that at least one of the discovered (unqueried) vertices becomes a neighbor of $q_t$, when we pair up the at most $d$ free half edges incident on $q_t$, is at most 
\[d\cdot\frac{|D_{t-1}\setminus Q_{t-1}|\cdot d}{d(n/2+2)-2d}\leq\frac{d\cdot|D_{t-1}|}{n/2}\leq\frac{2d^2n^{1/2+\delta}\ln n}{n}=\frac{2d^2\ln n}{n^{1/2-\delta}}\text{.}\]
\end{proof}

Our next lemma and proves a bound on the probability that the algorithm will throw an error in round $t$.

\begin{lemma}\label{lem_nonforest}
If event $\mathcal{E}_{t-1}$ happens then the probability that \textsf{InteractionWithClosure} throws an error in round $t$, that is, event $\text{Err}_t$ happens, is at most $(16d^4\ln^2n)/n^{7/8-2\delta}$.
\end{lemma}

\begin{proof}
Event $\text{Err}_t$ happens only in the following two cases.
\begin{enumerate}
\item \textsf{InteractionWithClosure} encounters an edge $(q_t,v)$ such that the distance between $q_t$ and $v$ in $F_{t-1}$ is at most $b\ln n$.
\item \textsf{InteractionWithClosure} encounters an edge $(q_t,v)$ which forms a cycle with edges in $F_{t-1}$, and while generating the ball of radius $b\ln n$, it encounters another edge which cannot be added to the forest.
\end{enumerate}
Let us bound the probabilities of the above two events separately. The number of undiscovered vertices is at least
\[n-|D_{t-1}|\geq n-dn^{1/2+\delta}\ln n\geq\frac{n}{2}+2n^{1/8}+2\geq\frac{n}{2}+2\text{.}\]

First, observe that $|B_{H_{t-1}}(q_t,b\ln n)|\leq d^{b\ln n}=n^{b\ln d}=n^{1/8}$, by Proposition \ref{prop_ball}. $q_t$ has at most $d$ free half-edges, the vertices $B_{H_{t-1}}(q_t,b\ln n)\setminus Q_{t-1}$ have at most $dn^{1/8}$ free half-edges, and we have at least $dn/2$ free half-edges every time we pair a half-edge incident on $q_t$. Thus, the probability of finding a new non-forest edge closing a short cycle, which is same as the probability that at least one of the vertices $B_{H_{t-1}}(q_t,b\ln n)\setminus Q_{t-1}$ gets an edge incident on $q_t$, is at most
\[\frac{2d^2n^{1/8}}{dn}=\frac{2d}{n^{7/8}}\text{.}\]

Suppose that in round $t$ we find a neighbor $v$ of $q_t$ such that the edge $(q_t,v)$ cannot be added to the forest. Let us construct the breadth-first search tree around $q_t$ of radius $b\ln n$ by taking each vertex already added to the tree at a time, and pairing up its free half-edges. Consider the processing of some such vertex $u$, and let $W$ be the vertices already added to the BFS tree at the time $u$ is processed. Since $W\subseteq B_G(q,b\ln n)$, $|W|\leq d^{b\ln n}=n^{b\ln d}=n^{1/8}$, and therefore, the number of edges in the BFS tree is at most $n^{1/8}$. Now, the probability that $u$ gets a new edge to some vertex in $W\cup D_{t-1}$ is at most
\[d\cdot\frac{|W\cup D_{t-1}|\cdot d}{d(n/2+2n^{1/8}+2)-2n^{1/8}-2d}\leq\frac{2(n^{1/8}+dn^{1/2+\delta}\ln n)\cdot d}{n}\text{.}\leq\frac{4d^2n^{1/2+\delta}\ln n}{n}=\frac{4d^2\ln n}{n^{1/2-\delta}}\text{.}\]
Note that this must happen for some $u$ for \textsf{InteractionWithClosure} to find a non-forest edge close to the edge $(q,v)$ and throw the error. Since the number of such $u$'s is at most $|B_G(q_t,b\ln n)|\leq n^{b\ln d}=n^{1/8}$, the probability that \textsf{InteractionWithClosure} fails 
is bounded from above by
\[n^{1/8}\cdot\frac{4d^2\ln n}{n^{1/2-\delta}}=\frac{4d^2\ln n}{n^{3/8-\delta}}\text{.}\]
The above holds when conditioned on at least one of the vertices in $D_{t-1}\setminus Q_{t-1}$ being a neighbor of $q_t$. Unconditioning and using Lemma \ref{lem_prob_nonforest}, we get that the probability that \textsf{InteractionWithClosure} fails due to the second reason above is bounded by
\[\frac{2d^2\ln n}{n^{1/2-\delta}}\cdot\frac{4d^2\ln n}{n^{3/8-\delta}}=\frac{8d^4\ln^2n}{n^{7/8-2\delta}}\text{.}\]
Adding to this the probability of failure due to the first reason specified above, we have that the probability that event $\text{Err}_t$ happens is at most
\[\frac{2d}{n^{7/8}}+\frac{8d^4\ln^2n}{n^{7/8-2\delta}}\leq\frac{16d^4\ln^2n}{n^{7/8-2\delta}}\text{.}\]
\end{proof}

The next two lemmas essentially prove that if the event $\mathcal{E}_{t-1}$ happens, then it is very likely that $\mathcal{E}_t$ happens too. We then put together these claims and prove that the event $\mathcal{E}_T$ happens with high probability, where $T=n^{1/2+\delta}$ is the number of queries.

\begin{lemma}\label{lem_rt}
For every $t$ the following holds: if $\mathcal{E}_{t-1}$ happens, then $\Pr\left[|R_t|>(4d^2\ln n)\cdot(1+t/n^{1/2-\delta})\right]\leq n^{-2d^2/3}$.
\end{lemma}

\begin{proof}
For every $j\leq t$, conditioned on $\mathcal{E}_{j-1}$, we have that $|R_j\setminus R_{j-1}|$ is one with probability at most $(2d^2\ln n)/n^{1/2-\delta}$, and zero otherwise, by Lemma \ref{lem_prob_nonforest}. Let $r_1\ldots,r_t$ be independent Bernoulli random variables, each taking value one with probability $(2d^2\ln n)/n^{1/2-\delta}$, and zero otherwise. Then for each $j$, $|R_j \setminus R_{j-1}|=|R_j|-|R_{j-1}|$ is stochastically dominated by $r_j$. Let us use the Chernoff bound to upper bound $\Pr[\sum_{j=1}^tr_j>(4d^2\ln n)\cdot(1+t/n^{1/2-\delta})]$, which will also give an upper bound on $\Pr[|R_j|>(4d^2\ln n)\cdot(1+t/n^{1/2-\delta})]$. For this, observe that $\mathbb{E}[\sum_{j=1}^tr_j]=(2d^2t\ln n)/n^{1/2-\delta}$.

First, consider the case where $t<n^{1/2-\delta}$. Using Chernoff bound, we have,
\begin{eqnarray*}
\Pr\left[\sum_{j=1}^tr_j>4d^2\ln n\right] & \leq & \Pr\left[\sum_{j=1}^tr_j>\left(1+\frac{n^{1/2-\delta}}{t}\right)\cdot\frac{2d^2t\ln n}{n^{1/2-\delta}}\right]\\
 & \leq & \exp\left(-\frac{n^{1-2\delta}}{3t^2}\cdot\frac{2d^2t\ln n}{n^{1/2-\delta}}\right)=\exp\left(-\frac{(2d^2\ln n)\cdot n^{1/2-\delta}}{3t}\right)\text{.}
\end{eqnarray*}
Using the upper bound on $t$, we have
\begin{equation}\label{eqn_t_small}
\Pr\left[\sum_{j=1}^tr_j>(4d^2\ln n)\cdot\left(1+\frac{t}{n^{1/2-\delta}}\right)\right]\leq\Pr\left[\sum_{j=1}^tr_j>4d^2\ln n\right]\leq n^{-2d^2/3}\text{.}
\end{equation}

Next, consider the case where $t\geq n^{1/2-\delta}$. Using the Chernoff bound again, we get,
\[\Pr\left[\sum_{j=1}^tr_j>\frac{4d^2t\ln n}{n^{1/2-\delta}}\right]\leq\exp\left(-\frac{2d^2t\ln n}{3n^{1/2-\delta}}\right)\leq n^{-2d^2/3}\text{.}\]
Thus,
\begin{equation}\label{eqn_t_large}
\Pr\left[\sum_{j=1}^tr_j>(4d^2\ln n)\cdot\left(1+\frac{t}{n^{1/2-\delta}}\right)\right]\leq\Pr\left[\sum_{j=1}^tr_j>\frac{4d^2t\ln n}{n^{1/2-\delta}}\right]\leq n^{-2d^2/3}\text{.}
\end{equation}

Equations (\ref{eqn_t_small}) and (\ref{eqn_t_large}) together imply that $\Pr\left[|R_t|>(4d^2\ln n)\cdot(1+t/n^{1/2-\delta})\right]\leq n^{-2d^2/3}$.
\end{proof}

\begin{lemma}\label{lem_dt}
For every $t$, if $\mathcal{E}_{t-1}$ happens and moreover, if $|R_t|\leq(4d^2\ln n)\cdot(1+t/n^{1/2-\delta})$, then $|D_t|\leq dn^{1/2+\delta}\ln n$.
\end{lemma}

\begin{proof}
If $\mathcal{E}_{t-1}$ happens then every round $j\leq t$ which did not discover a non-forest edge (that is, $|R_j|=|R_{j-1}|$) discovered at most $d$ new vertices. On the other hand, every round $j\leq t$ which discovered a new non-forest edge (that is, $|R_j|=|R_{j-1}|+1$) discovered at most $n^{b\ln d}=n^{1/8}$ new vertices, as it discovered $B_G(q_j,b\ln n)$, whose size is at most $n^{b\ln d}$, by Proposition \ref{prop_ball}. Therefore,
\[|D_t|\leq dt+n^{1/8}\cdot|R_t|\leq dt+n^{1/8}\cdot(4d^2\ln n)\cdot(1+t/n^{1/2-\delta})\text{.}\]
Since $t\leq n^{1/2+\delta}$ and $\delta<1/16$, we have,
\[|D_t|\leq dn^{1/2+\delta}+4d^2n^{1/8}\ln^2n\cdot(1+n^{2\delta})\leq dn^{1/2+\delta}\ln n\text{.}\]
\end{proof}

\begin{lemma}\label{lem_good}
The event $\mathcal{E}_T$ happens with probability $1-o(1)$.
\end{lemma}

\begin{proof}
Let $p^{\text{err}}_t$ be the probability that event $\text{Err}_t$ happens. Then we prove by induction that there is an absolute constant $c$ such that for each $t$, event $\mathcal{E}_t$ happens with probability at least 
\[1-\frac{32d^4\cdot\ln^2n}{n^{7/8-2\delta}}\cdot t\text{.}\]
The claim is obvious for $t=0$. For $t>0$, let us upper bound the probability that $\mathcal{E}_t$ does not happen, given $\mathcal{E}_{t-1}$ happens. The reasons for $\mathcal{E}_t$ not happening are the following.
\begin{enumerate}
\item Event $\text{Err}_t$ happens. This happens with probability $p^{\text{err}}_t$.
\item $|R_t|>(4d^2\ln n)\cdot(1+t/n^{1/2-\delta})$. (If $|R_t|\leq(4d^2\ln n)\cdot(1+t/n^{1/2-\delta})$ then $dn^{1/2+\delta}\ln n$ is guaranteed by Lemma \ref{lem_dt}.)
\end{enumerate}
By Lemma \ref{lem_nonforest}, the probability that \textsf{InteractionWithClosure} throws an error in round $t$ is at most
\[p^{\text{err}}_t\leq\frac{16d^4\cdot\ln^2n}{n^{7/8-2\delta}}\text{.}\]
By Lemma \ref{lem_rt}, the event $|R_t|>(4d^2\ln n)\cdot(1+t/n^{1/2-\delta})$ happens with probability at most $n^{-2d^2/3}<n^{-6}$, because we assumed $d\geq 3$ in the definition of the \noisyparities problem. By induction hypothesis, $\mathcal{E}_{t-1}$ itself happens with probability at least
\[1-\frac{32d^4\cdot\ln^2n}{n^{7/8-2\delta}}\cdot(t-1)\text{.}\]
Thus, $\mathcal{E}_t$ happens with probability at least
\[1-\frac{32d^4\cdot\ln^2n}{n^{7/8-2\delta}}\cdot(t-1)-\frac{16d^4\cdot\ln^2n}{n^{7/8-2\delta}}-n^{-6}\geq1-\frac{32d^4\cdot\ln^2n}{n^{7/8-2\delta}}\cdot t\text{,}\]
as required.

As a consequence, the event $\mathcal{E}_T$ happens with probability at least
\[1-\frac{32d^4\cdot\ln^2n}{n^{7/8-2\delta}}\cdot T\geq1-\frac{32d^4\cdot\ln^2n}{n^{7/8-2\delta}}\cdot n^{1/2+\delta}\geq1-\frac{32d^4\cdot\ln^2n}{n^{3/8-3\delta}}\text{.}\]
Using the fact $\delta<1/16$, we conclude that $\mathcal{E}_T$ happens with probability $1-o(1)$.
\end{proof}


\subsubsection{Bounding TVD in each Round}\label{sec:fourier}

Recall that our goal is to prove Theorem \ref{thm_main_aux_lb}, which states that an algorithm which makes at most $n^{1/2+\delta}$ queries is unable to determine whether \textsf{InteractionWithClosure} is executing the YES or the NO case, assuming $\delta$ is less than some constant times $\varepsilon$. For this, we crucially use Corollary \ref{cor_coupling} as follows. The random variable $X_t$ consists of the $t^{\text{\tiny{th}}}$ query of the algorithm and its result in a YES instance, whereas the random variable $X_t'$ consists of the $t^{\text{\tiny{th}}}$ query of the algorithm and its result in a NO instance. Thus, the realization of the random variable $(X_1,\ldots,X_t)$ (resp.\ $(X_1',\ldots,X_t')$) captures the snapshot of the run of \textsf{InteractionWithClosure} until the $t^{\text{\tiny{th}}}$ query in the YES (resp.\ NO) case. The events $\mathcal{E}_t$ are as defined in Definition \ref{def_E}, and they satisfy the requirements of Corollary \ref{cor_coupling}.

Our goal is to prove that if $T\leq n^{1/2+\delta}$, then
\begin{equation}\label{eqn_tvd_claim}
\text{TVD}((X_1,\ldots,X_T),(X_1',\ldots,X_T'))=o(1)\text{.}
\end{equation}
Since the answer of the algorithm is a function of the realization of $(X_1,\ldots,X_T)$ (resp.\ $(X_1',\ldots,X_T')$) in the YES (resp.\ NO) case, the above statement implies that the total variation distance between the algorithm's answer in the YES case and the algorithm's answer in the NO case is only $o(1)$. Therefore, the algorithm's answer is correct with probability $1/2+o(1)$.

In order to establish (\ref{eqn_tvd_claim}), by Corollary \ref{cor_coupling}, it is sufficient to prove that
\begin{eqnarray}
\sum_{t=1}^T\sum_{(s_1,\ldots,s_{t-1})\in\mathcal{E}_{t-1}}\Pr\left[\bigwedge_{j=1}^{t-1}X_j=s_j\right]\cdot\text{TVD}\left(\left(X_t\mid\bigwedge_{j=1}^{t-1}X_j=s_j\right),\left(X_t'\mid\bigwedge_{j=1}^{t-1}X_j'=s_j\right)\right) & + & \nonumber\\
\Pr[(s_1,\ldots,s_T)\notin\mathcal{E}_T] & = & o(1)\text{.}\label{eqn_split}
\end{eqnarray}
Here, we already proved in Lemma \ref{lem_good} that $\Pr[(s_1,\ldots,s_T)\notin\mathcal{E}_T]=o(1)$. Therefore, it is sufficient to prove that
\begin{equation}\label{eqn_claim1}
\sum_{t=1}^T\sum_{(s_1,\ldots,s_{t-1})\in\mathcal{E}_{t-1}}\Pr\left[\bigwedge_{j=1}^{t-1}X_j=s_j\right]\cdot\text{TVD}\left(\left(X_t\mid\bigwedge_{j=1}^{t-1}X_j=s_j\right),\left(X_t'\mid\bigwedge_{j=1}^{t-1}X_j'=s_j\right)\right)=o(1)\text{.}
\end{equation}
Informally, the above claim states the following. Suppose \textsf{InteractionWithClosure} executes on a YES instance and a NO instance in parallel, and for the first $t-1$ rounds of these executions, the queries and the responses to the queries match. Then the probability distributions of the responses to the query in the $t^{\text{\tiny{th}}}$ round are $o(1)$-close in total variation distance.


Recall that the executions of \textsf{InteractionWithClosure} on YES and NO instances differ only in the following situation: the current edge whose label is to be generated forms a cycle with edges in $F$. Therefore, in such a situation, if the forced label in the NO case does not match the uniformly random label in the YES case, then this is responsible for some TVD between $X_t$ and $X_t'$ conditioned on the snapshot of the run of \textsf{InteractionWithClosure} until round $t-1$.
Apart from this step, the executions of \textsf{InteractionWithClosure} in the YES case and the NO case are identical. Moreover, the event $\mathcal{E}_T$ ensures that the number of rounds in which an edge closing a cycle is encountered is at most $O(d^2n^{2\delta}\ln n)$. Therefore, it is sufficient to prove that for all $t\leq T$ and for all $(s_1,\ldots,s_{t-1})\in\mathcal{E}_{t-1}$, we have
\begin{equation}\label{eqn_claim}
\text{TVD}\left(\left(X_t\mid\bigwedge_{j=1}^{t-1}X_j=s_j\right),\left(X_t'\mid\bigwedge_{j=1}^{t-1}X_j'=s_j\right)\right)=o(n^{-2b\varepsilon})\text{,}
\end{equation}
where $b=1/(8\ln d)$, as defined earlier. From this, as long as $\delta<b\varepsilon$, (\ref{eqn_claim1}) follows. We devote the rest of this subsection to prove claim (\ref{eqn_claim}).

Let $e$ be an edge such that when $e$ arrives, the forest $F$ maintained by \textsf{InteractionWithClosure} already contains a path $P$ between the endpoints of $e$. In the YES case, the label $Y(e)$ of $e$ is $0$ or $1$ uniformly at random, whereas in the NO case, the label is $Z(e)+\sum_{e'\in P}(Y(e')+Z(e'))$. 
We are, therefore, interested in bounding the TVD between the distribution of $Z(e)+\sum_{e'\in P}(Y(e')+Z(e'))$ conditioned on the labels of the previous edges, and the uniform distribution on $\{0,1\}$. Since we are conditioning on the labels of all the previous edges, inclusive of edges $e'\in P$, this distance is same as the TVD between the distribution of $Z(e)+\sum_{e'\in P}Z(e')$ conditioned on the labels, and the uniform distribution. Furthermore, observe that $Z_e$ itself is independent of the labels of the previous edges, and is $1$ with probability $\varepsilon<1/2$ and $0$ otherwise. Therefore, the TVD between $Z(e)+\sum_{e'\in P}Z(e')$ conditioned on the labels and the uniform distribution is at most the TVD between $\sum_{e'\in P}Z(e')$ conditioned on the labels and the uniform distribution.

In order to bound the TVD between $\sum_{e'\in P}Z(e')$ conditioned on the labels and the uniform distribution, we need to determine the distribution of $\sum_{e'\in P}Z(e')$ conditioned on the labels in the first place. We use the Fourier transform to achieve this. We use Bayes' rule and write the posterior distribution of $Z$, conditioned on the labels $Y=y$, as being proportional to the product of the prior distribution of $Z$ and the probability of labels $Y$ being $y$ conditioned on $Z$. Then we use the convolution theorem to get the Fourier transform of the posterior distribution of $Z$. An appropriate Fourier coefficient then gives us the \textit{bias} of $\sum_{e'\in P}Z(e')$ conditioned on the labels.

\begin{definition}
The \textit{bias} of a binary random variable $X$ is the TVD between its distribution and the uniform distribution over $\{0,1\}$. Equivalently, the bias of $X$ is equal to $|\Pr[X=0]-1/2|=|\Pr[X=1]-1/2|$.
\end{definition}

Let $H=(V,E_H)$ be the graph formed by the edges which arrived before $e$. Let $F_H$ be the forest maintained by \textsf{InteractionWithClosure} when $e$ arrives (so that $F_H$ is a spanning forest of $H$). Let the random variable $Y$ and $Z$, both taking values in $\{0,1\}^{E_H}$, denote the random labels and the random noise of the edges in $E_H$ respectively. Fix $y\in\{0,1\}^{E_H}$. We are interested in the distribution of $Y(e)$, the label on edge $e$, conditioned on $Y=y$. We want to prove that if the graph $H$ and the spanning forest $F_H$ satisfy certain properties, then the distribution of $Y(e)$ conditioned on $Y=y$ is close to uniform.

\begin{theorem}\label{thm_Fourier}
Suppose the graph $H$ and the spanning forest $F_H$ are such that the endpoints of the edges in $(E_H\cup\{e\})\setminus F$ are pairwise at least a distance $\Delta$ apart in $F_H$. Then the bias of the distribution of the NO-case label of the new edge $e$ conditioned on the labels of the previous edges is at most
\[\frac{(1-2\varepsilon)^{\Delta-1}(1+(1-2\varepsilon)^{\Delta})^{|E_H\setminus F_H|}}{2-(1+(1-2\varepsilon)^{\Delta})^{|E_H\setminus F_H|}}\text{.}\]
\end{theorem}

\begin{proof}
Let $P$ be the path in $F_H$ between the endpoints of $e$, and let $C=P\cup\{e\}$ be the cycle in $F\cup\{e\}$. Since $Y(e)=Z(e)+\sum_{e'\in P}(Y(e')+Z(e'))$, the distribution of $Y(e)$ conditioned on $Y=y$, has the same bias as the distribution of $Z(e)+\sum_{e'\in P}Z(e')=\sum_{e'\in C}Z(e)$ conditioned on $Y=y$. Here $Z(e)$ is independent of the previous labels $Y$, and hence, the bias of $\sum_{e'\in C}Z(e')$ conditioned on $Y=y$ is at most the bias of $\sum_{e'\in P}Z(e')$ conditioned on $Y=y$. It is, therefore, sufficient to bound from above the bias of $\sum_{e'\in P}Z(e')$ conditioned on $Y=y$.

The posterior distribution of the random noise $Z$ given the labels $Y=y$ is given by
\[\Pr[Z=z\text{ }\mid\text{ }Y=y]=\frac{\Pr[Y=y\text{ }\mid\text{ }Z=z]\cdot\Pr[Z=z]}{\Pr[Y=y]}=\frac{f(z)\cdot g_y(z)}{\sum_{z'\in\{0,1\}^{E_{t-1}}}f(z')\cdot g_y(z')}=\frac{h_y(z)}{\sum_{z'\in\{0,1\}^{E_{t-1}}}h_y(z')}\text{,}\]
where the functions $f$, $g_y$ and $h_y$ are defined as $f(z)=\Pr[Z=z]$, $g_y(z)=\Pr[Y=y\text{ }\mid\text{ }Z=z]$, and $h_y=f\cdot g_y$. The Fourier transforms of these functions are as follows. Since $f(z)=\varepsilon^{|z|}(1-\varepsilon)^{|E_H|-|z|}$, by Proposition \ref{prop_noise}, we have for all $\alpha\in\{0,1\}^{E_H}$,
\[\widehat{f}(\alpha)=2^{-|E_H|}(1-2\varepsilon)^{|\alpha|}\text{.}\]
Next let us consider the function $g_y$. Let $E^*$ denote the nullspace of the incidence matrix of $H_{t-1}$, that is, the set of indicator vectors of Eulerian subgraphs of $H_{t-1}$. We say that $y$ and $z$ are compatible if for every cycle $C$ in $H_{t-1}$ we have $\sum_{e\in C}y(e)=\sum_{e\in C}z(e)$, that is, $\gamma\cdot y=\gamma\cdot z$ for all $\gamma\in E^*$. Since the dimension of $E^*$ is $|E_H|-|F_H|$, there are exactly $2^{|F_H|}$ compatible $y$'s for every $z$, one for each of the $2^{|F_H|}$ labelings of edges in $F$. Moreover, each of the $2^{|F_H|}$ labelings are realized with equal probability, because the edges of $F$ are labeled independently with a $0$ or a $1$ with probability $1/2$ each. Therefore we have,
\[g_y(z)=\begin{cases}
          2^{-|F_H|} & \text{if }\gamma\cdot y=\gamma\cdot z\text{ for all }\gamma\in E^*\\
          0 & \text{otherwise.}          
         \end{cases}
\]
Then by Proposition \ref{prop_affine}, the Fourier transform of $g_y$ is given by
\begin{eqnarray*}
\widehat{g}_y(\alpha) & = & \begin{cases}
        2^{-|E_H|}(-1)^{\alpha\cdot y} & \text{if }\alpha\in E^*\\
        0 & \text{otherwise.}
       \end{cases}
\end{eqnarray*}
Using convolution theorem (Proposition \ref{prop_convolution}), we have,
\[\widehat{h}_y(\alpha)=\sum_{\beta\in\{0,1\}^{E_H}}\widehat{g}_y(\beta)\widehat{f}(\alpha+\beta)=2^{-2|E_H|}\sum_{\beta\in E^*}(-1)^{\beta\cdot y}(1-2\varepsilon)^{|\alpha+\beta|}\text{.}\]

Recall that our goal was to bound the bias of $\sum_{e'\in P}Z(e')$ conditioned on the labels $y$, where $e$ is the edge whose label is being generated, and $P$ is the unique path in $F$ between the endpoints of $e$. Let $\pi\in\{0,1\}^{E_H}$ be the indicator vector of $P$. Then the bias of $\sum_{e'\in P}Z(e')=\pi\cdot Z$ conditioned on $y$ is expressed as follows.
\begin{eqnarray*}
\widehat{h}_y(\pi) & = & 2^{-|E_H|}\sum_{z\in\{0,1\}^{E_{t-1}}}h_y(z)(-1)^{\pi\cdot z}\text{,}\\
\widehat{h}_y(0) & = & 2^{-|E_H|}\sum_{z\in\{0,1\}^{E_{t-1}}}h_y(z)\text{.}
\end{eqnarray*}
Therefore,
\begin{eqnarray*}
\frac{|\widehat{h}_y(\pi)|}{\widehat{h}_y(0)} & = & \frac{1}{\sum_{z'\in\{0,1\}^{E_{t-1}}}h_y(z')}\cdot\left\lvert \sum_{z\in\{0,1\}^{E_{t-1}}\text{, }\pi\cdot z=0}h_y(z)-\sum_{z\in\{0,1\}^{E_{t-1}}\text{, }\pi\cdot z=0}h_y(z)\right\rvert\\
 & = & \lvert\Pr\left[\pi\cdot Z=0\text{ }\mid\text{ }Y=y\right]-\Pr\left[\pi\cdot Z=1\text{ }\mid\text{ }Y=y\right]\rvert\\
 & = & \text{bias}\left(\pi\cdot Z\text{ }\mid\text{ }Y=y\right)\text{.}
\end{eqnarray*}
It is thus sufficient to upper bound $|\widehat{h}_y(\pi)|/\widehat{h}_y(0)$. We now bound $|\widehat{h}_y(\pi)|$ and $\widehat{h}_y(0)$ separately. We have
\[\widehat{h}_y(\pi)=2^{-2|E_H|}\sum_{\beta\in E^*}(-1)^{\beta\cdot y}(1-2\varepsilon)^{\beta+\pi}\leq2^{-2|E_H|}\sum_{\beta\in E^*}(1-2\varepsilon)^{\beta+\pi}\text{.}\]

For $\beta\in E^*$, the indicator vector of an Eulerian subgraph of $H$, consider the set $P'$ of edges whose indicator vector is $\beta+\pi$. Then $P'\cup\{e\}$ is Eulerian. Conversely, if $P'\cup\{e\}$ is Eulerian for some $P'\subseteq E_H$, then its indicator vector is $\beta+\pi$ for some $\beta\in E^*$. Since we assumed that the endpoints of the edges in $(E_H\cup\{e\})\setminus F_H$ are pairwise at least a distance $\Delta$ apart in $F_H$, by Lemma \ref{lem_sep}, we have 
\[|P'\cup\{e\}|\geq\Delta|(P'\cup\{e\})\setminus F_H|=\Delta(|P'\setminus F_H|+1)\text{.}\]
For any $\gamma\in\{0,1\}^{E_H}$, let $\overline{\gamma}$ denote the projection of $\beta$ onto the span of the indicator vectors of the edges not in $F$. Then the above statement can be rewritten as,
\[|\beta+\pi|+1\geq\Delta(|\overline{\beta+\pi}|+1)=\Delta(|\overline{\beta}|+1)\text{,}\]
where the last equality holds because $\pi$, being the indicator vector of a path in $F_H$, has zero projection onto the span of the indicator vectors of the edges not in $F$. Therefore,
\[\widehat{h}_y(\pi)\leq2^{-2|E_H|}\sum_{\beta\in E^*}(1-2\varepsilon)^{\Delta(|\overline{\beta}|+1)-1}=2^{-2|E_H|}(1-2\varepsilon)^{\Delta-1}\sum_{\beta\in E^*}(1-2\varepsilon)^{\Delta|\overline{\beta}|}\text{.}\]
By Lemma \ref{lem_Eulerian}, as $\beta$ varies over the indicator vectors of Eulerian subgraphs of $H$, its projection $\overline{\beta}$ varies over $\{0,1\}^{E_H\setminus F_H}$. Therefore,
\[\widehat{h}_y(\pi)\leq2^{-2|E_H|}(1-2\varepsilon)^{\Delta-1}\sum_{\overline{\beta}\in\{0,1\}^{E_H\setminus F_H}}(1-2\varepsilon)^{\Delta|\overline{\beta}|}=2^{-2|E_H|}(1-2\varepsilon)^{\Delta-1}(1+(1-2\varepsilon)^{\Delta})^{|E_H\setminus F_H|}\text{.}\]

We also have,
\[\widehat{h}_y(0)=2^{-2|E_H|}\sum_{\beta\in E^*}(-1)^{\beta\cdot y}(1-2\varepsilon)^{|\beta|}\geq2^{-2|E_H|}\left(2-\sum_{\beta\in E^*}(1-2\varepsilon)^{|\beta|}\right)\text{.}\]
Again, by Lemma \ref{lem_sep} we have $|\beta|\geq\Delta\cdot|\overline{\beta}|$. Therefore,
\[\widehat{h}_y(0)\geq2^{-2|E_H|}\left(2-\sum_{\beta\in E^*}(1-2\varepsilon)^{\Delta\cdot|\overline{\beta}|}\right)\text{.}\]
As before, as $\beta$ varies over the indicator vectors of Eulerian subgraphs of $H$, its projection $\overline{\beta}$ varies over $\{0,1\}^{E_H\setminus F_H}$. Therefore,
\[\widehat{h}_y(0)\geq2^{-2|E_H|}\left(2-\sum_{\overline{\beta}\in\{0,1\}^{E_H\setminus F_H}}(1-2\varepsilon)^{\Delta\cdot|\overline{\beta}|}\right)=2^{-2|E_H|}\left(2-(1+(1-2\varepsilon)^{\Delta})^{|E_H\setminus F_H|}\right)\text{.}\]

The upper bound on $|\widehat{h}_y(P)|$ and the lower bound on $\widehat{h}_y(\emptyset)$ together imply
\[\text{bias}\left(\sum_{e'\in P}Z(e')\text{ }\mid\text{ }Y=y\right)=\frac{|\widehat{h}_y(P)|}{\widehat{h}_y(\emptyset)}\leq\frac{(1-2\varepsilon)^{\Delta-1}(1+(1-2\varepsilon)^{\Delta})^{|E_H\setminus F_H|}}{2-(1+(1-2\varepsilon)^{\Delta})^{|E_H\setminus F_H|}}\text{.}\]
\end{proof}

\subsubsection{Wrapping Up}\label{subsec_wrapping}

\begin{proof}[Proof of Theorem \ref{thm_main_aux_lb}]
As a consequence of Lemma \ref{lem_good} and Lemma \ref{lem_rt}, at any point of time during the execution of \textsf{InteractionWithClosure}, with probability $1-o(1)$ we have that the endpoints of the edges in $E_H\setminus F_H$ are pairwise separated in $F_H$ by a distance at least $b\ln n$, and moreover,
\[|E_H\setminus F_H|=|R|\leq(4d^2\ln n)\cdot(1+t/n^{1/2-\delta})=O(d^2n^{2\delta}\ln n)\text{,}\]
because $t\leq n^{1/2+2\delta}$. At the time of labeling a new edge $e$ which forms a cycle with edges in $F$, let us apply Theorem \ref{thm_Fourier}, with $\Delta=b\ln n$. This gives that the bias in the label of $e$ in the NO case is at most
\[\frac{(1-2\varepsilon)^{\Delta-1}(1+(1-2\varepsilon)^{\Delta})^{|E_H\setminus F_H|}}{2-(1+(1-2\varepsilon)^{\Delta})^{|E_H\setminus F_H|}}=\frac{(1-2\varepsilon)^{b\ln n-1}(1+(1-2\varepsilon)^{b\ln n})^{|E_H\setminus F_H|}}{2-(1+(1-2\varepsilon)^{b\ln n})^{|E_H\setminus F_H|}}\text{.}\]
Since $|E_H\setminus F_H|=O(d^2n^{2\delta}\ln n)$, we have
\[(1+(1-2\varepsilon)^{b\ln n})^{|E_H\setminus F_H|}\leq(1+n^{-2b\varepsilon})^{|E_H\setminus F_H|}\leq(1+n^{-2b\varepsilon})^{O(d^2n^{2\delta}\ln n)}=1+o(1)\text{,}\]
because $\delta<b\varepsilon$. Therefore, the bias in the label of $e$ in the NO case is at most
\[\frac{n^{-2b\varepsilon}}{1-2\varepsilon}\cdot(1+o(1))=O(n^{-2b\varepsilon})\text{.}\]
This is the required bound on the TVD between the snapshots of the executions of \textsf{InteractionWithClosure} in the YES and the NO case, in a generic round, given that the snapshots until the end of the previous round were the same. This proves claim (\ref{eqn_claim}), and hence, Theorem \ref{thm_main_aux_lb}.
%
\end{proof}

%% file: appc.tex
\section{Proof of Lemmas from Section \ref{sec_liftingoracle}}\label{app:matrix-estimation}

\noindent\textbf{Lemma~\ref{lem_matrix_estimation} (restated).} 
Let $G=(V_G,E_G)$ be a graph. Let $0<\sigma\leq1$ $t>0$, $\mu_{\text{err}}>0$, $k$ be an integer, and let $S$ be a multiset of $s$ vertices, all whose elements are $(\sigma,t)$-good. Let
\[R=\max\left(\frac{100s^2\sigma^{1/2}}{\mu_{\text{err}}},\frac{200s^4\sigma^{3/2}}{\mu_{\text{err}}^2}\right)\text{.}\]
For each $a\in S$ and each $b\in V_G$, let $\mathbf{q}_a(b)$ be the random variable which denotes the fraction out of the $R$ random walks starting from $a$, which end in $b$. Let $Q$ be the matrix whose columns are $(D^{-\frac{1}{2}} \mathbf{q}_a)_{a\in S}$. Then with probability at least $49/50$,
$|\mu_{k+1}(Q^{\top}Q)-\mu_{k+1}((D^{-\frac{1}{2}}M^tS)^{\top}(D^{-\frac{1}{2}}M^tS))|\leq\mu_{\text{err}}$.

\begin{proof}
Let $X_{a,r}^i$ be a random variable which is $\frac{1}{\sqrt{\text{deg}(i)}}$ if the $r^{\text{th}}$ random walk starting from $a$, ends at vertex $i$, and $0$ otherwise. Thus, $\mathbb{E}[X_{a,r}^i]=\frac{\mathbf{p}^t_a(i)}{\sqrt{\text{deg}(i)}}$. Let $Z=Q^{\top}Q$. For any two vertices $a,b\in S$, observe that the entry $Z_{a,b}$ is a random variable given by
\[Z_{a,b}=\frac{1}{R^2} \sum_{i\in V_G} (\sum_{r_1=1}^R X_{a,r_1}^i)(\sum_{r_2=1}^R X_{b,r_2}^i).\]
Thus,

\begin{align} 
\mathbb{E}[Z_{a,b}] &=\frac{1}{R^2} \sum_{i\in V_G} (\sum_{r_1=1}^R \mathbb{E}[X_{a,r_1}^i])(\sum_{r_2=1}^R \mathbb{E}[X_{b,r_2}^i])  \nonumber \\
&= \sum_{i\in V_G} \frac{\mathbf{p}^t_a(i)}{\sqrt{\text{deg}(i)}}\cdot \frac{\mathbf{p}^t_b(i)}{\sqrt{\text{deg}(i)}} = (D^{-\frac{1}{2}}M^t\mathds{1}_a)^\top (D^{-\frac{1}{2}}M^t\mathds{1}_b)\text{.}  \label{eq:expz}
\end{align}

We know that $\text{Var}(Z_{a,b}) = \mathbb{E}[Z^2_{a,b}]-\mathbb{E}[Z_{a,b}]^2$. Let us first compute $\mathbb{E}[Z^2_{a,b}]$. 
\begin{align*}
 \mathbb{E}[Z^2_{a,b}] &= \mathbb{E}\left[\frac{1}{R^4} \sum_{i\in V_G} \sum_{j\in V_G} \sum_{r_1=1}^R \sum_{r_2=1}^R \sum_{r^{\prime}_1=1}^R \sum_{r^{\prime}_2=1}^R X_{a,r_1}^i  X_{b,r_2}^i X_{a,r^{\prime}_1}^j  X_{b,r^{\prime}_2}^j\right] \\
 &= \frac{1}{R^4} \sum_{i\in V_G} \sum_{j\in V_G} \sum_{r_1=1}^R \sum_{r_2=1}^R \sum_{r^{\prime}_1=1}^R \sum_{r^{\prime}_2=1}^R \mathbb{E}[X_{a,r_1}^i  X_{b,r_2}^i X_{a,r^{\prime}_1}^j  X_{b,r^{\prime}_2}^j]  
\end{align*}
To compute $\mathbb{E}[X_{a,r_1}^i  X_{b,r_2}^i X_{a,r^{\prime}_1}^j  X_{b,r^{\prime}_2}^j] $, we need to consider the following cases.
\begin{enumerate}
\item $i\neq j $: 
$\mathbb{E}[X_{a,r_1}^i  X_{b,r_2}^i X_{a,r^{\prime}_1}^j  X_{b,r^{\prime}_2}^j] \leq 
\frac{\mathbf{p}^t_a(i)}{\sqrt{\text{deg}(i)}} \cdot 
\frac{\mathbf{p}^t_b(i)}{\sqrt{\text{deg}(i)}} \cdot 
\frac{\mathbf{p}^t_a(j)}{\sqrt{\text{deg}(j)}} \cdot 
\frac{\mathbf{p}^t_b(j)}{\sqrt{\text{deg}(j)}} $. 
(This is an equality if $r_1\neq r^{\prime}_1$ and $r_2\neq r^{\prime}_2$. Otherwise, the expectation is zero.)

\item $i=j, \quad r_1=r_1^\prime, \quad r_2=r_2^\prime$: 
$\mathbb{E}[X_{a,r_1}^i  X_{b,r_2}^i X_{a,r^{\prime}_1}^j  X_{b,r^{\prime}_2}^j] = 
\frac{\mathbf{p}^t_a(i)}{\sqrt{\text{deg}(i)}} \cdot 
\frac{\mathbf{p}^t_b(i)}{\sqrt{\text{deg}(i)}} \cdot 
\frac{1}{\sqrt{\text{deg}(i)}} \cdot 
\frac{1}{\sqrt{\text{deg}(i)}} $.
\item $i=j, \quad r_1=r_1^\prime, \quad r_2 \neq r_2^\prime$: 
$\mathbb{E}[X_{a,r_1}^i  X_{b,r_2}^i X_{a,r^{\prime}_1}^j  X_{b,r^{\prime}_2}^j] = 
\frac{\mathbf{p}^t_a(i)}{\sqrt{\text{deg}(i)}} \cdot 
\frac{\mathbf{p}^t_b(i)}{\sqrt{\text{deg}(i)}} \cdot 
\frac{1}{\sqrt{\text{deg}(i)}} \cdot 
\frac{\mathbf{p}^t_b(i)}{\sqrt{\text{deg}(i)}}$.
\item $i=j, \quad r_1 \neq r_1^\prime, \quad r_2 = r_2^\prime$: 
$\mathbb{E}[X_{a,r_1}^i  X_{b,r_2}^i X_{a,r^{\prime}_1}^j  X_{b,r^{\prime}_2}^j] = 
\frac{\mathbf{p}^t_a(i)}{\sqrt{\text{deg}(i)}} \cdot 
\frac{\mathbf{p}^t_b(i)}{\sqrt{\text{deg}(i)}} \cdot 
\frac{\mathbf{p}^t_a(i)}{\sqrt{\text{deg}(i)}} \cdot 
\frac{1}{\sqrt{\text{deg}(i)}}$.

\item $i=j, \quad r_1 \neq r_1^\prime, \quad r_2 \neq r_2^\prime$: 
$\mathbb{E}[X_{a,r_1}^i  X_{b,r_2}^i X_{a,r^{\prime}_1}^j  X_{b,r^{\prime}_2}^j] = 
\frac{\mathbf{p}^t_a(i)}{\sqrt{\text{deg}(i)}} \cdot 
\frac{\mathbf{p}^t_b(i)}{\sqrt{\text{deg}(i)}} \cdot 
\frac{\mathbf{p}^t_a(i)}{\sqrt{\text{deg}(i)}} \cdot 
\frac{\mathbf{p}^t_b(i)}{\sqrt{\text{deg}(i)}}$.

\end{enumerate}
Thus we have,
\begin{align*}
 \mathbb{E}[Z^2_{a,b}] &= \frac{1}{R^4} \sum_{i\in V_G} \sum_{j\in V_G} \sum_{r_1=1}^R \sum_{r_2=1}^R \sum_{r^{\prime}_1=1}^R \sum_{r^{\prime}_2=1}^R \mathbb{E}[X_{a,r_1}^i  X_{b,r_2}^i X_{a,r^{\prime}_1}^j  X_{b,r^{\prime}_2}^j]   \\
 &\leq \sum_{i\in V_G} \sum_{j\in V_G\setminus\{i\}} 
 \frac{\mathbf{p}^t_a(i) \cdot \mathbf{p}^t_a(j) \cdot \mathbf{p}^t_b(i) \cdot \mathbf{p}^t_b(j)}{\text{deg}(i)\cdot \text{deg}(j)} + \sum_{i\in V_G} \frac{{\mathbf{p}^t_a(i)}^2\cdot {\mathbf{p}^t_b(i)}^2}{\text{deg}(i)^2} \\
&+ \frac{1}{R^2} \sum_{i\in V_G} \frac{\mathbf{p}^t_a(i)\cdot \mathbf{p}^t_b(i)}{\text{deg}(i)^2} + \frac{1}{R} \sum_{i\in V_G} \frac{\mathbf{p}^t_a(i)\cdot {\mathbf{p}^t_b(i)}^2}{\text{deg}(i)^2} + \frac{1}{R} \sum_{i\in V_G} \frac{ \mathbf{p}^t_a(i)^2\cdot \mathbf{p}^t_b(i)}{\text{deg}(i)^2}  \\
&= \sum_{i,j\in V_G} \frac{\mathbf{p}^t_a(i) \cdot \mathbf{p}^t_a(j) \cdot \mathbf{p}^t_b(i) \cdot \mathbf{p}^t_b(j)}{\text{deg}(i)\cdot \text{deg}(j)} + \frac{1}{R^2} \sum_{i\in V_G} \frac{\mathbf{p}^t_a(i)\cdot \mathbf{p}^t_b(i)}{\text{deg}(i)^2} \\
 &+ \frac{1}{R} \sum_{i\in V_G} \frac{\mathbf{p}^t_a(i)\cdot \mathbf{p}^t_b(i) \cdot ( \mathbf{p}^t_a(i)+ \mathbf{p}^t_b(i))}{\text{deg}(i)^2}.
\end{align*}
Therefore we get, 
\begin{align}
\text{Var}(Z_{a,b}) & = \mathbb{E}[Z^2_{a,b}]-\mathbb{E}[Z_{a,b}]^2 \nonumber \\
& \leq \sum_{i,j\in V_G} \frac{\mathbf{p}^t_a(i) \cdot \mathbf{p}^t_a(j) \cdot \mathbf{p}^t_b(i) \cdot \mathbf{p}^t_b(j)}{\text{deg}(i)\cdot \text{deg}(j)} + \frac{1}{R^2} \sum_{i\in V_G} \frac{\mathbf{p}^t_a(i)\cdot \mathbf{p}^t_b(i)}{\text{deg}(i)^2} \nonumber\\
 &+ \frac{1}{R} \sum_{i\in V_G} \frac{\mathbf{p}^t_a(i)\cdot \mathbf{p}^t_b(i) \cdot ( \mathbf{p}^t_a(i)+ \mathbf{p}^t_b(i))}{\text{deg}(i)^2} - \left(\sum_{i\in V_G} \frac{\mathbf{p}^t_a(i)\cdot \mathbf{p}^t_b(i)}{\text{deg}(i)}\right)^2 \nonumber\\
 &=  \frac{1}{R^2} \sum_{i\in V_G} \frac{\mathbf{p}^t_a(i)\cdot \mathbf{p}^t_b(i)}{\text{deg}(i)^2} + \frac{1}{R} \sum_{i\in V_G} \frac{\mathbf{p}^t_a(i)^2\cdot \mathbf{p}^t_b(i)}{\text{deg}(i)^2}  + \frac{1}{R} \sum_{i\in V_G} \frac{\mathbf{p}^t_a(i)\cdot \mathbf{p}^t_b(i)^2}{\text{deg}(i)^2} \nonumber\\
&\leq \frac{1}{R^2} \sum_{i\in V_G} \frac{\mathbf{p}^t_a(i)}{\sqrt{\text{deg}(i)}} \cdot \frac{\mathbf{p}^t_b(i)}{\sqrt{\text{deg}(i)}} + \frac{1}{R} \sum_{i\in V_G} \left(\frac{\mathbf{p}^t_a(i)}{\sqrt{\text{deg}(i)}}\right)^2 \cdot \frac{ \mathbf{p}^t_b(i)}{\sqrt{\text{deg}(i)}}  + \frac{1}{R} \sum_{i\in V_G}  \frac{ \mathbf{p}^t_a(i)}{\sqrt{\text{deg}(i)}} \cdot \left(\frac{\mathbf{p}^t_b(i)}{\sqrt{\text{deg}(i)}}\right)^2 \nonumber\\
&\leq \frac{1}{R^2} ||D^{-\frac{1}{2}} \mathbf{p}^t_a||_2\cdot ||D^{-\frac{1}{2}} \mathbf{p}^t_b||_2 + \frac{1}{R} ||D^{-\frac{1}{2}} \mathbf{p}^t_a||_4^2\cdot ||D^{-\frac{1}{2}} \mathbf{p}^t_b||_2 + \frac{1}{R} || D^{-\frac{1}{2}} \mathbf{p}^t_a||_2\cdot ||D^{-\frac{1}{2}} \mathbf{p}^t_b||_4^2 \nonumber\\
 &\leq \frac{1}{R^2} ||D^{-\frac{1}{2}} \mathbf{p}^t_a||_2\cdot ||D^{-\frac{1}{2}} \mathbf{p}^t_b||_2 + \frac{1}{R} ||D^{-\frac{1}{2}} \mathbf{p}^t_a||_2^2\cdot ||D^{-\frac{1}{2}} \mathbf{p}^t_b||_2 + \frac{1}{R} ||D^{-\frac{1}{2}} \mathbf{p}^t_a||_2\cdot ||D^{-\frac{1}{2}} \mathbf{p}^t_b||_2^2 \label{eq:varz}
 \end{align}
 
Notice that all vertices in $S$ are $(\sigma,t)$-good, therefore we get, \[\text{Var}(Z_{a,b}) \leq \frac{\sigma}{R^2}+\frac{2\sigma^{3/2}}{R}\text{.}\]

Then by Chebyshev's inequality, we get,
\[\Pr\left[|Z_{a,b}-\mathbb{E}[Z_{a,b}]|>\frac{\mu_{\text{err}}}{s}\right]<\frac{\text{Var}[Z_{a,b}]}{(\frac{\mu_{\text{err}}}{s})^2}\leq\frac{s^2}{\mu_{\text{err}}^2}\left(\frac{\sigma}{R^2}+\frac{2\sigma^{3/2}}{R}\right)\leq\frac{1}{50s^2}\text{,}\]
where the last inequality follows by our choice of $R$. By the union bound, with probability at least $49/50$, we have for all $a,b\in S$,
\[|(Q^{\top}Q)_{a,b}-((D^{-\frac{1}{2}}M^t)^\top (D^{-\frac{1}{2}}M^t))_{a,b}|=|Z_{a,b}-\mathbb{E}[Z_{a,b}]|\leq\frac{\mu_{\text{err}}}{s}\text{,}\]
which implies $\lVert Q^{\top}Q-(D^{-\frac{1}{2}}M^tS)^{\top}(D^{-\frac{1}{2}}M^tS)\rVert_F\leq\mu_{\text{err}}$. This, in turn, implies
\[|\mu_{k+1}(Q^{\top}Q)-\mu_{k+1}((D^{-\frac{1}{2}}M^tS)^{\top}(D^{-\frac{1}{2}}M^tS))|\leq\mu_{\text{err}}\text{,}\]
due to Weyl's inequality and the fact that the Frobenius norm of a matrix bounds its maximum eigenvalue from above.
\end{proof}

\noindent\textbf{Lemma \ref{lem_CPS4.3} (restated).} For all $0<\alpha<1$, and all $G=(V_G,E_G)$ which is $(k,\varphi_{\text{in}})$-clusterable, there exists $V'_G\subseteq V_G$ with $\text{vol}(V'_G)\geq (1-\alpha) \text{vol}(V_G)$ such that for any $t\geq \frac{2\ln (\text{vol}(V_G))}{\varphi_{\text{in}}^2}$, every $u\in V'_G$ is $\left(\frac{2k}{\alpha \cdot \text{vol}(V_G)},t\right)$-good.

\begin{proof}
Recall that we say that vertex $u$ is \textit{$(\sigma,t)$-good} if $\lVert D^{-\frac{1}{2}}\mathbf{p}^t_u\rVert_2^2\leq\sigma$. We can write $D^{-\frac{1}{2}}\mathbf{p}^t_u$ as
\[D^{-\frac{1}{2}}\mathbf{p}^t_u=D^{-\frac{1}{2}}M^t\mathds{1}_u = D^{-\frac{1}{2}} (D^{\frac{1}{2}} \overline{M}^tD^{-\frac{1}{2}}) \mathds{1}_u = \overline{M}^tD^{-\frac{1}{2}} \mathds{1}_u\] 
Recall from section \ref{sec_prelim} that $1-\frac{\lambda_1}{2}\geq\cdots\geq 1-\frac{\lambda_n}{2}$, are eigenvalues of $\overline{M}$, and $v_1,\ldots,v_n$ are the corresponding orthonormal eigenvectors. We write $D^{-\frac{1}{2}}\mathds{1}_u$ in the eigenbasis of $\overline{M}$ as $D^{-\frac{1}{2}}\mathds{1}_u=\sum_{i=1}^n \alpha_i(u)\cdot v_i$ where $\alpha_i(u) = (D^{-\frac{1}{2}}\mathds{1}_u)^\top v_i = \frac{v_i(u)}{\sqrt{\text{deg}(u)}}  $. Therefore we get, 
\begin{align*}
||D^{-\frac{1}{2}}\mathbf{p}^t_u||_2^2 &= ||\overline{M}^tD^{-\frac{1}{2}} \mathds{1}_u||_2^2 \\
&=\sum_{i=1}^n \alpha_i(u) ^2 \left(1-\frac{\lambda_i}{2}\right)^{2t} \\
&=\sum_{i=1}^{k} \alpha_i(u) ^2 \left(1-\frac{\lambda_i}{2}\right)^{2t} + \sum_{i=k+1}^n \alpha_i(u) ^2 \left(1-\frac{\lambda_i}{2}\right)^{2t} \\
&\leq \sum_{i=1}^{k} \alpha_i(u) ^2 + \left(1-\frac{\lambda_{k+1}}{2}\right)^{2t}  \sum_{i=k+1}^n \alpha_i(u) ^2 \\
&\leq \sum_{i=1}^{k} \alpha_i(u) ^2 +\left(1- \frac{\varphi_{\text{in}}^2}{4}\right)^{2t}  .
\end{align*}
The last inequality follows from Lemma \ref{lem_Cheeger}, and the fact that $\sum_{i=k+1}^n \alpha_i(u) ^2 \leq ||v_i||_2^2 \leq 1$.
We now bound $h(u):=\sum_{i=1}^{k} \alpha_i(u) ^2$. Let $\mathcal{D}$ denote the degree distribution of $G$ (i.e., $\mathcal{D}(v) = \frac{\deg(v)}{\text{vol}(G)} $). Observe that 
\begin{align*}
\mathbb{E}_{\mathcal{D}} \left[ h(u) \right] &=\sum_{u\in V_G}\frac{\text{deg}(u)}{\text{vol}(V_G)}\cdot \left(\sum_{i=1}^k {\alpha_i(u)}^2\right)  = \sum_{u\in V_G}\frac{\text{deg}(u)}{\text{vol}(V_G)}\cdot \left(\sum_{i=1}^k \frac{v_i(u)^2}{\text{deg}(u)}\right) \\
&=\frac{1}{\text{vol}(V_G)} \sum_{i=1}^k \sum_{u\in V_G}  {v_i(u)}^2 = \frac{1}{\text{vol}(V_G)} \sum_{i=1}^k ||v_i||_2^2 = \frac{k}{\text{vol}(V_G)} 
\end{align*}

Thus by Markov's inequality there exists a set $V'_G\subseteq V_G$ with $\text{vol}(V'_G)\geq (1-\alpha) \text{vol}(V_G)$ such that for any $u \in V^\prime_G$, 
\[h(u) \leq \frac{1}{\alpha} \cdot \frac{k}{\text{vol}(V_G)} \text{ .}\]
Thus if $t\geq \frac{2\ln (\text{vol}(V_G))}{\varphi_{\text{in}}^2}$ for any $u \in V^\prime_G$ we have
\[||D^{-\frac{1}{2}}\mathbf{p}^t_u||_2^2 \leq \frac{k}{\alpha \cdot \text{vol}(V_G)} + (1- \frac{\varphi_{\text{in}}^2}{4})^{2t} \leq \frac{2k}{\alpha \cdot \text{vol}(V_G)}\text{,}\]
therefore every $u\in V'_G$ is $\left(\frac{2k}{\alpha \cdot \text{vol}(V_G)},t\right)$-good.
\end{proof}

\noindent\textbf{Lemma~\ref{lem:l2norm-tester} (restated).} 
Let $G=(V_G,E_G)$. Let $a\in V_G$, $\sigma>0$, $0<\delta<1$, and $R\geq \frac{16\sqrt{\text{vol}(G)}}{\delta} $. Let $t\geq1$, and $\mathbf{p}^t_a$ be the probability distribution of the endpoints of a $t$-step random walk starting from $a$. There exists an algorithm, denoted by \textbf{$\ell_2^2$-norm tester}($G,a,\sigma,R$), that outputs accept if $\lVert D^{-\frac{1}{2}} \mathbf{p}^t_a\rVert_2^2\leq \frac{\sigma}{4}$, and outputs reject if $\lVert D^{-\frac{1}{2}} \mathbf{p}^t_a\rVert_2^2>\sigma$, with probability at least $1-\delta$. The running time of the tester is $O(R\cdot t)$.

\begin{proof}
The description of the algorithm \textbf{$\ell_2^2$-norm tester}($G,a,\sigma,r$) is simple:
\begin{enumerate}
\item Run $2R$ random walks of length $t$ starting from $a$.
\item Let $X_{a,r}^i$ be a random variable which is $\frac{1}{\sqrt{\text{deg}(i)}}$ if the $r^{\text{th}}$ random walk starting from $a$, ends at vertex $i$, and $0$ otherwise. 
\item Let $Z$ be a random variable given by
$Z=\frac{1}{R^2} \sum_{i\in V_G} (\sum_{r_1=1}^R X_{a,r_1}^i)(\sum_{r_2=R+1}^{2R} X_{a,r_2}^i)$. 
\item Reject if and only if $\lVert D^{-\frac{1}{2}} \mathbf{p}^t_a\rVert_2^2>\frac{\sigma}{2}$.

By equation \ref{eq:expz}, and inequality \ref{eq:varz}, in the proof of Lemma \ref{lem_matrix_estimation}, we have $\mathbb{E}[Z] = \lVert D^{-\frac{1}{2}} \mathbf{p}^t_a\rVert_2^2$, and  
\begin{align*}
\text{Var}(Z) &\leq \frac{1}{R^2} ||D^{-\frac{1}{2}} \mathbf{p}^t_a||_2\cdot ||D^{-\frac{1}{2}} \mathbf{p}^t_a||_2 + \frac{1}{R} ||D^{-\frac{1}{2}} \mathbf{p}^t_a||_2^2\cdot ||D^{-\frac{1}{2}} \mathbf{p}^t_a||_2 + \frac{1}{R} ||D^{-\frac{1}{2}} \mathbf{p}^t_a||_2\cdot ||D^{-\frac{1}{2}} \mathbf{p}^t_a||_2^2 \\
&\leq \frac{1}{R^2} ||D^{-\frac{1}{2}} \mathbf{p}^t_a||_2^2 + \frac{2}{R} ||D^{-\frac{1}{2}} \mathbf{p}^t_a||_2^3 \\
&= \frac{1}{R^2}\mathbb{E}[Z] + \frac{2}{R}\mathbb{E}[Z]^\frac{3}{2} \text{.}
\end{align*}

Then by Chebyshev's inequality, we get,
\[\Pr\left[|Z-\mathbb{E}[Z]|>\frac{\mathbb{E}[Z]}{2}\right]<\frac{\text{Var}[Z]}{(\frac{\mathbb{E}[Z]}{2})^2}\leq \frac{ \frac{4}{R^2}\mathbb{E}[Z] + \frac{8}{R}\mathbb{E}[Z]^\frac{3}{2}}{\mathbb{E}[Z]^2}=  \frac{4}{R^2\cdot \mathbb{E}[Z] }+ \frac{8}{R\cdot \mathbb{E}[Z]^\frac{1}{2}}.\]

Now Observe that $\mathbb{E}[Z]=\sum_{i=1}^n\frac{\left(\mathbf{p}^t_a(i)\right)^2}{\text{deg}(i)}$, is a convex funtion which is minimized when for all $1\leq i \neq j \leq n$, $\frac{\mathbf{p}^t_a(i)}{\text{deg}(i)}=\frac{\mathbf{p}^t_a(j)}{\text{deg}(i)}=\frac{1}{\text{vol}(V_G)}$. Thus we have
\[\mathbb{E}[Z] \geq \sum_{i=1}^n\frac{\left(\mathbf{p}^t_a(i)\right)^2}{\text{deg}(i)} \geq \sum_{i=1}^n
\frac{ \left( \frac{\text{deg}(i)}{\text{vol}(V_G)} \right)^2 }{\text{deg}(i)} = \frac{1}{\text{vol}(V_G)}.\]

Hence, we get,
\[\Pr\left[|Z-\mathbb{E}[Z]|>\frac{\mathbb{E}[Z]}{2}\right] \leq \frac{4\cdot \text{vol}(V_G)}{R^2 }+ \frac{8\cdot \text{vol}(V_G)^{\frac{1}{2}}}{R} \leq \delta.\]

The last inequality holds since $R\geq \frac{16\sqrt{\text{vol}(G)}}{\delta}$. 

Thus if $\lVert D^{-\frac{1}{2}} \mathbf{p}^t_a\rVert_2^2\leq \frac{\sigma}{4}$, then $\mathbb{E}[Z] \leq \frac{\sigma}{4}$, and  hence, with probability at least $1-\delta$, we have $Z\leq \frac{\sigma}{4}+\frac{\sigma}{8} < \frac{\sigma}{2}$. And if $\lVert D^{-\frac{1}{2}} \mathbf{p}^t_a\rVert_2^2\geq \sigma$, then with probability at least $1-\delta$, we have $Z\geq \frac{\mathbb{E}[Z]}{2} \geq \frac{\sigma}{2}$. Therefore, the tester outputs the correct anwer with probability at least $1-\delta$.
\end{enumerate}

\end{proof}

%% file: mainfile.bbl
\newcommand{\etalchar}[1]{$^{#1}$}
\begin{thebibliography}{CGR{\etalchar{+}}14}

\bibitem[ALM13]{ZhuLM_ICML13}
Zeyuan {Allen Zhu}, Silvio Lattanzi, and Vahab~S. Mirrokni.
\newblock A local algorithm for finding well-connected clusters.
\newblock In {\em {ICML}}, pages 396--404, 2013.

\bibitem[Bol80]{Bollobas80}
B{\'{e}}la Bollob{\'{a}}s.
\newblock A probabilistic proof of an asymptotic formula for the number of
  labelled regular graphs.
\newblock {\em Eur. J. Comb.}, 1(4):311--316, 1980.

\bibitem[Bol88]{Bollobas88}
B{\'{e}}la Bollob{\'{a}}s.
\newblock The isoperimetric number of random regular graphs.
\newblock {\em Eur. J. Comb.}, 9(3):241--244, 1988.

\bibitem[CGR{\etalchar{+}}14]{CzumajGRSSS_RSA14}
Artur Czumaj, Oded Goldreich, Dana Ron, C.~Seshadhri, Asaf Shapira, and
  Christian Sohler.
\newblock Finding cycles and trees in sublinear time.
\newblock {\em Random Struct. Algorithms}, 45(2):139--184, 2014.

\bibitem[CPS15a]{CzumajPS_STOC15}
Artur Czumaj, Pan Peng, and Christian Sohler.
\newblock Testing cluster structure of graphs.
\newblock In {\em {STOC}}, pages 723--732, 2015.

\bibitem[CPS15b]{CzumajPS_arXiv15}
Artur Czumaj, Pan Peng, and Christian Sohler.
\newblock Testing cluster structure of graphs.
\newblock {\em CoRR}, abs/1504.03294, 2015.

\bibitem[CS10]{CzumajS_CPC10}
Artur Czumaj and Christian Sohler.
\newblock Testing expansion in bounded-degree graphs.
\newblock {\em Combinatorics, Probability {\&} Computing}, 19(5-6):693--709,
  2010.

\bibitem[DMS15]{DemboMS_arXiv15}
Amir Dembo, Andrea Montanari, and Subhabrata Sen.
\newblock Extremal cuts of sparse random graphs.
\newblock {\em CoRR}, abs/1503.03923, 2015.

\bibitem[ELR18]{EdenLR_SODA18}
Talya Eden, Reut Levi, and Dana Ron.
\newblock Testing bounded arboricity.
\newblock In {\em {SODA}}, pages 2081--2092, 2018.

\bibitem[ELRS17]{EdenLRS_SIAMJC17}
Talya Eden, Amit Levi, Dana Ron, and C.~Seshadhri.
\newblock Approximately counting triangles in sublinear time.
\newblock {\em {SIAM} J. Comput.}, 46(5):1603--1646, 2017.

\bibitem[ER18]{EdenR_SOSA18}
Talya Eden and Will Rosenbaum.
\newblock On sampling edges almost uniformly.
\newblock In {\em {SOSA}}, pages 7:1--7:9, 2018.

\bibitem[ERS17a]{EdenRS_arXiv17}
Talya Eden, Dana Ron, and C.~Seshadhri.
\newblock On approximating the number of {\textdollar}k{\textdollar}-cliques in
  sublinear time.
\newblock {\em CoRR}, abs/1707.04858, 2017.

\bibitem[ERS17b]{EdenRS_ICALP17}
Talya Eden, Dana Ron, and C.~Seshadhri.
\newblock Sublinear time estimation of degree distribution moments: The
  degeneracy connection.
\newblock In {\em {ICALP}}, pages 7:1--7:13, 2017.

\bibitem[GR00]{GoldreichR_ECCC00}
Oded Goldreich and Dana Ron.
\newblock On testing expansion in bounded-degree graphs.
\newblock {\em Electronic Colloquium on Computational Complexity {(ECCC)}},
  7(20), 2000.

\bibitem[GR02]{GoldreichR_Algorithmica02}
Oded Goldreich and Dana Ron.
\newblock Property testing in bounded degree graphs.
\newblock {\em Algorithmica}, 32(2):302--343, 2002.

\bibitem[GR10]{GonenR_Algorithmica10}
Mira Gonen and Dana Ron.
\newblock On the benefits of adaptivity in property testing of dense graphs.
\newblock {\em Algorithmica}, 58(4):811--830, 2010.

\bibitem[GR11]{GoldreichR_SIAMJC11}
Oded Goldreich and Dana Ron.
\newblock Algorithmic aspects of property testing in the dense graphs model.
\newblock {\em {SIAM} J. Comput.}, 40(2):376--445, 2011.

\bibitem[GS13]{GishbolinerS_ECCC13}
Lior Gishboliner and Asaf Shapira.
\newblock Deterministic vs non-deterministic graph property testing.
\newblock {\em Electronic Colloquium on Computational Complexity {(ECCC)}},
  20:59, 2013.

\bibitem[GT14]{GharanT_SODA14}
Shayan~Oveis Gharan and Luca Trevisan.
\newblock Partitioning into expanders.
\newblock In {\em {SODA}}, pages 1256--1266, 2014.

\bibitem[HJ90]{HornJ}
Roger~A. Horn and Charles~R. Johnson.
\newblock {\em Matrix analysis}.
\newblock Cambridge University Press, 1990.

\bibitem[KKSV17]{KapralovKSV_SODA17}
Michael Kapralov, Sanjeev Khanna, Madhu Sudan, and Ameya Velingker.
\newblock {$1+\Omega(1)$}-approximation to {MAX-CUT} requires linear space.
\newblock In {\em {SODA}}, pages 1703--1722, 2017.

\bibitem[KS11]{KaleS_SIAMJC11}
Satyen Kale and C.~Seshadhri.
\newblock An expansion tester for bounded degree graphs.
\newblock {\em {SIAM} J. Comput.}, 40(3):709--720, 2011.

\bibitem[KVV04]{KannanVV_JACM04}
Ravi Kannan, Santosh Vempala, and Adrian Vetta.
\newblock On clusterings: Good, bad and spectral.
\newblock {\em J. {ACM}}, 51(3):497--515, 2004.

\bibitem[LV13]{LovaszV_CPC13}
L{\'{a}}szl{\'{o}} Lov{\'{a}}sz and Katalin Vesztergombi.
\newblock Non-deterministic graph property testing.
\newblock {\em Combinatorics, Probability {\&} Computing}, 22(5):749--762,
  2013.

\bibitem[NS10]{NachmiasS_IC10}
Asaf Nachmias and Asaf Shapira.
\newblock Testing the expansion of a graph.
\newblock {\em Inf. Comput.}, 208(4):309--314, 2010.

\bibitem[O'D14]{ODonnell}
Ryan O'Donnell.
\newblock {\em Analysis of Boolean Functions}.
\newblock Cambridge University Press, 2014.

\bibitem[PS18]{PengS_SODA18}
Pan Peng and Christian Sohler.
\newblock Estimating graph parameters from random order streams.
\newblock In {\em {SODA}}, pages 2449--2466, 2018.

\bibitem[Ses15]{seshadhri2015simpler}
C~Seshadhri.
\newblock A simpler sublinear algorithm for approximating the triangle count.
\newblock {\em arXiv preprint arXiv:1505.01927}, 2015.

\bibitem[Yos10]{Yoshida_arXiv10}
Yuichi Yoshida.
\newblock Lower bounds on query complexity for testing bounded-degree csps.
\newblock {\em CoRR}, abs/1007.3292, 2010.

\bibitem[Yos11]{Yoshida_CCC11}
Yuichi Yoshida.
\newblock Lower bounds on query complexity for testing bounded-degree csps.
\newblock In {\em {CCC}}, pages 34--44, 2011.

\end{thebibliography}
